\newif\iffigs\figstrue
\documentclass[a4paper,11pt]{article}
\usepackage{latexsym,amssymb,lscape,graphics}
\usepackage{graphicx}        
\usepackage{longtable}
\usepackage{youngtab}
\usepackage{multirow}
\usepackage{color}
\usepackage{slashed,epsfig}
\usepackage{amsfonts}

\newtheorem{definizione}{Definition}[section]

\newcommand{\bd}{\begin{definizione}}
\newcommand{\ed}{\end{definizione}}
\def\Gslat{\relax{\slash\kern-.58em G}}
\def\Fslat{\relax{\slash\kern-.68em F}}
\def\Phislat{\relax{\slash\kern-.65em \Phi}}

\def\IC{\relax\,\hbox{$\inbar\kern-.3em{\rm C}$}}
\def\IG{\relax\,\hbox{$\inbar\kern-.3em{\rm G}$}}
\def\IB{\relax{\rm I\kern-.18em B}}
\def\ID{\relax{\rm I\kern-.18em D}}
\def\IL{\relax{\rm I\kern-.18em L}}
\def\IF{\relax{\rm I\kern-.18em F}}
\def\IH{\relax{\rm I\kern-.18em H}}
\def\II{\relax{\rm I\kern-.17em I}}
\def\IN{\relax{\rm I\kern-.18em N}}
\def\IP{\relax{\rm I\kern-.18em P}}
\def\IQ{\relax\,\hbox{$\inbar\kern-.3em{\rm Q}$}}
\def\bfzero{\relax\,\hbox{$\inbar\kern-.3em{\rm 0}$}}
\def\IK{\relax{\rm I\kern-.18em K}}
\def\IG{\relax\,\hbox{$\inbar\kern-.3em{\rm G}$}}
 \font\cmss=cmss10 \font\cmsss=cmss10 at 7pt
\def\IR{\relax{\rm I\kern-.18em R}}
\def\ZZ{\relax\ifmmode\mathchoice
{\hbox{\cmss Z\kern-.4em Z}}{\hbox{\cmss Z\kern-.4em Z}}
{\lower.9pt\hbox{\cmsss Z\kern-.4em Z}} {\lower1.2pt\hbox{\cmsss
Z\kern-.4em Z}}\else{\cmss Z\kern-.4em Z}\fi}
\def\bfone{\relax{\rm 1\kern-.35em 1}}

\def\inbar{\vrule height1.5ex width.4pt depth0pt}
\def\bfzero{\relax{\rm I\kern-.18em 0}}
\def\bfone{\relax{\rm 1\kern-.35em 1}}

\DeclareFontFamily{U}{rsf}{} \DeclareFontShape{U}{rsf}{m}{n}{
  <5> <6> rsfs5 <7> <8> <9> rsfs7 <10-> rsfs10}{}
\DeclareMathAlphabet\Scr{U}{rsf}{m}{n}

\newcommand{\ft}[2]{{\textstyle\frac{#1}{#2}}}
\def\tilde{\widetilde}

\def\1bar{1\hskip -.275cm -}
\def\2bar{2\hskip -.275cm -}
\def\3bar{3\hskip -.275cm -}

\newsavebox{\uuunit}
\sbox{\uuunit}
                 {\setlength{\unitlength}{0.825em}
                      \begin{picture}(0.6,0.7)
                                      \thinlines
                                      \put(0,0){\line(1,0){0.5}}
                                      \put(0.15,0){\line(0,1){0.7}}
                                      \put(0.35,0){\line(0,1){0.8}}
                                     \multiput(0.3,0.8)(-0.04,-0.02){10}{\rule{0.5pt}{0.5pt}}
                      \end {picture}}

\makeatletter \@addtoreset{equation}{section} \makeatother

\def\bfone{\relax{\rm 1\kern-.35em 1}}

\def\bfone{\relax{\rm 1\kern-.35em 1}}
\font\cmss=cmss10 \font\cmsss=cmss10 at 7pt

\begin{document}
\begin{titlepage}
\begin{center}
\vskip 0.2cm
\vskip 0.2cm {\Large\sc
$2$-branes  with Arnold-Beltrami Fluxes \\
\vskip 0.2cm
from  Minimal $D=7$ Supergravity}\\[1cm]
{\sc
P.~Fr\'e${}^{\; a1,a2,c}$ and  A.S.~Sorin$^{\; b,c}$ }\\[10pt]
{${}^{a1}$\sl\small Dipartimento di Fisica\footnote{Prof. Fr\'e is
presently fulfilling the duties of Scientific Counselor of the
Italian Embassy in the Russian Federation, Denezhnij pereulok, 5,
121002 Moscow, Russia.
\emph{e-mail:} \quad {\small {\tt pietro.fre@esteri.it}}}, Universit\'a di Torino\\${}^{a2}$INFN -- Sezione di Torino \\
via P. Giuria 1, \ 10125 Torino \ Italy\\}
\emph{e-mail:} \quad {\small {\tt fre@to.infn.it}}\\
\vspace{5pt}
{{\em $^{b}$\sl\small Bogoliubov Laboratory of Theoretical Physics and}}\\
{{\em Veksler and Baldin Laboratory of High Energy Physics}}\\
{{\em Joint Institute for Nuclear Research,}}\\
{\em 141980 Dubna, Moscow Region, Russia}~\quad\\
\emph{e-mail:}\quad {\small {\tt sorin@theor.jinr.ru}}\\
\vspace{5pt}
{{\em $^{c}$\sl\small  National Research Nuclear University MEPhI\\ (Moscow Engineering Physics Institute),}}\\
{\em Kashirskoe shosse 31, 115409 Moscow, Russia}~\quad\\
\quad \vspace{6pt}
\vspace{15pt}
\begin{abstract}
We describe this paper as a Sentimental Journey from Hydrodynamics to Supergravity. Beltrami equation in three
dimensions that plays a key role in the hydrodynamics of incompressible fluids has an unsuspected relation with minimal
supergravity in seven dimensions. We show that just $D=7$ supergravity and no other theory with the same field content
but different coefficients in the lagrangian, admits exact two-brane solutions where  Arnold-Beltrami fluxes in the
transverse directions have been switched on. The rich variety of discrete groups that classify the solutions of
Beltrami equation, namely the eigenfunctions of the $\star d$ operator on a three-torus, are by this newly discovered
token injected into the brane world. A new quite extensive playing ground opens up for supergravity and for its dual
gauge theories in three dimensions, where all classical fields and all quantum composite operators will be assigned to
irreducible representations of discrete crystallographic groups $\Gamma$.
\end{abstract}
\end{center}
\end{titlepage}
\tableofcontents \noindent {}
\newpage
\section{Introduction}
The canvas of this paper can be provocatively described as a \textit{Sentimental Journey from Hydrodynamics to
Supergravity}. The main character of this play is a simple first order differential equation written in the XIX century
by the great Italian Mathematician Eugenio Beltrami\cite{beltramus}:  an equation that bears his name and can be cast
in the following modern notation:
\begin{equation}\label{Beltramoide}
    \star \,\mathrm{d} \mathbf{Y}_{[1]} \, = \, \mu \,  \mathbf{Y}_{[1]}
\end{equation}
That above is an eigenvalue problem for a $1$-form $\mathbf{Y}_{[1]}$ and makes sense only on three-manifolds
$\mathcal{M}_3$. If $\mathcal{M}_3$ is compact, the spectrum of the $\star \,\mathrm{d}$ operator is discrete and
encodes topological properties of the manifold. In particular if $\mathcal{M}_3$ is a flat torus $T^3$, all the
spectrum of eigenvalues and eigenfunctions can be constructed with simple algorithms and it can be organized into
irreducible representations of a rich variety of crystallographic groups that were recently explored and classified by
the two of us \cite{arnolderie}. The hydrodynamical viewpoint on eq.(\ref{Beltramoide}) arises from the trivial
observation that a $1$-form $\mathbf{Y}_{[1]}$ is dual to a vector field $ {\mathbf{V}}$ and that any vector field
in three-dimensions can be interpreted as the velocity field of some fluid. This hydrodynamical interpretation of
eq.(\ref{Beltramoide}) is boosted by the existence of a very important theorem proved by V. Arnold
\cite{arnoldorussopapero}: \textit{on compact manifolds $\mathcal{M}_3$, streamlines of a steady flow have a chance of
displaying a chaotic behavior only if the one-form dual to the vector-field of the flow satisfies Beltrami
equation}.
\par
Yet one-forms can also be interpreted as gauge fields and one can  conceive the idea of using the solutions of
eq.(\ref{Beltramoide}) in their primary capacity, namely as ingredients in classical solutions of some gauge-theory.
Due to the strictly euclidian signature of the metric utilized in  eq.(\ref{Beltramoide}) one is naturally led to
imagine that the manifold $\mathcal{M}_3$ is either part of the internal compact variety in a spontaneous
compatification of a higher dimensional theory, typically supergravity, or part of the transverse manifold in a
brane-solution of the same. Ah! Here we are: \textit{flux-branes}! This is the word! Arnold-Beltrami fields can change
their profession and from \textit{flows} they can be turned into \textit{fluxes}. Once the first seed of this change of
perspective is planted the tree grows fast and the idea develops along logical lines. If our target are $p$-branes,
then we need to decide how large is $p$ and our final goal will be the world-volume gauge--theory $\mathcal{GT}_{p+1}$
in $p+1$-dimensions. The lowest reasonable choice is $p=2$, leading to three-dimensional world-volume gauge-theories
that can be Maxwell Chern Simons. The challenging perspective is the following. If we are able to find exact
supergravity solutions of the $2$-brane type that have \textit{Arnold-Beltrami fluxes} in the transverse directions,
then the discrete crystallographic symmetry group $\Gamma$ of the fluxes will be transmitted to the $2$-brane classical
supergravity solution and from the latter to the $\mathcal{GT}_{3}$ on the world volume. This scenario is quite
attractive since it envisages, for the first time, a systematic and rich injection of discrete group symmetries into
the brane--world: the journey from hydrodynamics to supergravity starts being quite interesting if not sentimental! In
order to proceed we have to count dimensions carefully. Three dimensions are occupied by the world volume, another
three by a $T^3$ torus transverse to the brane. This makes already six. Hence we have to look at six-dimensional
supergravity or higher. A guiding line comes from another constraint. If we want a two-brane, in the
bosonic spectrum of the considered supergravity there should be  a gauge three-form $\mathbf{B}^{[3]}$ that will couple to the
world-volume of the brane. Then six-dimensional supergravity is not sufficient since it contains only gauge two-forms
$\mathbf{B}^{[2]}$ (for $D=6$ supergravities see
\cite{romans6},\cite{Andrianopoli:2001rs},\cite{D'Auria:2002fh},\cite{D'Auria:2000ay},\cite{D'Auria:2000ad}). The first
favorable case is $D=7$: here we have minimal supergravity, that contains 16 supercharges and it is usually named
$\mathcal{N}=2$ since the 16 supercharges are arranged into a pair of pseudo-Majorana spinors. Supergravities in seven
dimensions were constructed (up to four fermion terms) in the mid eighties in several papers \cite{PvNT},\cite{SalamSezgin},
\cite{bershoffo1},\cite{salametal},\cite{Pilch:1984xy},\cite{Pernici:1984xx}.
The Poincar\'e (ungauged) version of the minimal theory was independently   constructed
 by  Townsend and van Nieuwenhuizen \cite{PvNT} and by Salam and Sezgin  \cite{SalamSezgin} in two different
formulations that use respectively a three-form gauge field $\mathbf{B}^{[3]}_{\mu\nu\rho}$ and a two-form gauge
field $\mathbf{B}^{[2]}_{\mu\nu}$, in addition to the graviton $g_{\mu\nu}$, the gravitino $\Psi_{A|\mu}^{\alpha}$
($\alpha=1,\dots,8$, $\mu\,=\,0,1,\dots,6$, $A\, = \, 1,2$), the gravitello $\chi_{A}^\alpha$, three gauge fields
$\mathcal{A}^\Lambda_\mu$ ($\Lambda \, = \, 1,2,3$) and the dilaton $\phi$, that are common to both formulations.
From the on-shell point of view the number of degrees of freedom described by either
$\mathbf{B}^{[3]}_{\mu\nu\rho}$ or $\mathbf{B}^{[2]}_{\mu\nu}$ is the same and the two types of gauge
fields are electric-magnetic dual to each other.
\par
This field content constitutes excellent news for our $2$-brane plans.
Either in an electric or in a magnetic formulation we have at our disposal a $\mathbf{B}^{[3]}$ form which
can couple to the $2$-brane world volume. In addition the triplet of gauge fields $\mathcal{A}^\Lambda$
is a very encouraging starting point for Arnold-Beltrami fluxes. Indeed the theory has a global
symmetry $\mathrm{SO(3)}$ under which the three vector fields $\mathcal{A}^\Lambda$ transform in
the defining representation (which in this case coincides with the adjoint): hence they are specially
prepared to be identified with triplets of Arnold-Beltrami one-forms transforming in any three dimensional
representation of any discrete subgroup $\Gamma \subset \mathrm{SO(3)}$. If such fluxes can be consistently
switched on within the setup of a $2$-brane solution, such solution will be invariant under $\Gamma$ and this
symmetry will descend to the gauge theory on the world volume.
\par
There is only one question that remains open: what about the $7$th dimension?
At first sight it seems a sort of uninvited guest that hangs around without purpose,
yet we know that in supergravity and supersymmetry nothing is ever superfluous,
nothing sits there without a deep reason: on the contrary, like in a well built swiss watch,
all the wheels, larger or smaller are equally essential to the proper working of the whole thing.
A suggestion comes from our previous experience with fractional $D3$-branes \cite{noialtrilast},
\cite{nuovesoluzie},\cite{Billo.M}.
 \par
Considering in particular the smooth realization \cite{noialtrilast} of  the fractional
D3-brane as a $3$-brane solution of  $D=10$ type IIB supergravity  in which the transverse space
to the brane world--volume is of the form:
\begin{equation}\label{trasversone}
    \mbox{transverse space to the D3-brane} \, = \, \mbox{ALE}_4 \, \times \, \mathbb{R}^2
\end{equation}
we see something similar to what we are faced with in $D=7$. The fractional D3-brane is a flux-brane
where the doublet of gauge two-forms $\mathbf{B}^\Lambda_{[2]}$ develop geometrical fluxes,
being identified with linear combinations of the non trivial cohomology two-cycles $\omega^I_{[2]}$
that leave on the four-dimensional $\mathrm{ALE}$-space:
$\mathbf{B}^\Lambda_{[2]}\, = \, \gamma^\Lambda_I \, \omega^I_{[2]}$.
At first sight also in this case the extra flat dimensions associated with $\mathbb{R}^2$
seem unnecessary spectators. Actually this is not true.
The coefficients $\gamma^\Lambda_I$ introduced one line above have to be functions of
the extra coordinates $x,y$ on $\mathbb{R}^2$ and using the natural complex structure
$z\, = \, x+{\rm i}y$ they happen to be holomorphic functions $\gamma^\Lambda_I \, = \,\gamma^\Lambda_I(z)$
of the coordinate $z$. These holomorphic functions play an essential role in establishing the overall
supergravity solution.
\par
\begin{figure}[!hbt]
\begin{center}
\iffigs
\includegraphics[height=80mm]{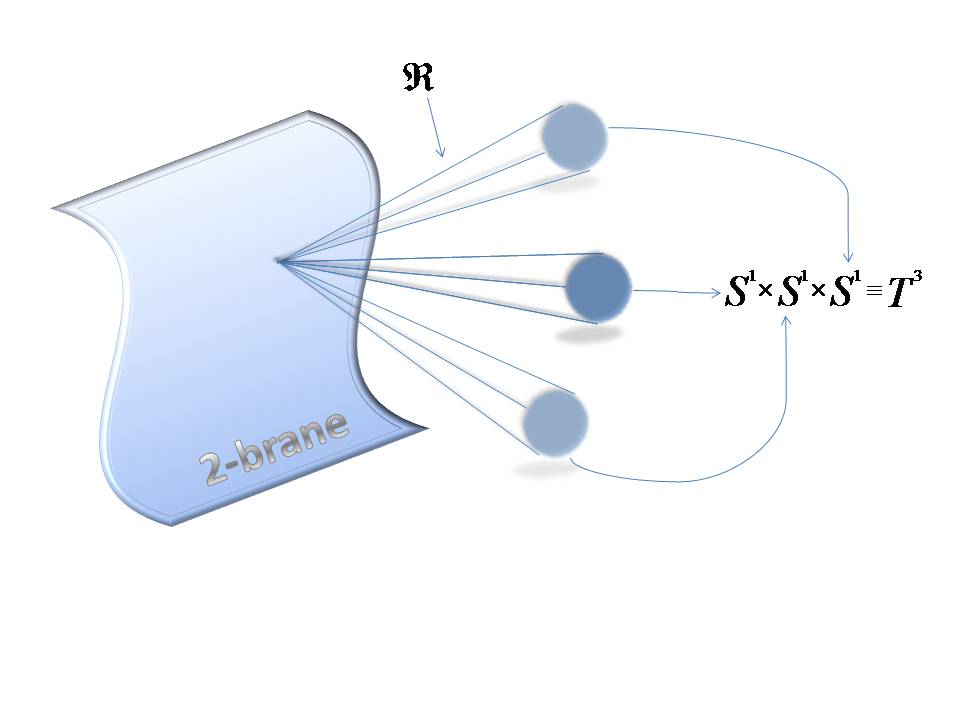}
\else
\end{center}
 \fi
\caption{\it  A metaphoric view of the two-brane structure in $D=7$. The transverse space to the
two-brane world-volume is the direct product of a three torus $T^3$, homeomorphic to three circles
$\mathbb{S}^1\times \mathbb{S}^1 \times \mathbb{S}^1$ with a straight line $\mathbb{R}$.
The Arnold Beltrami fluxes live on $\mathrm{T}^3$ but are embedded into supergravity
with coefficients that have a predetermined exponential dependence from the coordinate $U$ of $\mathbb{R}$.
In some sense $U$ measures the distance from the $2$-brane that is a boundary for $D=7$ space-time.} \label{metafora}
 \iffigs
 \hskip 1cm \unitlength=1.1mm
 \end{center}
  \fi
\end{figure}
Hence, \textit{mutatis mutandis},  we are lead to consider a similar situation where the
transverse space to our candidate $2$-brane is the following one:
\begin{equation}\label{trasversoneDue}
    \mbox{transverse space to the 2-brane} \, = \, T^3 \, \times \, \mathbb{R}
\end{equation}
The torus $T^3$ is the compact manifold which replaces the $\mathrm{ALE}$-space and the
Arnold-Beltrami one-forms $\mathbf{Y}^I_{[1]}$, leaving on the torus, play the role
played by the cohomology two-cycles leaving on $\mathrm{ALE}$.
The triplet of gauge fields $\mathcal{A}^\Lambda$ play in $D=7$
the same role that was played in $D=10$ by the doublet of two forms $\mathbf{B}^\Lambda_{[2]}$,
namely they develop fluxes being identified with linear combinations of the Arnold-Beltrami one-forms:
\begin{equation}\label{identifiuciona}
    \mathcal{A}^\Lambda \, = \, \gamma^\Lambda_I \, \mathbf{Y}^I_{[1]}
\end{equation}
The catch is that the coefficients $\gamma^\Lambda_I $ have to be functions of the unique coordinate $U$ on $\mathbb{R}$:
\begin{equation}\label{spilutta}
    \gamma^\Lambda_I \, = \, \gamma^\Lambda_I(U)
\end{equation}
It remains to be understood which functional condition on the $\gamma^\Lambda_I(U)$ replaces the
holomorphicity pertaining to the D3-brane case. We will see that the $\gamma^\Lambda_I(U)$ are constrained
to have an exponential dependence:
\begin{equation}\label{esponato}
    \gamma^\Lambda_I(U) \, = \, e^{\mu \, U} \, \mathcal{E}^{\Lambda}_{\phantom{\Lambda}I}
\end{equation}
where $\mu$ is the eigenvalue of $\star \, \mathrm{d}$-operator in Beltrami equation (\ref{Beltramoide})
and $\mathcal{E}^{\Lambda}_{\phantom{\Lambda}I}$ denotes a constant embedding matrix whose group-theoretical
structure we discuss in later sections. The overall conception of the proposed $2$-branes with Arnold-Beltrami
fluxes is graphically and metaphorically summarized in  fig.\ref{metafora}.
\par
All what we have discussed so far materializes into a definite ansatz for all the bosonic
fields of minimal $D=7$ supergravity and the question is whether such an ansatz does or does
not satisfy the field equations of supergravity. In full analogy with the case of the D3-brane
we expect that all the field equations should reduce, upon use of the advocated ansatz,
to a unique differential equation of the following form:
\begin{equation}\label{coltello}
    \Box_{T^3 \times \mathbb{R}} \, H(U,\mathbf{X}) \, = \, \mathbf{j}(U,\mathbf{X})
\end{equation}
where $H(U,\mathrm{X})$ is a scalar function of the transverse coordinates that enters the $D=7$ brane-like metric:
\begin{equation}\label{matrullona}
    ds^2 \,=\,   H(U,\mathbf{X})^{-\frac{8}{5 \, \Delta}}\, d\xi^\mu\otimes d\xi^\nu \, \eta_{\mu\nu}
    \, - \, H(U,\mathbf{X})^{\frac{12}{5\, \Delta}}
  \, \left(dU^2\,+ \, dX^2 \, + \, dY^2 \, + \, dZ^2 \right)
\end{equation}
The source function $\mathbf{j}(U,\mathbf{X})$ appearing in eq.(\ref{coltello}) should be uniquely
defined, as in the D3-brane case by the fluxes and should vanish at zero fluxes. In that case $H(U,\mathbf{X})$
is a harmonic function.
\par
In the present paper we show that the above expectations are
indeed fulfilled and that $2$-branes with Arnold-Beltrami fluxes
are exact solutions of minimal $D=7$ supergravity.
Actually we show something even stronger.
While $2$-brane-solutions without fluxes do exist for any bosonic theory that has the same field content
as minimal supergravity but not necessarily the specific coefficients imposed by supersymmetry,
Arnold-Beltrami flux $2$-branes are a specific feature of supergravity.
All the field equations reduce to equation (\ref{coltello}) \textit{if and only}
if the lagrangian coefficients are in the precise ratios predicted
by the supergravity construction of \cite{PvNT},\cite{bershoffo1}.
This implies, in particular, that $\Delta \, = \, 4$ in eq. (\ref{matrullona}).
\par
Hence quite unexpectedly Beltrami equation (\ref{Beltramoide}) has a hidden and deep relation with supersymmetry that
is unveiled by the existence of the  flux-branes presented in this paper. The injection of discrete symmetries into the
brane--world turns out to be a successful operation and the \textit{Sentimental Journey from Hydrodynamics to
Supergravity} has a happy starting. However, we must stress that  Rev. Yorick has just disembarked in Calais and that
he has only exchanged snuff boxes with his monk acquaintance: the road to Paris and to the South is still long. We need
to derive equations for the Killing spinors and to determine the  supersymmetries preserved by the flux-branes, we need
to discuss their fate in the gauged version of the theory and their analogue in curved backgrounds. All that requires a
firm control on the lagrangian, the transformation rules and the gaugings.
\par
The gauging of $D=7$ minimal supergravity was also independently considered both in \cite{PvNT} and in
\cite{SalamSezgin}. The coupling of minimal $D=7$ supergravity to $n$ vector multiplets was constructed by Bergshoeff
et al in \cite{bershoffo1} on the basis of the two-form formulation and shown to be founded on the use of the coset
manifold:
\begin{equation}\label{targuccio}
    \mathcal{M}_{3n+1} \, = \, \mathrm{SO(1,1)} \,\times \,\frac{\mathrm{SO(3,n)}}{\mathrm{SO(3)\times SO(n)}}
\end{equation}
as scalar manifold that encodes the spin zero degrees of freedom of
the theory.
\par
In all the quoted references the construction was done using the Noether coupling procedure, up to four-fermion terms in the
Lagrangian and up to two-fermion and three-fermion terms in the transformation rules.
Correspondingly the on-shell closure of the
supersymmetry algebra was also checked only up to such terms.
Furthermore the possible addition of new topological interaction terms was proposed but never proved.
\par
In consideration of  the renewed interest in this particular supergravity theory in relation with the Arnold-Beltrami
flux-branes, a separate collaboration involving one of us \cite{d7collabo} is presently  reconsidering the
reconstruction of minimal $D=7$ supergravity and its gauging in the approach based on Free Differential Algebras and
rheonomy (for reviews see \cite{castdauriafre} and also the second volume of \cite{pietrobook}). The goal is that of
clarifying the algebraic structure underlying the theory and perfectioning its construction to all fermion orders. The
issue, as we will demonstrate, is particularly relevant in connection with gauging since there the FDA structure
becomes essential and comes into contact with the formalism of the embedding tensor \cite{embedtensor1,embedtensor2}.
\par
We postpone the discussion of Killing spinors and of the preserved supersymmetries to the moment when the results of
\cite{d7collabo} will be available.
\subsection{Organization of the paper}
\label{organisazia} Since some of the concepts, of the definitions and of the mathematical techniques heavily used in
\cite{arnolderie} are not common in Particle Physics and Supergravity, we devote section \ref{elementi} to a
comprehensive summary of these topics, introducing here and there in our presentation a change of perspective   which
takes into account the different goals pursued by this paper. Particulary important for the understanding of what will
follow is sect. \ref{ottostruttura} and its subsection \ref{universalone}. In the latter we recall the notion of the
Universal Classifying Group which has been invented by the two us in \cite{arnolderie} and plays a key role both in the
Hydrodynamical and in the Supergravity interpretation of Beltrami equation.
\par
Sect.\ref{classificazia} summarizes the classification of Arnold-Beltrami one-forms obtained in \cite{arnolderie}
showing its bearing on the issue of flux-branes.
\par
Sect. \ref{giffone192} contains a detailed discussion of the space group $\mathrm{GF_{192}}\subset \mathrm{G_{1536}}$
which is
 the $\Gamma$-symmetry group of the explicit examples of Arnold-Beltrami flux branes presented in this paper. Let us also stress that this section contains the precise discussion of how the discrete symmetry groups of Beltrami flows are transmitted to supergravity.
  \par
Sect.s \ref{k100gf192},\ref{k100gs24},\ref{k200d7},\ref{k110d9} present the explicit construction of the triplets of
Arnold-Beltrami one-forms utilized in the afore mentioned examples. The transformation of these triplets under
$\mathrm{GF_{192}}$ or one of its subgroups are carefully discussed here.
\par
Sect. \ref{twobranastoria} and \ref{compaTPvN} contain the core result of this paper announced in the introduction,
namely the derivation of the $2$-brane solutions of minimal $D=7$ supergravity having  Arnold Beltrami fluxes in the
transverse space.
\par
Sect. \ref{concludone} contains our conclusions.
\par
In the appendices we provide tables of the conjugacy classes of the group $\mathrm{GF_{192}}$ and of its subgroup
$\mathrm{GS_{24}}$.
\par
A part of the material presented in this paper repeats that presented in \cite{arnolderie}. We did these repetitions to
make the present  paper self-consistent both conceptually and technically. Furthermore we have discarded all those
items of \cite{arnolderie} that are not pertinent to our present goals and that  might even be source of confusion in
the present interpretation of Arnold-Beltrami one forms.
\section{Crystallographic Lattices, the Torus $T^3$ and Discrete Groups}
\label{elementi}
As we explained in the introduction, we are interested in $2$-brane  solutions of a gravitational gauge theory,
identifiable with minimal $D=7$ Supergravity, where the vector fields develop fluxes that are Beltrami fields on a
three torus transverse to brane world-volume:in this way the discrete symmetry groups of such fluxes will be
transmitted to the brane solution and to the brane gauge- theory. In the present section, in order to fix notations and
to clarify our working setup,  we summarize some essential facts about  crystallographic lattices and about the
algorithmic construction of Beltrami fields on the three-tours; in this  we closely follow our previous paper
\cite{arnolderie}.
\par
Topologically the three torus is  defined as the product of three
circles, namely:
\begin{equation}\label{tritorustop}
    \mathrm{T}^3 \, \equiv \, \mathbb{S}^1 \, \times \, \mathbb{S}^1  \, \times \, \mathbb{S}^1  \, \equiv \,
    \frac{\mathbb{R}}{\mathbb{Z}} \, \times \, \frac{\mathbb{R}}{\mathbb{Z}}\, \times \, \frac{\mathbb{R}}{\mathbb{Z}}
\end{equation}
Alternatively we can define the three-torus by modding
$\mathbb{R}^3$ with respect to a three dimensional lattice. In this
case the three-torus comes automatically equipped with a flat
constant metric:
\begin{equation}\label{metricT3}
     \mathrm{T}^3_g \, = \, \frac{\mathbb{R}^3}{\Lambda}
\end{equation}
According to (\ref{metricT3}) the  flat Riemaniann space
$\mathrm{T}^3_g$ is defined as the set of equivalence classes with
respect to the following equivalence relation:
\begin{equation}\label{equivalenza}
     {\mathbf{r}}^\prime \, \sim \,  {\mathbf{r}} \quad \mbox{iff} \quad  {\mathbf{r}}^\prime \,
     - \,  {\mathbf{r}} \, \in \, \Lambda
\end{equation}
The metric (\ref{gmunu}) defined on $\mathbb{R}^3$ is inherited by
the quotient space and therefore it endows  the topological torus
(\ref{tritorustop}) with a flat Riemaniann structure. Seen from
another point of view the space of flat metrics on  $\mathrm{T}^3$
is just the coset manifold $\mathrm{SL(3,\mathbb{R})}/\mathrm{O(3)}$
encoding all possible symmetric matrices, alternatively all possible
space lattices, each lattice being spanned by an arbitrary triplet
of basis vectors (\ref{sospirone}). So let us consider the standard
$\mathbb{R}^3$ manifold and introduce a basis of three  linearly
independent 3-vectors that are not necessarily orthogonal to each
other and of equal length:
\begin{equation}\label{sospirone}
     {\mathbf{w}}_\mu \, \in \, \mathbb{R}^3 \quad \mu \, = \, 1, \dots \, 3
\end{equation}
Any vector in $\mathbb{R}$ can be decomposed along such a basis and
we have: $ {\mathbf{r}} \, = \,  r^\mu {\mathbf{w}}_\mu$. The
flat (constant) metric on $\mathbb{R}^3$ is defined by:
\begin{equation}\label{gmunu}
    g_{\mu\nu} \, = \, \langle   {\mathbf{w}}_\mu \, , \,   {\mathbf{w}}_\nu \rangle
\end{equation}
where $\langle \, ,\, \rangle$ denotes the standard euclidian scalar
product. The space lattice $\Lambda$ consistent with the metric
(\ref{gmunu}) is the free abelian group (with respect to sum)
generated by the three basis vectors (\ref{sospirone}), namely:
\begin{equation}\label{reticoloLa}
 \mathbb{R}^3 \, \ni \,    {\mathbf{q}}  \, \in \, \Lambda \, \Leftrightarrow \,  {\mathbf{q}}
 \, = \, q^\mu \,  {\mathbf{w}}_\mu \quad \mbox{where} \quad q^\mu \, \in \, \mathbb{Z}
\end{equation}
The momentum lattice is the dual lattice $\Lambda^\star$ defined by
the property:
\begin{equation}\label{reticoloLastar}
    \mathbb{R}^3 \, \ni \,    {\mathbf{p}}  \, \in \, \Lambda^\star \, \Leftrightarrow \,
    \langle  {\mathbf{p}} \, , \,  {\mathbf{q}}\rangle \, \in \, \mathbb{Z} \quad \forall \,
      {\mathbf{q}}\, \in \, \Lambda
\end{equation}
A basis for the dual lattice is provided by a set of three
\textit{dual vectors} $ {\mathbf{e}}^\mu$ defined by the
relations\footnote{In the sequel for the scalar product of two
vectors we utilize also the equivalent shorter notation
$ {\mathbf{a}}\, \cdot  {\mathbf{b}} \, = \, \langle
 {\mathbf{a}}\, \cdot  {\mathbf{b}}\rangle $}:
\begin{equation}\label{dualvecti}
    \langle  {\mathbf{w}}_\mu \, , \,  {\mathbf{e}}^\nu \rangle \, = \, \delta^\nu_\mu
\end{equation}
so that
\begin{equation}\label{pcompi}
    \forall \,  {\mathbf{p}} \, \in \, \Lambda^\star \quad  {\mathbf{p}} \, =
    \, p_\mu \,  {\mathbf{e}}^\mu \quad \mbox{where } \quad p_\mu \, \in \, \mathbb{Z}
\end{equation}
Every lattice $\Lambda$ yields a metric $g$ and every metric $g$
singles out an isomorphic copy $\mathrm{SO_g(3)}$ of the continuous
rotation group $\mathrm{SO(3)}$, which leaves it invariant:
\begin{equation}\label{copiaso3}
    M \, \in \, \mathrm{SO_g(3)} \quad \Leftrightarrow \quad M^T \, g \, M \, = \, g
\end{equation}
By definition $\mathrm{SO_g(3)}$ is  the conjugate of the standard
$\mathrm{SO(3)}$ in $\mathrm{GL(3,\mathbb{R})}$:
\begin{equation}\label{Sconiugo}
    \mathrm{SO_g(3)} \, = \, \mathcal{S} \, \mathrm{SO(3)} \, \mathcal{S}^{-1}
\end{equation}
with respect to the matrix $\mathcal{S}\,  \in
\,\mathrm{GL(3,\mathbb{R})}$ which reduces the metric $g$ to the
Kronecker delta:
\begin{equation}\label{Smatrucca}
    \mathcal{S}^T \, g \, \mathcal{S} \, = \, \mathbf{1}
\end{equation}
Notwithstanding this a generic lattice $\Lambda$  is not invariant
with respect to any proper subgroup of the rotation group
$\mathrm{G} \, \subset \, \mathrm{SO_g(3)}\, \equiv \,
\mathrm{SO(3)}$. Indeed by invariance of the lattice one understands
the following condition:
\begin{equation}\label{GruppoReticolo}
    \forall \, \gamma \, \in \, \mathrm{G} \quad \mbox{and} \quad \forall \,
     {\mathbf{q}}\, \in \, \Lambda \,\, : \quad \quad \gamma\,\cdot \,  {\mathbf{q}} \, \in \, \Lambda
\end{equation}
Lattices that have a non trivial symmetry group $\mathrm{G} \subset
\mathrm{SO(3)}$ are those relevant to Solid State Physics and
Crystallography. There are 14 of them grouped in 7 classes  that
were already classified in the XIX century by Bravais. The symmetry
group $\mathrm{G}$ of each of these Bravais lattices $\Lambda_B$ is
necessarily one of the well known  finite  subgroups of the
three-dimensional rotation group $\mathrm{O(3)}$. In the language
universally adopted by Chemistry and Crystallography for each
Bravais lattice $\Lambda_B$ the corresponding invariance group
$\mathrm{G_B}$ is named the \textit{Point Group}. For purposes
different from our present one, the point group  can be taken as the
lattice invariance subgroup within $\mathrm{O(3)}$ that, besides
rotations, contains also improper rotations and reflections. Since
we are interested in Beltrami equation, which is covariant only
under proper rotations, of interest to us are only those point
groups that are subgroups of $\mathrm{SO(3)}$.
\par
According to a standard nomenclature the $7$ classes of Bravais
lattices are respectively named \textit{Triclinic, Monoclinic,
Orthorombic, Tetragonal, Rhombohedral, Hexagonal and Cubic}. Such
classes are specified by giving the lengths of the basis vectors
$ {\mathbf{w}}_\mu$ and the three angles between them, in other
words, by specifying the 6 components of the metric (\ref{gmunu}).
\subsection{The proper Point Groups}
\label{puntigruppi} 
\begin{figure}[!hbt]
\begin{center}
\iffigs
\includegraphics[height=70mm]{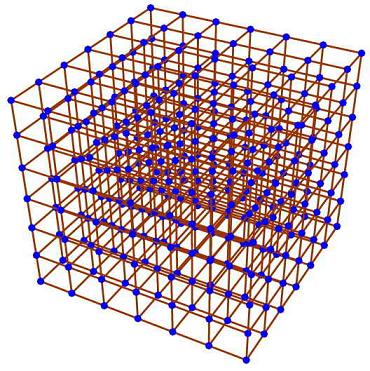}
\else
\end{center}
 \fi
\caption{\it  A view of the self-dual cubic lattice}
\label{cubicogenerale}
 \iffigs
 \hskip 1cm \unitlength=1.1mm
 \end{center}
  \fi
\end{figure}

Restricting one's attention to proper rotations, the proper point groups that appear in the $7$
lattice classes are either the cyclic groups $\mathbb{Z}_h$ with $h=2,3,4$ or the dihedral groups $D_h$ with $h=3,4,6$
or the tetrahedral group $\mathrm{T}$ or the octahedral  group $\mathrm{O_{24}}$. Here we restrict our attention to
lattice with the largest possible Point Group, namely to  Cubic Lattice with $\mathrm{O_{24}}$ symmetry (see
fig.\ref{cubicogenerale}).
\par
\subsection{Group Characters}
\label{gruppicaratteri} A fundamental ingredient for our present goals and for those that were pursued in
\cite{arnolderie}  are the characters of the point group  and of other classifying groups that have emerged in the
constructions we performed in \cite{arnolderie}.
\par
Given a finite group $\mathrm{G}$, according to standard theory and
notations \cite{mieibukki} one defines its order and the order of
its conjugacy classes as follows:
\begin{eqnarray}\label{ordinari}
    g & = & \left|\mathrm{G} \right| \, = \, \mbox{$\#$ of group elements} \nonumber\\
    g_i & = & \left|\mathcal{C}_i \right| \, = \, \mbox{$\#$ of group
    elements in the conjugacy class $\mathcal{C}_i$} \quad i\, = \, i,\dots,r
\end{eqnarray}
If there are $r$ conjugacy classes one knows from first principles
that there are exactly $r$ inequivalent irreducible representation
$D^{\mu}$ of dimensions $n_\mu \, = \, \mbox{dim} \, D^\mu$, such
that:
\begin{equation}\label{dimensiali}
    \sum_{\mu \, = \, 1}^r \, n_\mu^2 \, = \, g
\end{equation}
For any reducible or irreducible representation of dimension $d$:
\begin{equation}\label{RRepra}
    \forall \, \gamma \, \in \, \mathrm{G} \quad : \gamma \, \rightarrow \,
    \quad \mathfrak{R}\left[\gamma\right]  \, \in \, \mbox{Hom}\left[\mathbb{R}^d \, ,\, \mathbb{R}^d\right]
\end{equation}
the character vector is defined as:
\begin{equation}\label{caratdefi}
    \chi^{\mathfrak{R}}\, = \, \left\{ \mbox{Tr}\, \left(\mathfrak{R}\left[\gamma_1\right] \right)
     \, , \, \mbox{Tr}\, \left(\mathfrak{R}\left[\gamma_2\right] \right)\, , \,\dots ,\,
     \mbox{Tr}\, \left(\mathfrak{R}\left[\gamma_r\right] \right)\right \} \, , \quad \gamma_i \, \in \, \mathcal{C}_i
\end{equation}
The choice of a representative $\gamma_i$ within each conjugacy
class $\mathcal{C}_i$ is irrelevant since all representatives have
the same trace. In particular one can calculate the characters of
the irreducible representations:
\begin{equation}\label{fundamentalcarat}
   \chi^\mu \, = \,  \chi\left[D^\mu\right]\, \, = \, \left\{ \mbox{Tr}\, \left(D^\mu\left[\gamma_1\right] \right) \, , \, \mbox{Tr}\, \left(D^\mu\left[\gamma_2\right] \right)\, , \,\dots \, , \,\mbox{Tr}\, \left(D^\mu\left[\gamma_r\right] \right)\right \} \, , \quad \gamma_i \, \in \, \mathcal{C}_i
\end{equation}
that are named \textit{fundamental characters} and constitute the
\textit{character table}. We stick to the widely adopted convention
that the first conjugacy class is that of the identity element
$\mathcal{C}_1 \, = \, \left\{\mathbf{e}\right\}$, containing only
one member. In this way the first entry of the character vector is
always the dimension $d$ of the considered representation. In the
same way we order the irreducible representation starting always
with the identity one dimensional representation which associates to
each group element simply the number $1$. It is well known that for
any finite group $\mathrm{G}$, the character vectors satisfy the
following two fundamental relations:
\begin{equation}\label{idechar1}
    \sum_{\mu \, = \, 1}^r \, \chi_i ^\mu \, \chi_j^\mu \, = \, \frac{g}{g_i} \, \delta_{ij} \quad ; \quad
    \sum_{i \, = \, 1}^r \, g_i \,\chi_i ^\mu \, \chi_i^\nu \, = \, g \, \delta^{\mu\nu}
\end{equation}
Utilizing these identities one can immediately retrieve the
decomposition of any given reducible representation $\mathcal{R}$
into its irreducible components. Suppose that the considered
representation is the following direct sum of irreducible ones:
\begin{equation}\label{multippi}
    \mathfrak{R} \, = \, \oplus_{\mu =1}^r \, a_\mu \, D^\mu
\end{equation}
Where $a_\mu$ denotes the number of times the irrep $D^\mu$ is
contained in the direct sum and it is named the
\textit{multiplicity}. Given the character vector of any considered
representation $\mathfrak{R}$ the vector of its multiplicities is
immediately obtained by use of (\ref{idechar1}):
\begin{equation}\label{multipvector}
    a_\mu \, = \, \frac{1}{g} \, \sum_{i}^r \, g_i \chi_i^{\mathfrak{R}} \, \chi_i^\mu
\end{equation}
Furthermore one can construct the projectors onto the invariant
subspaces $a_\mu \, D^\mu$ by means of another classical formula
that we will extensively use in the sequel:
\begin{equation}\label{proiettori}
    \Pi^\mu_{\mathfrak{R}} \, = \, \frac{g_i}{g} \, \sum_{k=1}^r \, \chi^\mu_k \,
    \sum_{\ell \,=\, 1}^{g_k} \, \underbrace{\mathfrak{R}\left[{\gamma_\ell}\right]}_{\gamma_\ell \in \, \mathcal{C}_k}
\end{equation}
\subsection{The spectrum of the $\star \mathbf{d}$ operator on $\mathrm{T}^3$ and Beltrami equation}
\label{fantasmabeltrami} As we explained in the introduction, the main ingredient of the present paper are the Beltrami
one-forms defined over the three-torus $\mathrm{T}^3$. By definition these  are eigenstates of the $\star_g \mathrm{d}$
operator, namely of solutions of the following equation:
\begin{eqnarray}
    \star_g \mathrm{d} \mathbf{Y}^{(n;I)} &=& \, \mu_{(n)} \, \mathbf{Y}^{(n;I)}   \label{formaduale}
\end{eqnarray}
where $\mathrm{d}$ is the exterior differential, and $\star_g$ is
the Hodge-duality operator which, differently from the exterior
differential, can be defined only with reference to a given metric
$g$. By $\mathbf{Y}^{(n;I)}$ we denote a one-form:
\begin{equation}\label{omegas}
    \mathbf{Y}^{(n;I)} \, = \, {\mathbf{Y}}^{(n;I)}_i \,dX^i \quad ; \quad (i \, = \, 1,2,3)
\end{equation}
where the composite index $(n;I)$  makes reference to the quantized
eigenvalues $\mu_{(n)}$ of the $\star_g \mathrm{d}$ operator (ordered
in increasing magnitude $|\mu_{(n)}|$) and to a basis of the
corresponding eigenspaces
\begin{equation}\label{superpongomega}
    \star_g \mathrm{d} \mathbf{Y}^{(n)}\, = \, \, \mu_{(n)} \,  \mathbf{Y}^{(n)}
    \quad \Rightarrow \quad  \mathbf{Y}^{(n)}\, = \,
    \sum_{I=1}^{d_n} \,c_I \,  \mathbf{Y}^{(n;I)}
\end{equation}
the symbol $d_n$ denoting the degeneracy of $|\mu_{(n)}|$ and $c_I$
being constant coefficients. Indeed, since $T^3$ is a compact
manifold, the eigenvalues $\mu_{(n)}$ form a discrete set. We recall
the general procedure introduced in \cite{arnolderie} to construct
the eigenfunctions of $\star_g \mathrm{d}$ and to determine  their
degeneracies. In tensor notation, equation (\ref{formaduale}) has
the following appearance:
\begin{equation}\label{tensoBeltra}
    \frac{1}{2} \, g_{ij} \, \epsilon^{jk\ell} \partial_k \mathbf{Y}_\ell \, = \, m \, \mathbf{Y}_i
\end{equation}
The equation written above is named Beltrami equation since it was
already considered by the great italian mathematician Eugenio
Beltrami in 1881 \cite{beltramus}, who presented one of its periodic
solutions previously constructed by  Gromeka in 1881\cite{balubbo}.
In the case of the cubic lattice, which is our main goal in this
paper, the metric is simply given by the Kronecker delta and it can
be deleted from the equation, upper and lower indices coinciding. In
this way we can rewrite eq.(\ref{formaduale}) in the equivalent way:
\begin{equation}\label{tensoBeltra2}
    \frac{1}{2} \,  \epsilon_{ijk} \partial_j \mathbf{Y}_k \, = \, \mu  \, \mathbf{Y}_i
\end{equation}
The next task is that of constructing an ansatz for the vector
harmonics $\mathbf{Y}_i( {\mathbf{X}})$ that are eigenfunctions
of  $\star_g \mathrm{d}$. Since such eigenfunctions have to be well
defined on $\mathrm{T}^3$, their general form is necessarily the
following one:
\begin{eqnarray}\label{harmogen}
   \mathbf{ Y}_i\left( {\mathbf{k}}\, | \, {\mathbf{X}}\right)
   & = & v_i\left( {\mathbf{k}}\right) \,\cos\left( 2\,\pi \,  {\mathbf{k}}\cdot  {\mathbf{X}}\right)
   \, +\, \omega_i\left( {\mathbf{k}}\right) \,\sin\left( 2\,\pi \,  {\mathbf{k}}\cdot  {\mathbf{X}}\right) \nonumber\\
     {\mathbf{k}} & \in & \Lambda^\star
\end{eqnarray}
The condition that the momentum $ {\mathbf{k}}$ lies in the dual
lattice guarantees that $\mathbf{Y}_i( {\mathbf{X}})$ is periodic
with respect to the space lattice $\Lambda$: $\forall \,
 {\mathbf{q}} \, \in \, \Lambda \, : \quad
Y_i\left( {\mathbf{X}}\, + \,  {\mathbf{q}}\right) \, = \,
Y_i\left( {\mathbf{X}}\right)$. Considering next eq.
(\ref{tensoBeltra2}) we immediately see that it implies the further
condition $\partial^i \, Y_i \, = \, 0$. Imposing such a condition
on the general ansatz (\ref{harmogen}) we obtain:
$ {\mathbf{k}}\, \cdot \,
 {\mathbf{v}}\left( {\mathbf{k}}\right) \, = \, 0 \quad ;
\quad  {\mathbf{k}}\, \cdot \,
 {\mathbf{\omega}}\left( {\mathbf{k}}\right) \, = \, 0$ which
reduces the 6 parameters contained in the general ansatz
(\ref{harmogen}) to 4. Imposing next the very equation
(\ref{tensoBeltra2}) we get the following two conditions:
\begin{eqnarray}
  \mu \, v_i \left( {\mathbf{k}}\right) &=& \pi \, \epsilon_{ij\ell} \, k_j \,
  \omega_\ell \left( {\mathbf{k}}\right) \quad ; \quad
  \mu \, \omega_i \left( {\mathbf{k}}\right) \, =\, -\pi \, \epsilon_{ij\ell}
  \, k_j \, v_\ell \left( {\mathbf{k}}\right) \label{curlo1}
\end{eqnarray}
The two equations are self consistent if and only if the following
condition is verified: $\mu^2 \, = \, \pi^2 \,
\langle {\mathbf{k}}\, , \,  {\mathbf{k}}\rangle$. This
trivial elementary calculation completely determines the spectrum of
the operator $\star_g \, \mathrm{d}$ on $\mathrm{T}_g^3$ endowed
with the metric fixed by the choice of a lattice $\Lambda$. The
possible eigenvalues are provided by:
\begin{equation}\label{autovalore}
    m_\mathbf{k} \, = \, \pm \, \pi \, \mbox{det}(w) \, \sqrt{\langle {\mathbf{k}}
    \, , \,  {\mathbf{k}}\rangle}  \quad \quad  {\mathbf{k}} \, \in \, \Lambda^\star
\end{equation}
The degeneracy of each eigenvalue is geometrically provided by
counting the number of intersection points of the dual lattice
$\Lambda^\star$ with a sphere whose center is in the origin and
whose radius is:
\begin{equation}\label{radius}
    r \, = \,  \sqrt{\langle {\mathbf{k}}\, , \,  {\mathbf{k}}\rangle}
\end{equation}
For a generic lattice the number of solutions of equation
(\ref{radius}) namely the number of intersection points of the
lattice with the sphere is just two: $\pm  {\mathbf{k}}$, so that
the typical degeneracy of each eigenvalue is just $2$. If the
lattice $\Lambda$ is one of the Bravais lattices admitting a non
trivial point group  $\mathrm{G}$, then the number of solutions of
eq.(\ref{radius})  increases since all lattice vectors
$ {\mathbf{k}}$ that sit  in one orbit of  $\mathrm{G}$ have the
same norm and therefore are located on the same spherical surface.
The degeneracy of the $\star_g \, \mathrm{d}$ eigenvalue is
precisely the order of the corresponding $\mathrm{G}$-orbit in the
dual lattice $\Lambda^\star$.
\subsection{The algorithm to construct Arnold Beltrami one-forms}
\label{algoritmo} What we explained in the previous section provides
a well defined algorithm to construct a series of Arnold Beltrami
one-forms quite suitable for a systematic  computer aided
implementation.
\par
The steps of the algorithm adapted to the case of the cubic lattice
$\Lambda \, = \Lambda_{cubic}$ are the following ones:
\begin{description}
  \item[a)] Consider the character table and the irreducible
  representations of the Point Group $\mathfrak{P}_\Lambda \, = \, \mathrm{O} \, \simeq \, \mathrm{S}_4$.
  \item[b)] Analyze the structure of orbits of $\mathfrak{P}_\Lambda$
  on the lattice and determine the number  of lattice points
  contained in each spherical layer $\mathfrak{S}_n$ of the dual lattice $\Lambda^\star \, = \, \Lambda$
  of quantized radius $r_n$:
      \begin{eqnarray}\label{spherlayer}
     && {\mathbf{k}}_{(n)} \, \in \,  \mathfrak{S}_n \quad \Leftrightarrow
     \quad  \langle  {\mathbf{k}}_{(n)} \, , \,  {\mathbf{k}}_{(n)}\rangle
     \, = \, r_n^2\quad: \quad 2\,P_n  \, \equiv \, \left|  \mathfrak{S}_n \right|
      \end{eqnarray}
      The number of lattice points in each spherical layer is always
      even since if $ {\mathbf{k}} \, \in \, \Lambda$ also $- {\mathbf{k}}
      \, \in \, \Lambda$ and obviously any vector and its negative have the same norm.
      The spherical layer $\mathfrak{S}_n$ can be composed of one
      or of more $\mathfrak{P}_\Lambda$-orbits.
      In any case it corresponds to a fixed eigenvalue $m_n \, = \, \pi \, r_n$ of the $\star \, \mathrm{d}$-operator.
  \item[c)] Construct the most general solution
  of the Beltrami equation with eigenvalue $m_n$ by using the individual harmonics
  constructed in eq. (\ref{harmogen}):
  \begin{equation}\label{beltramusgen}
    \mathbf{Y}_i\left( {\mathbf{X}}\right) \, = \, \sum_{ {\mathbf{k}}
    \, \in \,\mathfrak{S}_n } \,\mathbf{Y}_i\left( {\mathbf{k}}\, | \, {\mathbf{X}}\right)
  \end{equation}
  Hidden in each harmonic $\mathbf{Y}_i\left( {\mathbf{k}}\, | \, {\mathbf{X}}\right)$
  there are two parameters that are the remainder of the six parameters $v_i\left( {\mathbf{k}}\right)$
  and $\omega_i\left( {\mathbf{k}}\right)$ after conditions (\ref{curlo1}) have been imposed.
  This amount to a total of $4\, P_n$ parameters, yet, since the trigonometric
  functions $\cos(\theta)$ and $\sin(\theta)$ are mapped into plus or minus themselves
  under change of sign of their argument if the  spherical layer $\mathfrak{S}_n $ contains
  lattice vectors in pairs $\pm  {\mathbf{k}}$,  which always happens except in one case, than it follows that the number of independent parameters is  reduced to $2 P_n$. In the unique case of orbits where the  lattice vectors do not appear in pairs $\pm  {\mathbf{k}}$ the number of parameters is $4 P_n$. Hence, at the end of the construction encoded in eq. (\ref{beltramusgen}),
  we have a Beltrami vector depending on a set of $2 P_n$ (or $4 P_n$ )parameters that we can call $F_I$ and consider as the
  components of a $2 \, P_n$-component vector $\mathbf{F}$ ($4 \, P_n$-component vector $\mathbf{F}$). Ultimately we have an object of the following form:
  \begin{equation}\label{gomez}
     {\mathbf{Y}}\left( {\mathbf{X}}\, | \, \mathbf{F}\right)
  \end{equation}
 which under the point group   $\mathfrak{P}_\Lambda$ necessarily transforms in the following way:
 \begin{equation}\label{Rtrasformogen}
    \forall \, \gamma \, \in \, G_\Lambda \, : \quad \gamma^{-1} \,\cdot \,
     {\mathbf{Y}}\left(\gamma \,\cdot \, {\mathbf{X}}\, | \, \mathbf{F}\right)
    \, = \,  {\mathbf{Y}}\left( {\mathbf{X}}\, | \, \mathfrak{R}[\gamma] \,\cdot \, \mathbf{F}\right)
 \end{equation}
 where $\mathfrak{R}[\gamma]$ are $2\, P_n \times 2\, P_n$ matrices
 that form a representation of $\mathfrak{P}_\Lambda$. Eq.(\ref{Rtrasformogen}) is necessarily true because
 any rotation $\gamma \, \in \, \mathrm{G}$ permutes the elements of $\mathfrak{S}_n$ among themselves.
   \item[e)] Decompose the representation $\mathfrak{R}[\gamma]$ into irreducible representations of $\mathfrak{P}_\Lambda$:
       \begin{equation}\label{rutilante}
        \mathbf{F} \, = \, \bigoplus_{\mbox{irr}\in \mathfrak{R}}\, \mathbf{A}_{\mbox{irr}}
       \end{equation}
  where $\mathbf{A}_{\mbox{irr}}$ denotes a set of parameters spanning
  the considered irreducible representation.     Finally writing:
       \begin{equation}\label{siluetta}
       \mathbf{Y}^{[1]}\left(\mathbf{A}_{\mbox{irr}}\right) \,
       \equiv \,  {\mathbf{Y}}\left( {\mathbf{X}}\, | \, \mathbf{A}_{\mbox{irr}}\right) \, \cdot \, d {\mathbf{X}}
       \, = \, \sum_{i=1}^3 \mathbf{Y}_i\left( {\mathbf{X}}\, | \, \mathbf{A}_{\mbox{irr}}\right) \, dX^i
       \end{equation}
 we obtain a multiplet of $1$-forms   $\mathbf{Y}^{[1]}\left( \mathbf{A}_{\mbox{irr}}\right)$
 that satisfy  Beltrami equation (\ref{formaduale})
and transform  in the considered irreducible representation of the point group.
       \end{description}
An obvious question which arises in connection with such a constructive algorithm is the following: how many
Arnold--Beltrami one-forms are there? At first sight it seems that there is an infinite number of such systems since we
can arbitrarily increase the radius of the spherical layer and on each new layer it seems that we have new solutions of
Beltrami equation. However, as we have demonstrated in \cite{arnolderie} the number of different types of
Arnold-Beltrami forms is finite: $48$. This follows from two facts that have a common origin: a) there is a
$\mathbb{Z}_4$ periodicity in the lattice spherical layers $\mathfrak{S}_n$, b) the solutions of Beltrami equation fall
into irreducible representations of a Universal Classifying Group that has $1536$-elements, $37$ conjugacy classes and
$37$ irreducible representations. We refer the reader to our paper \cite{arnolderie} for a detailed discussion of the
classification, yet we dwell on the construction of the Universal Classifying Group $\mathrm{G_{1536}}$ since together
with its subgroups it plays a key role in the use of Beltrami one-forms within the context of supergravity.
\subsection{The Octahedral Group $\mathrm{O_{24}}$ and its extension to the
Universal Classifying Group $\mathrm{G_{1536}}$ }
\label{ottostruttura} Abstractly the octahedral Group
$\mathrm{O_{24}}\sim \mathrm{S_{24}}$ is isomorphic to the symmetric
group of permutations of 4 objects. It is defined by the following
generators and relations:
\begin{equation}\label{octapresa}
 T, \, S \quad : \quad    T^3 \, = \, \mbox{\bf e} \quad ; \quad S^2 \, = \, \mbox{\bf e}
 \quad ; \quad (S\,T)^4 \, = \, \mbox{\bf e}
\end{equation}
On the other hand $\mathrm{O_{24}}$ is a finite, discrete subgroup
of the three-dimensional rotation group and any $\gamma \, \in \,
\mathrm{O_{24}}\, \subset \, \mathrm{SO(3)}$ of its 24 elements can
be uniquely identified by its action on the coordinates
$ {\mathbf{X}}\, = \{X,Y,Z\}$,  as it is displayed below:
\begin{equation}\label{nomiOelemen}
\begin{array}{cc}
\begin{array}{|c|rcl|}
\hline
\mbox{\bf e} & 1_1 & = & \{X,Y,Z\} \\
 \hline
\null & 2_1 & = & \{-Y,-Z,X\} \\
\null &  2_2 & = & \{-Y,Z,-X\} \\
\null & 2_3 & = & \{-Z,-X,Y\} \\
C_3 & 2_4 & = & \{-Z,X,-Y\} \\
 \null &2_5 & = & \{Z,-X,-Y\} \\
\null & 2_6 & = & \{Z,X,Y\} \\
\null & 2_7 & = & \{Y,-Z,-X\} \\
\null & 2_8 & = & \{Y,Z,X\} \\
 \hline
\null & 3_1 & = & \{-X,-Y,Z\} \\
C_4^2 & 3_2 & = & \{-X,Y,-Z\} \\
\null & 3_3 & = & \{X,-Y,-Z\} \\
 \hline
\end{array} & \begin{array}{|c|rcl|}
\hline
\null & 4_1 & = & \{-X,-Z,-Y\} \\
\null & 4_2 & = & \{-X,Z,Y\} \\
C_2 &4_3 & = & \{-Y,-X,-Z\} \\
\null & 4_4 & = & \{-Z,-Y,-X\} \\
\null & 4_5 & = & \{Z,-Y,X\} \\
\null & 4_6 & = & \{Y,X,-Z\} \\
 \hline
\null & 5_1 & = & \{-Y,X,Z\} \\
\null & 5_2 & = & \{-Z,Y,X\} \\
C_4 & 5_3 & = & \{Z,Y,-X\} \\
\null & 5_4 & = & \{Y,-X,Z\} \\
\null & 5_5 & = & \{X,-Z,Y\} \\
\null & 5_6 & = & \{X,Z,-Y\}\\
 \hline
\end{array} \\
\end{array}
\end{equation}
As one sees from the above list the 24 elements are distributed into
5 conjugacy classes mentioned in the first column of the table,
according to a nomenclature which is standard in the chemical
literature on crystallography. The relation between the abstract and
concrete presentation of the octahedral  group is obtained by
identifying in the list (\ref{nomiOelemen}) the generators $T$ and
$S$ mentioned in eq. (\ref{octapresa}). Explicitly we have:
\begin{equation}\label{generatiTS}
    T \, = \, 2_8 \, = \, \left(
\begin{array}{lll}
 0 & 1 & 0 \\
 0 & 0 & 1 \\
 1 & 0 & 0
\end{array}
\right)\quad ; \quad S \, = \, 4_6 \, =\left(
\begin{array}{lll}
 0 & 1 & 0 \\
 1 & 0 & 0 \\
 0 & 0 & -1
\end{array}
\right)
\end{equation}
All other elements are reconstructed from the above two using the
multiplication table of the group which is displayed below: {\small
\begin{equation}\label{Omultipla}
\begin{array}{|l|llllllllllllllllllllllll|}
\hline
 \null &   1_1 & 2_1 & 2_2 & 2_3 & 2_4 & 2_5 & 2_6 & 2_7 & 2_8 & 3_1 & 3_2 & 3_3 & 4_1 & 4_2 & 4_3 & 4_4 & 4_5 & 4_6 & 5_1 & 5_2 & 5_3 & 5_4 & 5_5 & 5_6 \\
 \hline
 1_1 &   1_1 & 2_1 & 2_2 & 2_3 & 2_4 & 2_5 & 2_6 & 2_7 & 2_8 & 3_1 & 3_2 & 3_3 & 4_1 & 4_2 & 4_3 & 4_4 & 4_5 & 4_6 & 5_1 & 5_2 & 5_3 & 5_4 & 5_5 & 5_6 \\
 2_1 &   2_1 & 2_5 & 2_4 & 3_3 & 3_2 & 1_1 & 3_1 & 2_6 & 2_3 & 2_7 & 2_2 & 2_8 & 5_3 & 4_4 & 5_6 & 4_6 & 5_4 & 4_2 & 4_1 & 4_3 & 5_1 & 5_5 & 4_5 & 5_2 \\
 2_2 &   2_2 & 2_6 & 2_3 & 1_1 & 3_1 & 3_3 & 3_2 & 2_5 & 2_4 & 2_8 & 2_1 & 2_7 & 4_5 & 5_2 & 5_5 & 5_4 & 4_6 & 4_1 & 4_2 & 5_1 & 4_3 & 5_6 & 5_3 & 4_4 \\
 2_3 &   2_3 & 3_2 & 1_1 & 2_2 & 2_8 & 2_7 & 2_1 & 3_3 & 3_1 & 2_4 & 2_6 & 2_5 & 4_6 & 5_1 & 5_3 & 5_6 & 4_1 & 4_5 & 5_2 & 4_2 & 5_5 & 4_4 & 4_3 & 5_4 \\
 2_4 &   2_4 & 3_1 & 3_3 & 2_1 & 2_7 & 2_8 & 2_2 & 1_1 & 3_2 & 2_3 & 2_5 & 2_6 & 5_4 & 4_3 & 4_5 & 5_5 & 4_2 & 5_3 & 4_4 & 4_1 & 5_6 & 5_2 & 5_1 & 4_6 \\
 2_5 &   2_5 & 1_1 & 3_2 & 2_8 & 2_2 & 2_1 & 2_7 & 3_1 & 3_3 & 2_6 & 2_4 & 2_3 & 5_1 & 4_6 & 5_2 & 4_2 & 5_5 & 4_4 & 5_3 & 5_6 & 4_1 & 4_5 & 5_4 & 4_3 \\
 2_6 &   2_6 & 3_3 & 3_1 & 2_7 & 2_1 & 2_2 & 2_8 & 3_2 & 1_1 & 2_5 & 2_3 & 2_4 & 4_3 & 5_4 & 4_4 & 4_1 & 5_6 & 5_2 & 4_5 & 5_5 & 4_2 & 5_3 & 4_6 & 5_1 \\
 2_7 &   2_7 & 2_3 & 2_6 & 3_1 & 1_1 & 3_2 & 3_3 & 2_4 & 2_5 & 2_1 & 2_8 & 2_2 & 5_2 & 4_5 & 4_2 & 5_1 & 4_3 & 5_6 & 5_5 & 5_4 & 4_6 & 4_1 & 4_4 & 5_3 \\
 2_8 &   2_8 & 2_4 & 2_5 & 3_2 & 3_3 & 3_1 & 1_1 & 2_3 & 2_6 & 2_2 & 2_7 & 2_1 & 4_4 & 5_3 & 4_1 & 4_3 & 5_1 & 5_5 & 5_6 & 4_6 & 5_4 & 4_2 & 5_2 & 4_5 \\
 3_1 &   3_1 & 2_8 & 2_7 & 2_6 & 2_5 & 2_4 & 2_3 & 2_2 & 2_1 & 1_1 & 3_3 & 3_2 & 5_6 & 5_5 & 4_6 & 5_3 & 5_2 & 4_3 & 5_4 & 4_5 & 4_4 & 5_1 & 4_2 & 4_1 \\
 3_2 &   3_2 & 2_7 & 2_8 & 2_5 & 2_6 & 2_3 & 2_4 & 2_1 & 2_2 & 3_3 & 1_1 & 3_1 & 5_5 & 5_6 & 5_4 & 4_5 & 4_4 & 5_1 & 4_6 & 5_3 & 5_2 & 4_3 & 4_1 & 4_2 \\
 3_3 &   3_3 & 2_2 & 2_1 & 2_4 & 2_3 & 2_6 & 2_5 & 2_8 & 2_7 & 3_2 & 3_1 & 1_1 & 4_2 & 4_1 & 5_1 & 5_2 & 5_3 & 5_4 & 4_3 & 4_4 & 4_5 & 4_6 & 5_6 & 5_5 \\
 4_1 &   4_1 & 5_4 & 4_6 & 4_5 & 5_3 & 5_2 & 4_4 & 5_1 & 4_3 & 5_5 & 5_6 & 4_2 & 1_1 & 3_3 & 2_8 & 2_6 & 2_3 & 2_2 & 2_7 & 2_5 & 2_4 & 2_1 & 3_1 & 3_2 \\
 4_2 &   4_2 & 4_6 & 5_4 & 5_3 & 4_5 & 4_4 & 5_2 & 4_3 & 5_1 & 5_6 & 5_5 & 4_1 & 3_3 & 1_1 & 2_7 & 2_5 & 2_4 & 2_1 & 2_8 & 2_6 & 2_3 & 2_2 & 3_2 & 3_1 \\
 4_3 &   4_3 & 5_3 & 5_2 & 5_6 & 4_2 & 5_5 & 4_1 & 4_5 & 4_4 & 4_6 & 5_1 & 5_4 & 2_6 & 2_4 & 1_1 & 2_8 & 2_7 & 3_1 & 3_2 & 2_2 & 2_1 & 3_3 & 2_5 & 2_3 \\
 4_4 &   4_4 & 4_2 & 5_5 & 5_1 & 5_4 & 4_6 & 4_3 & 5_6 & 4_1 & 5_2 & 4_5 & 5_3 & 2_8 & 2_1 & 2_6 & 1_1 & 3_2 & 2_5 & 2_3 & 3_1 & 3_3 & 2_4 & 2_2 & 2_7 \\
 4_5 &   4_5 & 5_6 & 4_1 & 4_6 & 4_3 & 5_1 & 5_4 & 4_2 & 5_5 & 5_3 & 4_4 & 5_2 & 2_2 & 2_7 & 2_4 & 3_2 & 1_1 & 2_3 & 2_5 & 3_3 & 3_1 & 2_6 & 2_8 & 2_1 \\
 4_6 &   4_6 & 4_4 & 4_5 & 4_1 & 5_5 & 4_2 & 5_6 & 5_2 & 5_3 & 4_3 & 5_4 & 5_1 & 2_3 & 2_5 & 3_1 & 2_1 & 2_2 & 1_1 & 3_3 & 2_7 & 2_8 & 3_2 & 2_4 & 2_6 \\
 5_1 &   5_1 & 4_5 & 4_4 & 5_5 & 4_1 & 5_6 & 4_2 & 5_3 & 5_2 & 5_4 & 4_3 & 4_6 & 2_5 & 2_3 & 3_3 & 2_7 & 2_8 & 3_2 & 3_1 & 2_1 & 2_2 & 1_1 & 2_6 & 2_4 \\
 5_2 &   5_2 & 4_1 & 5_6 & 4_3 & 4_6 & 5_4 & 5_1 & 5_5 & 4_2 & 4_4 & 5_3 & 4_5 & 2_7 & 2_2 & 2_5 & 3_3 & 3_1 & 2_6 & 2_4 & 3_2 & 1_1 & 2_3 & 2_1 & 2_8 \\
 5_3 &   5_3 & 5_5 & 4_2 & 5_4 & 5_1 & 4_3 & 4_6 & 4_1 & 5_6 & 4_5 & 5_2 & 4_4 & 2_1 & 2_8 & 2_3 & 3_1 & 3_3 & 2_4 & 2_6 & 1_1 & 3_2 & 2_5 & 2_7 & 2_2 \\
 5_4 &   5_4 & 5_2 & 5_3 & 4_2 & 5_6 & 4_1 & 5_5 & 4_4 & 4_5 & 5_1 & 4_6 & 4_3 & 2_4 & 2_6 & 3_2 & 2_2 & 2_1 & 3_3 & 1_1 & 2_8 & 2_7 & 3_1 & 2_3 & 2_5 \\
 5_5 &   5_5 & 4_3 & 5_1 & 4_4 & 5_2 & 5_3 & 4_5 & 4_6 & 5_4 & 4_1 & 4_2 & 5_6 & 3_2 & 3_1 & 2_2 & 2_4 & 2_5 & 2_8 & 2_1 & 2_3 & 2_6 & 2_7 & 3_3 & 1_1 \\
 5_6 &   5_6 & 5_1 & 4_3 & 5_2 & 4_4 & 4_5 & 5_3 & 5_4 & 4_6 & 4_2 & 4_1 & 5_5 & 3_1 & 3_2 & 2_1 & 2_3 & 2_6 & 2_7 & 2_2 & 2_4 & 2_5 & 2_8 & 1_1 & 3_3\\
 \hline
\end{array}
\end{equation}
} This observation is important in relation with representation theory. Any linear representation of the group is
uniquely specified by giving the matrix representation of the two generators $T=2_8$ and $S=4_6$.
\par
There are five conjugacy classes in $\mathrm{O}_{24}$ and therefore
according to theory there are five irreducible representations of
the same group, that we name $D_i$, $i\, =\, 1,\dots, 5$.
\begin{table}[!hbt]
  \centering
  \begin{eqnarray*}
   \begin{array}{||l||ccccc||}
    \hline
    \hline
{\begin{array}{cc} \null &\mbox{Class}\\
\mbox{Irrep} & \null\\
\end{array}} & \{\mbox{\bf e},1\} & \left\{C_3,8\right\} & \left\{C_4^2,3\right\}
& \left\{C_2,6\right\} & \left\{C_4,6\right\} \\
\hline
 \hline
 D_1 \, , \quad \chi_1 \, = \,& 1 & 1 & 1 & 1 & 1 \\
D_2 \, , \quad \chi_2 \, = \,& 1 & 1 & 1 & -1 & -1 \\
 D_3 \, , \quad \chi_3 \, = \, & 2 & -1 & 2 & 0 & 0 \\
 D_4 \, , \quad \chi_4 \, = \, & 3 & 0 & -1 & -1 & 1 \\
D_5 \, , \quad \chi_5 \, = \, & 3 & 0 & -1 & 1 & -1\\
 \hline
 \hline
\end{array}
\end{eqnarray*}
  \caption{Character Table of the proper Octahedral Group}\label{caratteriO}
\end{table}
The table of characters of the octahedral group  is summarized in
eq.(\ref{caratteriO}).
\subsubsection{Extension of the Point Group with Translations }
\label{maingruppo} We recall now  what is the main mathematical
point of our previous paper \cite{arnolderie}, namely the extension
of the point group  with appropriate discrete subgroups of the
compactified translation group $\mathrm{U(1)}^3$. This issue bears
on a classical topic dating back to the XIX century, which was
developed by crystallographers and in particular by the great
russian mathematician Fyodorov \cite{fyodorovcryst}. We refer here
to the issue of space groups which historically resulted into the
classification of the $230$ crystallographic groups, well known in
the chemical literature, for which an international system of
notations and conventions has been  established that is available in
numerous  encyclopedic tables and books. Although in
\cite{arnolderie} we utilized one key-point of the logic that leads
to  the classification of space groups, yet our goal happened to be
slightly different and what we aimed at was not the identification
of space groups, rather the construction of what we named a
\textit{Universal Classifying Group}, that is of a group which
contains all the existing space groups as subgroups. We showed in
\cite{arnolderie}  that such a Universal Classifying Group is the
one appropriate to organize the eigenfunctions of the $\star_g
\mathrm{d}$-operator into irreducible representations  and
eventually to uncover the available hidden symmetries of all
Arnold-Beltrami one-forms.
\subsubsection{The idea of Space Groups and Frobenius congruences}
\label{spaziogruppi} The idea of space groups arises naturally in
the following way. The covering manifold of the $T^3$ torus is
$\mathbb{R}^3$ which can be regarded as the following coset
manifold:
\begin{equation}\label{fantacosetto}
    \mathbb{R}^3 \, \simeq \, \frac{\mathbb{E}^3}{\mathrm{SO(3)}} \quad ;
    \quad  \, \mathbb{E}^3 \, \equiv \, \mathrm{ISO(3)} \, \doteq \, \mathrm{SO(3)} \ltimes \mathcal{T}^3
\end{equation}
where $\mathcal{T}^3$  is the three dimensional translation group
acting on $\mathbb{R}^3$ in the standard way:
\begin{equation}\label{trasluco}
    \forall \mathbf{t} \, \in \, \mathcal{T}^3 \, ,\, \forall \mathbf{X} \,
    \in \, \mathbb{R}^3\quad | \quad\mathbf{t} \, : \,   \mathbf{X} \, \rightarrow \, \mathbf{X}\, + \, \mathbf{t}
\end{equation}
and the Euclidian group $\mathbb{E}^3$ is the semi-direct product
of the proper rotation group $\mathrm{SO(3)}$ with the translation
group $\mathcal{T}^3$. Harmonic analysis on $\mathbb{R}^3$ is a
complicated matter of functional analysis since $\mathcal{T}^3$ is a
non-compact group and its unitary irreducible representations are
infinite-dimensional. The landscape changes drastically when we
compactify our manifold from $\mathbb{R}^3$ to the three torus
$T^3$. Compactification is obtained taking the  quotient of
$\mathbb{R}^3$ with respect to the lattice $\Lambda \, \subset \,
\mathcal{T}^3$. As a result of this quotient the manifold becomes
$\mathbb{S}^1 \times \mathbb{S}^1 \times \mathbb{S}^1$ but also the
isometry group is reduced. Instead of $\mathrm{SO(3)}$ as rotation
group we are left with its discrete subgroup $\mathfrak{P}_\Lambda
\, \subset \, \mathrm{SO(3)}$  which maps the lattice $\Lambda$ into
itself (the point group ) and instead of the translation subgroup
$\mathcal{T}^3$ we are left with the quotient group:
\begin{equation}\label{quozienTra}
    \mathfrak{T}^3_\Lambda \, \equiv \, \frac{\mathcal{T}^3}{\Lambda} \,
    \simeq \, \mathrm{U(1)} \times \mathrm{U(1) }\times \mathrm{U(1)}
\end{equation}
In this way we obtain a new group which replaces the Euclidian group
and which is the semidirect product of the point group
$\mathfrak{P}_\Lambda$ with $\mathfrak{T}^3_\Lambda $:
\begin{equation}\label{goticoG}
    \mathfrak{G}_{\Lambda}  \, \equiv \, \mathfrak{P}_\Lambda \,
    \ltimes \,  \mathfrak{T}^3_\Lambda
\end{equation}
The group $\mathfrak{G}_{\Lambda} $  is an exact symmetry of
Beltrami equation (\ref{formaduale}) and its action is naturally
defined on the parameter space of any of its solutions
$\mathbf{Y}^{[1]}\left( {\mathbf{x}} | \mathbf{F}\right)$  that
we can obtain by means of the algorithm  described in section
\ref{algoritmo}. To appreciate  this point let us recall that every
component of the one-form $\mathbf{Y}^{[1]}\left( {\mathbf{X}} |
\mathbf{F}\right)$ associated with a $\mathfrak{P}_\Lambda$
point--orbit $\mathcal{O}$ is a linear combinations of the functions
$\cos\left[2\pi \, \mathbf{k}_i \cdot  {\mathbf{X}}\right]$ and
$\sin\left[2\pi \, \mathbf{k}_i \cdot  {\mathbf{X}}\right]$,
where $\mathbf{k}_i \, \in \, \mathcal{O}$  are all the momentum
vectors contained in the orbit. Consider next the same functions in
a translated point of the three torus $ {\mathbf{X}}^\prime \, =
\, {\mathbf{X}}\,  + \, \mathfrak{c}$ where $\mathfrak{c} \, = \,
\{\xi_1\,,\,\xi_2\, , \, \xi_3\,\}$ is a representative of an
equivalence class $\mathfrak{c}$ of constant vectors defined modulo
the lattice:
\begin{equation}\label{cribulla}
\mathfrak{c} \, = \, \mathfrak{c} \, + \, \mathbf{Y} \quad ; \quad
\forall \mathbf{Y}\, \in \, \Lambda
\end{equation}
The above equivalence classes are the elements  of the quotient group $\mathfrak{T}^3_\Lambda $. Using standard
trigonometric identities, $\cos\left[2\pi \, \mathbf{k}_i \cdot  {\mathbf{X}}\, + \, 2\pi \, \mathbf{k}_i \cdot
\mathfrak{c} \right]$  can be reexpressed as a linear combination of the $\cos\left[2\pi \, \mathbf{k}_i \cdot
 {\mathbf{X}}\right]$ and $\sin\left[2\pi \, \mathbf{k}_i \cdot  {\mathbf{X}}\right]$ functions with coefficients
that depend on trigonometric functions of $\mathfrak{c}$. The same is true of $\sin\left[2\pi \, \mathbf{k}_i \cdot
 {\mathbf{X}}\, + \, 2\pi \, \mathbf{k}_i \cdot \mathfrak{c} \right]$. Note also that because of the periodicity of
the trigonometric functions, the shift in their argument by a lattice translation is not-effective so that one deals
only with the equivalence classes (\ref{cribulla}). It follows that for each element $\mathfrak{c}\in
\mathfrak{T}^3_\Lambda$ we obtain a matrix representation $\mathcal{M}_\mathfrak{c}$ realized on the $\mathbf{F}$
parameters and defined by the following equation:
\begin{eqnarray}\label{traslazionaBuh}
   & \mathbf{Y}^{[1]}\left( {\mathbf{X}}+\mathfrak{c} |\mathbf{F}\right)\, =\,
   \mathbf{Y}^{[1]}\left( {\mathbf{X}} |\mathcal{M}_\mathfrak{c} \mathbf{F}\right) &
\end{eqnarray}
As we already noted in eq.(\ref{Rtrasformogen}), for any group
element $\gamma \, \in \, \mathfrak{P}_\Lambda$ we also have a
matrix representation induced on the parameter space by the same
mechanism:
\begin{equation}\label{RotazioneBuh}
    \forall \, \gamma \, \in \, \mathfrak{P}_\Lambda \, : \, \mathbf{Y}^{[1]}
    \left(\gamma \,\cdot \, {\mathbf{X}}\, | \, \mathbf{F}\right)
    \, = \, \mathbf{Y}^{[1]}\left( {\mathbf{X}}\, | \, \mathfrak{R}[\gamma] \,\cdot \, \mathbf{F}\right)
 \end{equation}
 Combining eq.s(\ref{traslazionaBuh}) and (\ref{RotazioneBuh}) we  obtain a
 matrix realization of the entire group $\mathfrak{G}_{\Lambda} $ in the following way:
 \begin{eqnarray}
  \mathbf{Y}^{[1]}\left(\gamma\, \cdot \,  {\mathbf{X}}+\mathfrak{c} |\mathbf{F}\right)&=&
  \mathbf{Y}^{[1]}\left( {\mathbf{X}} \,  | \, \mathfrak{R}[\gamma] \cdot \mathcal{M}_\mathfrak{c} \,\cdot \,\mathbf{F}\right) \\
   &\Downarrow&\nonumber\\
 \forall \left(\gamma \, , \, \mathfrak{c}\right) \, \in \,  \mathfrak{G}_{\Lambda}
 &\rightarrow& D\left[ \left(\gamma \, , \, \mathfrak{c}\right) \right] \, = \,
 \mathfrak{R}[\gamma] \cdot \mathcal{M}_\mathfrak{c}\label{direttocoriandolo}
 \end{eqnarray}
 Actually the construction of Beltrami one-forms in the lowest lying point-orbit,
 which usually yields a faithful matrix representation of all group elements,
 can be regarded as an automatic way of taking  the quotient (\ref{quozienTra})
 and the resulting representation can be considered the defining representation of the group $\mathfrak{G}_{\Lambda}$.
 \par
 The next point in the logic which leads to space groups is the following observation.
 $\mathfrak{G}_{\Lambda}$ is an unusual mixture of a discrete group (the point group )
 with a continuous one (the translation subgroup $\mathfrak{T}^3_\Lambda $). This latter is rather trivial,
 since its action corresponds to shifting the origin of coordinates in three-dimensional space.
 Yet there are in $\mathfrak{G}_{\Lambda}$ some discrete subgroups which can be isomorphic to
 the point group  $\mathfrak{G}_{\Lambda}$, or to one of its subgroups $\mathfrak{H}_\Lambda \, \subset \,
 \mathfrak{G}_{\Lambda}$,
 without being their conjugate. Such groups cannot be disposed of by shifting the origin of coordinates and consequently
 they can encode non-trivial hidden symmetries. The search of such non trivial discrete subgroups
 of $\mathfrak{G}_{\Lambda}$ is the mission accomplished by crystallographers the result of the mission being
 the classification of space groups.
 \par
 The simplest and most intuitive way of constructing Space-Groups relies on the
 so named Frobenius congruences
 \cite{Aroyo}\cite{Souvignier}. Let us outline this construction.
 Following classical approaches we use a $4\times 4$ matrix representation of the group $\mathfrak{G}_{\Lambda}$:
 \begin{equation}\label{quattroruote}
    \forall \left(\gamma \, , \, \mathfrak{c}\right) \, \in \,  \mathfrak{G}_{\Lambda} \,
    \rightarrow\,  \hat{D}\left[ \left(\gamma \, , \, \mathfrak{c}\right) \right] \, = \, \left( \begin{array}{c|c}
                             \gamma & \mathfrak{c} \\
                             \hline
                             0& 1
                           \end{array}
    \right)
 \end{equation}
 Performing the matrix product of two elements, in the translation block one has to take
 into account equivalence modulo lattice $\Lambda$, namely
 \begin{equation}\label{moduloLatte}
    \left( \begin{array}{c|c}
                             \gamma_1 & \mathfrak{c_1} \\
                             \hline
                             0& 1
                           \end{array}
    \right) \, \cdot \, \left( \begin{array}{c|c}
                             \gamma_2 & \mathfrak{c_2} \\
                             \hline
                             0& 1
                           \end{array}
    \right) \, = \,  \left( \begin{array}{c|c}
                             \gamma_1\, \cdot \, \gamma_2 & \gamma_1\,  \mathfrak{c_2} \, + \, \mathfrak{c_1}\, + \, \Lambda \\
                             \hline
                             0& 1
                           \end{array}
    \right)
 \end{equation}
 Utilizing this notation the next step consists of introducing translation deformations of the generators
 of the point group  searching for deformations that cannot be eliminated by conjugation.
 \subsubsection{Frobenius congruences for the Octahedral Group $\mathrm{O_{24}}$}
The octahedral  group is abstractly defined by the presentation
displayed in eq.(\ref{octapresa}). As a first step we parameterize
the candidate deformations of the two generators $T$  and $S$ in the
following way:
\begin{equation}\label{deformuccia}
    \hat{T} \, = \, \left(
\begin{array}{lll|l}
 0 & 1 & 0 & \tau _1 \\
 0 & 0 & 1 & \tau _2 \\
 1 & 0 & 0 & \tau _3 \\
 \hline
 0 & 0 & 0 & 1
\end{array}
\right) \quad ; \quad \hat{S} \, = \, \left(
\begin{array}{lll|l}
 0 & 0 & 1 & \sigma _1 \\
 0 & -1 & 0 & \sigma _2 \\
 1 & 0 & 0 & \sigma _3 \\
 \hline
 0 & 0 & 0 & 1
\end{array}
\right)
\end{equation}
which should be compared with eq.(\ref{generatiTS}). Next we try to
impose on the deformed generators the defining relations of
$\mathrm{O}_{24}$. By explicit calculation we find:
\begin{eqnarray}\label{finocchiona}
 \hat{T}^3 & = &    \left(
\begin{array}{lll|l}
 1 & 0 & 0 & \tau _1+\tau _2+\tau _3 \\
 0 & 1 & 0 & \tau _1+\tau _2+\tau _3 \\
 0 & 0 & 1 & \tau _1+\tau _2+\tau _3 \\
 \hline
 0 & 0 & 0 & 1
\end{array}
\right) \quad ; \quad \hat{S}^2 \, = \, \left(
\begin{array}{lll|l}
 1 & 0 & 0 & \sigma _1+\sigma _3 \\
 0 & 1 & 0 & 0 \\
 0 & 0 & 1 & \sigma _1+\sigma _3 \\
 \hline
 0 & 0 & 0 & 1
\end{array}
\right) \nonumber\\
\left(\hat{S}\hat{T}\right)^4 & = & \left(
\begin{array}{lll|l}
 1 & 0 & 0 & 4 \sigma _1+4 \tau _3 \\
 0 & 1 & 0 & 0 \\
 0 & 0 & 1 & 0 \\
 \hline
 0 & 0 & 0 & 1
\end{array}
\right)
\end{eqnarray}
so that we obtain the conditions:
\begin{equation}\label{frobeniale}
    \tau _1+\tau _2+\tau _3  \, \in \, \mathbb{Z}\quad ; \quad \sigma _1+\sigma _3 \, \in \,
    \mathbb{Z} \quad ; \quad 4 \sigma _1+4 \tau _3 \, \in \, \mathbb{Z}
\end{equation}
which are the Frobenius congruences for the present case. Next we
consider the effect of conjugation with the most general translation
element of the group $\mathfrak{G}_{cubic}$. Just for convenience we
parameterize the translation subgroup as follows:
\begin{equation}\label{tmatto}
    \mathfrak{t} \, = \, \left(
\begin{array}{lll|l}
 1 & 0 & 0 & a+c \\
 0 & 1 & 0 & b \\
 0 & 0 & 1 & a-c \\
 \hline
 0 & 0 & 0 & 1
\end{array}
\right)
\end{equation}
and we get:
\begin{equation}\label{giambatto}
    \mathfrak{t} \, \hat{T} \, \mathfrak{t}^{-1} \, = \, \left(
\begin{array}{lll|l}
 0 & 1 & 0 & a-b+c+\tau _1 \\
 0 & 0 & 1 & -a+b+c+\tau _2 \\
 1 & 0 & 0 & \tau _3-2 c \\
 \hline
 0 & 0 & 0 & 1
\end{array}
\right)\quad ; \quad \mathfrak{t} \, \hat{S} \, \mathfrak{t}^{-1} \,
= \,\left(
\begin{array}{lll|l}
 0 & 0 & 1 & 2 c+\sigma _1 \\
 0 & -1 & 0 & 2 b+\sigma _2 \\
 1 & 0 & 0 & \sigma _3-2 c \\
 \hline
 0 & 0 & 0 & 1
\end{array}
\right)
\end{equation}
This shows that by using the parameters $b,c$ we can always put
$\sigma_1 \, = \, \sigma_2 \, = \,0$, while using the parameter $a$
we can put $\tau_1 \, = \,0$ (obviously this is not the only
possible gauge choice, yet it is the most convenient) so that the
Frobenius congruences reduce to:
\begin{equation}\label{calugone}
     \tau _2+\tau _3  \, \in \, \mathbb{Z}\quad ; \quad \sigma _3 \, \in \, \mathbb{Z} \quad ;
     \quad 4 \tau _3 \, \in \, \mathbb{Z}
\end{equation}
Eq.(\ref{calugone}) is of great momentum. It tells us that any non
trivial subgroup of  $\mathfrak{G}_{cubic}$ which is not conjugate
to the point group  contains point group  elements extended with
rational translations of the form $\mathfrak{c} \, = \, \left\{
\ft{n_1}{4}\,  , \, \ft{n_2}{4}\,  , \, \ft{n_3}{4}\right\}$. Up to
this point our way and that of crystallographers was the same:
hereafter our paths separate. The crystallographers classify all
possible non trivial groups that extend the point group  with such
translation deformations: indeed looking at the crystallographic
tables one realizes that  all known space groups for the cubic
lattice have translation components of the form $\mathfrak{c} \, =
\, \left\{ \ft{n_1}{4}\,  , \, \ft{n_2}{4}\,  , \,
\ft{n_3}{4}\right\}$. On the other hand, we do something much
simpler which leads to a quite big group containing all possible
Space-Groups as subgroups, together with other subgroups that are
not space groups in the crystallographic sense.
\subsubsection{The Universal Classifying Group: $\mathrm{\mathrm{G_{1536}}}$}
\label{universalone} Inspired by the space group  construction and by Frobenius congruences, in \cite{arnolderie} we
just considered the subgroup of $\mathfrak{G}_{cubic}$ where translations are quantized in units of $\frac{1}{4}$. In
each direction and modulo integers there are just four translations $0, \, \ft 14, \, \ft 12, \, \ft 34$ so that the
translation subgroup reduces to $\mathbb{Z}_4 \, \otimes\,\mathbb{Z}_4\, \otimes \, \mathbb{Z}_4$  that has a total of
$64$ elements. In this way we singled out a discrete subgroup $\mathrm{G_{1536}} \, \subset \, \mathfrak{G}_{cubic}$ of
order $24 \times 64 \,  =  \, 1536$,  which is simply the semidirect product of the point group $\mathrm{O_{24}}$ with
$\mathbb{Z}_4 \, \otimes\,\mathbb{Z}_4\, \otimes \, \mathbb{Z}_4$:
\begin{equation}\label{1536defino}
    \mathfrak{G}_{cubic} \, \supset \, \mathrm{\mathrm{G_{1536}}} \, \simeq \, \mathrm{O_{24}} \,
    \ltimes \, \left (\mathbb{Z}_4 \, \otimes\,\mathbb{Z}_4\, \otimes \, \mathbb{Z}_4\right)
\end{equation}
We named $\mathrm{\mathrm{G_{1536}}}$ the universal classifying
group of the cubic lattice, and its elements were labeled by us as
follows:
\begin{equation}\label{elementando1536}
    \mathrm{\mathrm{G_{1536}}} \, \in \, \left\{ p_q \, , \, \ft{2 n_1}{4} \, , \, \ft{2 n_2}{4}
    \, , \, \ft{2 n_3}{4}\right\} \quad \Rightarrow\quad \left\{ \begin{array}{rcl}
   p_q & \in & \mathrm{O_{24}}\\
   \left\{ \ft{n_1}{4}\,  , \, \ft{n_2}{4}\,  , \, \ft{n_3}{4}\right\} & \in & \mathbb{Z}_4
   \, \otimes\,\mathbb{Z}_4\, \otimes \, \mathbb{Z}_4\end{array}\right.
\end{equation}
where for the elements of the point group we use the labels $p_q$
established in eq.(\ref{nomiOelemen}) while for the translation part
our notation encodes an equivalence class of  translation vectors
$\mathfrak{c} \, = \,  \left\{ \ft{n_1}{4}\,  , \, \ft{n_2}{4}\,  ,
\, \ft{n_3}{4}\right\}$. In view of eq.(\ref{direttocoriandolo}) we
could associate an explicit matrix to each group element of
$\mathrm{\mathrm{G_{1536}}}$, starting from the construction of the
Beltrami one-form field associated with the lowest lying
$6$-dimensional orbit  of the point group  in the momentum lattice.
We  considered such matrices as the defining representation having
verified that  the utilized representation is faithful.   Three
matrices are sufficient to characterize completely the defining
representation just as any other representation: the matrix
representing the generator $T$, the matrix representing the
generator $S$ and the matrix representing the translation $\left\{
\ft{n_1}{4}\,  , \, \ft{n_2}{4}\,  , \, \ft{n_3}{4}\right\}$. We
found:
\begin{equation}\label{TSindefi1536}
    \mathfrak{R}^{\mbox{defi}}[T] \, = \, \left(\begin{array}{llllll}
 0 & 0 & 0 & 0 & 1 & 0 \\
 0 & 0 & 0 & 0 & 0 & 1 \\
 0 & 1 & 0 & 0 & 0 & 0 \\
 1 & 0 & 0 & 0 & 0 & 0 \\
 0 & 0 & 0 & 1 & 0 & 0 \\
 0 & 0 & 1 & 0 & 0 & 0
\end{array}\right)\quad ; \quad \mathfrak{R}^{\mbox{defi}}[S]\, = \, \left(
\begin{array}{llllll}
 0 & 0 & 0 & 0 & 1 & 0 \\
 0 & 0 & 0 & 0 & 0 & 1 \\
 0 & 1 & 0 & 0 & 0 & 0 \\
 1 & 0 & 0 & 0 & 0 & 0 \\
 0 & 0 & 0 & 1 & 0 & 0 \\
 0 & 0 & 1 & 0 & 0 & 0
\end{array}
\right)
\end{equation}
\begin{eqnarray}
   \mathcal{M}_{\{\ft{2 n_1}{2},\ft{2 n_2}{2},\ft{2 n_3}{2}\}}^{\mbox{defi}} &= & \left(
\begin{array}{llllll}
 \cos \left(\ft{\pi}{2}  n _3\right) & 0 & \sin \left(\ft{\pi}{2}  n _3\right) & 0 & 0 & 0 \\
 0 & \cos \left(\ft{\pi}{2}  n _2\right) & 0 & 0 &
   -\sin \left(\ft{\pi}{2}  n _2\right) & 0 \\
 -\sin \left(\ft{\pi}{2}  n _3\right) & 0 & \cos
   \left(\ft{\pi}{2}  n _3\right) & 0 & 0 & 0 \\
 0 & 0 & 0 & \cos \left(\ft{\pi}{2}  n _1\right) & 0 &
   \sin \left(\ft{\pi}{2}  n _1\right) \\
 0 & \sin \left(\ft{\pi}{2}  n _2\right) & 0 & 0 & \cos
   \left(\ft{\pi}{2}  n _2\right) & 0 \\
 0 & 0 & 0 & -\sin \left(\ft{\pi}{2}  n _1\right) & 0 &
   \cos \left(\ft{\pi}{2}  n _1\right)
\end{array}
\right)\nonumber\\
\label{trasladefi1536}
\end{eqnarray}
Relying on the above matrices, any of the 1536 group elements obtains an explicit $6\times 6$ matrix representation
upon use of formula (\ref{direttocoriandolo}).
\subsubsection{Structure of the group $\mathrm{\mathrm{G_{1536}}}$ and description of its irreps}
The identity card of a finite group is given by the organization of
its elements into conjugacy classes, the list of its irreducible
representation and finally its character table. Since ours is not
any of the crystallographic groups, no explicit information is
available in the literature about its conjugacy classes, its irreps
and its character table. Hence in \cite{arnolderie} we were forced
to do everything from scratch by ourselves and we could accomplish
the task by means   of purposely written MATHEMATICA codes. Our
results were presented in the form of tables in the appendices
\cite{arnolderie}.
\paragraph{Conjugacy Classes and Irreps}
There are 37 conjugacy classes whose populations is distributed as
follows:
\begin{description}
  \item[1)]  2 classes of length 1
  \item[2)]  2  classes of length 3
  \item[3)]  2 classes of  length 6
  \item[4)] 1 class of length 8
  \item[5)] 7 classes of length 12
  \item[6)] 4 classes of length 24
  \item[7)] 13 classes of length 48
  \item[8)] 2 classes of length 96
  \item[9)] 4 classes of length 128
\end{description}
It follows that there must be $37$  irreducible representations
whose construction was performed in \cite{arnolderie} relying on the
method of induced representation and on the following  chain of
normal subgroups:
\begin{equation}\label{pernicinormaliText}
    \mathrm{\mathrm{G_{1536}} }\, \rhd \, \mathrm{G_{768}} \, \rhd \, \mathrm{G_{256}} \,
    \rhd \, \mathrm{G_{128}} \, \rhd \, \mathrm{G_{64}}
  \end{equation}
  where $\mathrm{G_{64}} \,  \sim \, \mathbb{Z}_4 \, \times \, \mathbb{Z}_4 \, \times \, \mathbb{Z}_4$
  is abelian and corresponds to the compactified translation group. The above chain leads to the following quotient groups:
  \begin{equation}\label{fagianirossiText}
    \frac{\mathrm{\mathrm{G_{1536}} }}{\mathrm{G_{768}}} \, \sim \,
    \mathbb{Z}_2 \quad ; \quad \frac{\mathrm{G_{768} }}{\mathrm{G_{256}}} \, \sim \,
    \mathbb{Z}_3 \quad ; \quad \frac{\mathrm{G_{256} }}{\mathrm{G_{128}}} \, \sim \, \mathbb{Z}_2
    \quad ; \quad \frac{\mathrm{G_{128} }}{\mathrm{G_{64}}} \, \sim \, \mathbb{Z}_2
  \end{equation}
The result for the irreducible representations, thoroughly described
in \cite{arnolderie}  is summarized here. The $37$ irreps are
distributed according to the following pattern:
\begin{description}
  \item[a)] 4 irreps of dimension $1$, namely $\mathrm{D}_1,\dots,\mathrm{D}_4$
  \item[b)] 2 irreps of dimension $2$, namely $\mathrm{D}_5,\dots,\mathrm{D}_6$
  \item[c)] 12 irreps of dimension $3$, namely $\mathrm{D}_6,\dots,\mathrm{D}_{18}$
  \item[d)] 10 irreps of dimension $6$, namely $\mathrm{D}_7,\dots,\mathrm{D}_{28}$
  \item[e)] 3 irreps of dimension $8$, namely $\mathrm{D}_{29},\dots,\mathrm{D}_{31}$
  \item[f)] 6 irreps of dimension $12$, namely $\mathrm{D}_{32},\dots,\mathrm{D}_{37}$
\end{description}
For the  character table we refer the reader to \cite{arnolderie}.
\section{The Classification of Arnold Beltrami one-forms is relevant to Supergravity/Brane Physics}
\label{classificazia}
In the previous sections we summarized, with a notation slightly adapted to the new perspective
of Spergravity/Brane Physics the main constructive points of \cite{arnolderie} whose perspective was instead that of
Hydrodynamics and Dynamical Systems. The core result of \cite{arnolderie} was the complete classification of
Arnold-Beltrami flows (one-forms in the new perspective) which turns out to be completely group theoretical. As we
already stressed in the introduction this classification happens to be relevant to Supergravity, since as we show in
section \ref{twobranastoria} and the following ones, each triplet of Arnold-Beltrami one-forms forming a
three-dimensional orthogonal representation $\mathrm{D}_x[\mathrm{G_{brane}},3]\subset \mathrm{O(3)}$ of a discrete
subgroup $\mathrm{G_{brane}} \subset \mathrm{G_{1536}}$ can be used as a multiplet of $1$-form fluxes in the transverse
directions within the framework of an exact $2$-brane solution of Minimal $D=7$ supergravity. The discrete group
$\mathrm{G_{brane}} \subset \mathrm{O(3)}$ will thus become an exact symmetry of such a solution and will be
transmitted as a discrete symmetry to the Maxwell-Chern Simons gauge theory leaving on the brane world volume.
\par
Indeed the irreducible representations of the universal classifying
group were a fundamental tool in our classification of the
Arnold-Beltrami one-forms. By choosing the various  point group
orbits of momentum vectors in the cubic lattice, according to their
classification presented below, and constructing the corresponding
Arnold-Beltrami fields we obtained all of the $37$ irreducible
representations of $\mathrm{\mathrm{G_{1536}}}$. Each representation
appears at least once and some of them appear several times.
Considering next the available subgroups $\mathrm{H}$ of
$\mathrm{\mathrm{G_{1536}}}$ and the branching rules of
$\mathrm{\mathrm{G_{1536}}}$ irreps with respect to $\mathrm{H}$ we
obtained an explicit algorithm to construct Arnold-Beltrami vector
fields with prescribed covariance groups $\mathrm{H}$. It suffices
to select a three dimensional representation of such subgroup in the
branching rules and, provided it is orthogonal, which is generically
true, one has the correct input for a $2$-brane solution of $D=7$
supergravity with $\mathrm{H}$ symmetry.
\subsection{Classification of the $48$ types of orbits}
\begin{figure}[!hbt]
\begin{center}
\iffigs
\includegraphics[height=70mm]{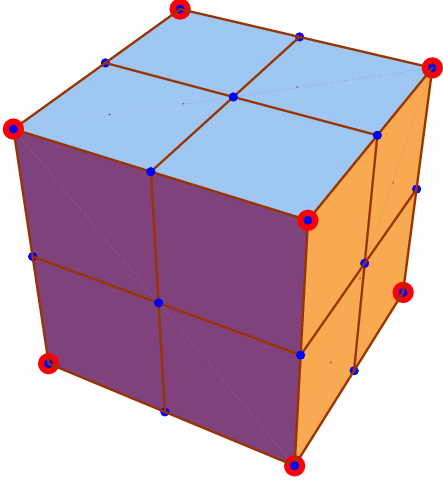}
\else
\end{center}
 \fi
\caption{\it The momenta in the cubic lattice
 forming an orbit of length $8$ under the octahedral group are of the form $\{\pm n,\pm n,\pm n\}$,
  and correspond to the vertices of a regular cube.}
 \label{cubamezzi}
 \iffigs
 \hskip 1cm \unitlength=1.1mm
 \end{center}
  \fi
\end{figure}

\begin{figure}[!hbt]
\begin{center}
\iffigs
\includegraphics[height=70mm]{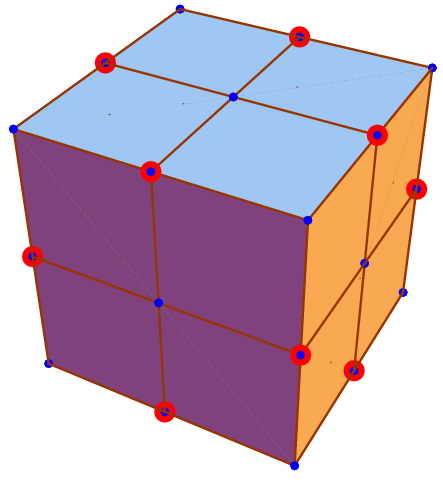}
\else
\end{center}
 \fi
\caption{\it The momenta in the cubic lattice
 forming an orbit of length $12$ under the octahedral group are of the form $\{\pm n,\pm n,0\}$,$\{0,\pm n,\pm n\}$,
 $\{\pm n,0, \pm n\}$
  and correspond to mid-points of the edges of a regular cube.}
 \label{cubavertici}
 \iffigs
 \hskip 1cm \unitlength=1.1mm
 \end{center}
  \fi
\end{figure}

\begin{figure}[!hbt]
\begin{center}
\iffigs
\includegraphics[height=70mm]{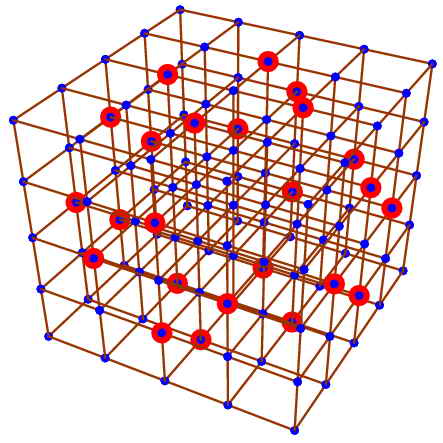}
\includegraphics[height=70mm]{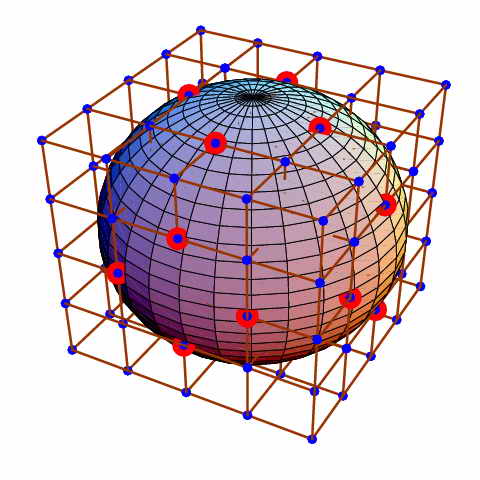}
\else
\end{center}
 \fi
\caption{\it  A view of an  orbit of length 24 in the cubic lattice: the lattice points are of the form $\{\pm a, \pm a,\pm b\}
$, $\{\pm a, \pm b,\pm a\}$,$\{\pm b, \pm a,\pm a\}$ and intersect the sphere of radius $r^2 \, = \, 2 a^2+b^2$}
\label{orbita24PS}
 \iffigs
 \hskip 1cm \unitlength=1.1mm
 \end{center}
  \fi
\end{figure}

The key observation is that the  group $\mathrm{\mathrm{G_{1536}}}$
has a finite number of irreducible representations so that,
irrespectively of the infinite number of spherical layers in the
momentum lattice, the number of different types of Arnold-Beltrami
one-forms has also got to be finite,  namely as many as the 37
irreps, times the number of different ways to obtain them from
orbits of length 6,8,12 or 24. The second observation was the key
role of the number $4$ introduced by Frobenius congruences which was
already the clue to the definition of $\mathrm{\mathrm{G_{1536}}}$.
What we found in \cite{arnolderie}  is that the various orbits
should be defined with integers modulo $4$.  In other words that we
should just consider the possible octahedral  orbits on a lattice
with coefficients in $\mathbb{Z}_4$ rather than $\mathbb{Z}$. The
easy guess, which was confirmed by computer calculations, is that
the pattern of $\mathrm{\mathrm{G_{1536}}}$ representations obtained
from the construction of Arnold-Beltrami vector fields according to
the algorithm of section \ref{algoritmo} depends only on the
equivalence classes of momentum orbits modulo $4$. Hence we found a
finite number of such orbits and a finite number of Arnold-Beltrami
one-form fields which we summarize. Let us stress that an embryo of
the exhaustive classification of orbits constructed in
\cite{arnolderie} was introduced by Arnold in his paper
\cite{arnoldorussopapero}. Arnold's  was only an embryo of the
complete classification for the following two reasons:
\begin{enumerate}
  \item The type of momenta orbits were partitioned  according to $odd$ and $even$
  (namely according to $\mathbb{Z}_2$, rather than $\mathbb{Z}_4$)
  \item The classifying group was taken to be the crystallographic $\mathrm{GS_{24}}$,
  as defined  \cite{arnolderie}, which is too small in comparison with the universal
  classifying group identified by us in $\mathrm{\mathrm{G_{1536}}}$.
\end{enumerate}
\begin{figure}[!hbt]
\begin{center}
\iffigs
\includegraphics[height=70mm]{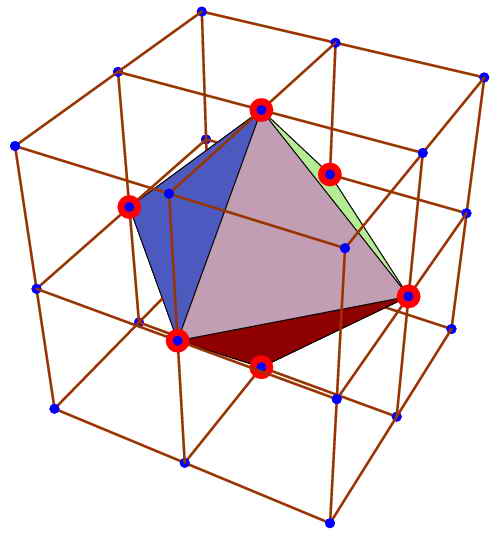}
\else
\end{center}
 \fi
\caption{\it The momenta in the cubic lattice
 forming an orbit of length $6$ under the octahedral group are of the form $\{\pm n,0,0\}$, $\{0,\pm n,0\}$,
 $\{0,0,\pm n,\}$ and correspond to the vertices of a regular octahedron.}
 \label{ottavertici}
 \iffigs
 \hskip 1cm \unitlength=1.1mm
 \end{center}
  \fi
\end{figure}
The complete classification of point orbits in the momentum lattice
is organized as follows. First we subdivide the momenta into five
groups:
\begin{description}
  \item[A)] Momenta of type $\{a,0,0\}$ which generate $\mathrm{O_{24}}$
  orbits of length 6 and representations of the universal group $\mathrm{G}_{1536}$ also of dimensions 6
  (see fig.\ref{ottavertici}).
  \item[B)]Momenta of type $\{a,a,a\}$ which generate $\mathrm{O_{24}}$ orbits of length 8
  and representations of the universal group $\mathrm{G}_{1536}$ also of dimensions 8 (see fig.\ref{cubavertici}).
  \item[C)] Momenta of type $\{a,a,0\}$ which generate $\mathrm{O_{24}}$ orbits of length 12
  and representations of the universal group $\mathrm{G}_{1536}$ also of dimensions 12 (see fig\ref{cubamezzi}).
  \item[D)] Momenta of type $\{a,a,b\}$ which generate $\mathrm{O_{24}}$ orbits of length 24
  and representations of the universal group $\mathrm{G}_{1536}$ also of dimensions 24 (see fig\ref{orbita24PS}).
  \item[E)] Momenta of type $\{a,b,c\}$ which generate $\mathrm{O_{24}}$ orbits
  of length 24 and representations of the universal group $\mathrm{G}_{1536}$ of dimensions 48.
\end{description}
The reason why in the cases A)\dots D) the dimension of the representation $\mathfrak{R}\left(\mathrm{\mathrm{G_{1536}}}\right)$
coincides with the dimension $|\mathcal{O}|$ of the orbit  is simple. For each momentum in the orbit
($\forall\mathbf{k}_i\, \in
\, \mathcal{O}$) also its negative is in the same orbit ($- \, \mathbf{k}_i\, \in \, \mathcal{O}$),
hence the number of arguments
$\Theta_i \, \equiv \, 2\pi \,\mathbf{k}_i \cdot \mathbf{X}$ of the independent trigonometric functions $\sin \left(
\Theta_i\right)$ and $\cos \left(\Theta_i\right)$ is $\ft 12 \times 2 |\mathcal{O}| \, = \, |\mathcal{O}|$ since $\sin \left(
\pm\Theta_i\right) \, = \, \pm  \sin \left( \Theta_i\right)$ and $\cos \left( \pm\Theta_i\right) \, = \, \cos \left(
\Theta_i\right)$. \par In case E), instead, the negatives of all the members of the orbit $\mathcal{O}$ are not in $\mathcal{O}$.
The number of independent trigonometric functions is therefore 48 and such is the dimension of the representation
$\mathfrak{R}\left(\mathrm{\mathrm{G_{1536}}}\right)$.
\par
In each of the five groups one still has  to reduce the entries to $\mathbb{Z}_4$,
namely to consider their equivalence class
$\mathrm{mod}\,4$. Each different choice of the pattern of $\mathbb{Z}_4$
classes appearing in an orbit leads to a different decomposition of
the representation into irreducible representation of $\mathrm{G}_{1536}$.
A simple consideration of the combinatorics leads to the conclusion
that there are in total $48$ cases to be considered. The very significant result is that all of the $37$
irreducible representations of
$\mathrm{G}_{1536}$ appear at least once in the list of these decompositions. Hence for all the \textit{irrepses}
of this group one can find a
corresponding Beltrami field and for some \textit{irrepses} such a
 Beltrami field admits a few inequivalent realizations.
For the list of the
$48$ distinct types of momentum vector classes and hence
of Arnold Beltrami one-forms we refer the reader to \cite{arnolderie}.
\par
\section{The Discrete Group $\mathrm{GF_{192}}$}
\label{giffone192}
 Among the various subgroups $\mathrm{H}\subset \mathrm{G_{1536}}$ that were listed in
\cite{arnolderie} and for which we provided there a full description together with their character tables, we have
chosen a particular one, namely $\mathrm{GF_{192}}$ as our main token in order  to illustrate by means of explicit
examples the construction of $2$-brane solutions of $D=7$ supergravity. This  choice is motivated by the fact that the
Arnold-Beltrami one-forms extracted from the $6$-dimensional orbits of lattice momenta split into three-dimensional
representations of this rather large space group. At the same time three-dimensional representations of the same space
group occur also in the branching of other irreducible representations of $\mathrm{G_{1536}}$ generated by other type
of momentum orbits, for instance by those of length $12$ (see \cite{arnolderie} for more details). In this section we
discuss the structure of this remarkable space group that, by means of supergravity, can be promoted to a discrete
symmetry group of three-dimensional Maxwell--Chern--Simons gauge theories.
\par
The precise definition of this group containing $192$-elements is provided in appendix \ref{DNAofGF192}, where we list
all of its elements, organized into 20 conjugacy classes. The nomenclature utilized to identify  the elements is
follows the notation introduced in eq.(\ref{elementando1536}). An intrinsic description of the group can be given in
terms of generators and relations. Let us choose the following three elements of the group:
\begin{equation}
\begin{array}{ccccc}
  S &=& \left\{4_6,0,0,\frac{1}{2}\right\} & = & \left(
\begin{array}{llllll}
 -\frac{1}{2} & \frac{1}{2} & 0 & -\frac{1}{2} &
   -\frac{1}{2} & 0 \\
 \frac{1}{2} & \frac{1}{2} & 0 & -\frac{1}{2} &
   \frac{1}{2} & 0 \\
 0 & 0 & -1 & 0 & 0 & 0 \\
 -\frac{1}{2} & -\frac{1}{2} & 0 & -\frac{1}{2} &
   \frac{1}{2} & 0 \\
 -\frac{1}{2} & \frac{1}{2} & 0 & \frac{1}{2} &
   \frac{1}{2} & 0 \\
 0 & 0 & 0 & 0 & 0 & 1
\end{array}
\right) \\
  Z &=& \left\{1_1,0,0,1\right\} & = & \left(
\begin{array}{llllll}
 0 & 0 & 0 & -1 & 0 & 0 \\
 0 & 0 & 0 & 0 & -1 & 0 \\
 0 & 0 & 1 & 0 & 0 & 0 \\
 -1 & 0 & 0 & 0 & 0 & 0 \\
 0 & -1 & 0 & 0 & 0 & 0 \\
 0 & 0 & 0 & 0 & 0 & 1
\end{array}
\right) \\
  T &=& \left\{2_8,\frac{1}{2},\frac{3}{2},1\right\} & = & \left(
\begin{array}{llllll}
 -\frac{1}{2} & -\frac{1}{2} & 0 & \frac{1}{2} &
   \frac{1}{2} & 0 \\
 -\frac{1}{2} & 0 & \frac{1}{2} & -\frac{1}{2} & 0 &
   \frac{1}{2} \\
 0 & -\frac{1}{2} & \frac{1}{2} & 0 & -\frac{1}{2} &
   -\frac{1}{2} \\
 -\frac{1}{2} & \frac{1}{2} & 0 & \frac{1}{2} &
   -\frac{1}{2} & 0 \\
 \frac{1}{2} & 0 & \frac{1}{2} & \frac{1}{2} & 0 &
   \frac{1}{2} \\
 0 & \frac{1}{2} & \frac{1}{2} & 0 & \frac{1}{2} &
   -\frac{1}{2}
\end{array}
\right)
\end{array}\label{GF192gen}
\end{equation}
where the $6\times 6$ representation is that provided by the $6\times 6$ defining representation of the parent group
$\mathrm{G_{1536}}$. By direct evaluation on can verify that all elements of the group $\mathrm{GF_{192}}$ can be
generated by multiple products of the generators (\ref{GF192gen}). These latter satisfy the following relations
\begin{eqnarray}
  S^2 \,=\, Z^2 \, = \, T^6 & = & \mathbf{1} \nonumber\\
 (S\,T)^4 \,=\, (Z\,T)^3 \, = \, (Z\,S)^2 & = & \mathbf{1}\label{relatoriGF192}
\end{eqnarray}
which can be taken as an intrinsic definition of the corresponding abstract group. The possibility of constructing all
the irreducible representations of $\mathrm{GF_{192} }$ by means of an induction procedure is related with its
solvability in terms of the following chain of normal subgroups:
\begin{equation}\label{pernicirossine}
    \mathrm{GF_{192} }\, \rhd \, \mathrm{GF_{96}} \, \rhd \, \mathrm{GF_{48}} \, \rhd \, \mathrm{GF_{16}}
    \end{equation}
    where the last element of the chain is abelian.
\par
The group $\mathrm{G_{192}}$ has 20 conjugacy classes and therefore it has  $20$  irreducible representations that are
distributed according to the following pattern:
\begin{description}
  \item[a)] 4 irreps of dimension $1$, namely $\mathrm{D_1,\dots,D_4}$
  \item[b)] 12 irreps of dimension $3$, namely $\mathrm{D_5,\dots,D_{16}}$
  \item[c)] 2 irreps of dimension $2$, namely $\mathrm{D_{17},D_{18}}$
  \item[d)] 2 irreps of dimension $6$, namely $\mathrm{D_{19},D_{20}}$
\end{description}
These representations were calculated in \cite{arnolderie} to which we refer for further details. The character table
is recalled below, where  by $\epsilon$ we have denoted the cubic root of unity $\epsilon = \exp\left[\frac{2\pi}{3} \,
{\rm i}\right]$. {\scriptsize
\begin{equation}\label{charto192}
\begin{array}{|l|llllllllllllllllllll|}
\hline
 0 & C_1 & C_2 & C_3 & C_4 & C_5 & C_6 & C_7 & C_8 & C_9 & C_{10} & C_{11} &
   C_{12} & C_{13} & C_{14} & C_{15} & C_{16} & C_{17} & C_{18} & C_{19} &
   C_{20} \\
   \hline
 D_1 & 1 & 1 & 1 & 1 & 1 & 1 & 1 & 1 & 1 & 1 & 1 & 1 & 1 & 1 & 1 & 1 & 1 & 1
   & 1 & 1 \\
 D_2 & 1 & 1 & 1 & 1 & 1 & 1 & 1 & 1 & 1 & 1 & -1 & -1 & -1 & -1 & -1 & -1 &
   -1 & -1 & 1 & 1 \\
 D_3 & 1 & -1 & -1 & 1 & 1 & -1 & 1 & -1 & -1 & 1 & 1 & -1 & -1 & 1 & 1 & -1
   & -1 & 1 & 1 & -1 \\
 D_4 & 1 & -1 & -1 & 1 & 1 & -1 & 1 & -1 & -1 & 1 & -1 & 1 & 1 & -1 & -1 & 1
   & 1 & -1 & 1 & -1 \\
 D_5 & 3 & -3 & -3 & 3 & -1 & 1 & -1 & 1 & 1 & -1 & -1 & 1 & 1 & -1 & 1 & -1
   & -1 & 1 & 0 & 0 \\
 D_6 & 3 & -3 & -3 & 3 & -1 & 1 & -1 & 1 & 1 & -1 & 1 & -1 & -1 & 1 & -1 & 1
   & 1 & -1 & 0 & 0 \\
 D_7 & 3 & 3 & 3 & 3 & -1 & -1 & -1 & -1 & -1 & -1 & -1 & -1 & -1 & -1 & 1 &
   1 & 1 & 1 & 0 & 0 \\
 D_8 & 3 & 3 & 3 & 3 & -1 & -1 & -1 & -1 & -1 & -1 & 1 & 1 & 1 & 1 & -1 & -1
   & -1 & -1 & 0 & 0 \\
 D_9 & 3 & 3 & -1 & -1 & -1 & 3 & 3 & -1 & -1 & -1 & -1 & 1 & -1 & 1 & 1 & 1
   & -1 & -1 & 0 & 0 \\
 D_{10} & 3 & 3 & -1 & -1 & -1 & 3 & 3 & -1 & -1 & -1 & 1 & -1 & 1 & -1 & -1
   & -1 & 1 & 1 & 0 & 0 \\
 D_{11} & 3 & -3 & 1 & -1 & -1 & -3 & 3 & 1 & 1 & -1 & -1 & -1 & 1 & 1 & 1 &
   -1 & 1 & -1 & 0 & 0 \\
 D_{12} & 3 & -3 & 1 & -1 & -1 & -3 & 3 & 1 & 1 & -1 & 1 & 1 & -1 & -1 & -1
   & 1 & -1 & 1 & 0 & 0 \\
 D_{13} & 3 & 3 & -1 & -1 & 3 & -1 & -1 & 3 & -1 & -1 & -1 & 1 & -1 & 1 & -1
   & -1 & 1 & 1 & 0 & 0 \\
 D_{14} & 3 & 3 & -1 & -1 & 3 & -1 & -1 & 3 & -1 & -1 & 1 & -1 & 1 & -1 & 1
   & 1 & -1 & -1 & 0 & 0 \\
 D_{15} & 3 & -3 & 1 & -1 & 3 & 1 & -1 & -3 & 1 & -1 & -1 & -1 & 1 & 1 & -1
   & 1 & -1 & 1 & 0 & 0 \\
 D_{16} & 3 & -3 & 1 & -1 & 3 & 1 & -1 & -3 & 1 & -1 & 1 & 1 & -1 & -1 & 1 &
   -1 & 1 & -1 & 0 & 0 \\
 D_{17} & 2 & 2 & 2 & 2 & 2 & 2 & 2 & 2 & 2 & 2 & 0 & 0 & 0 & 0 & 0 & 0 & 0
   & 0 & \epsilon  (\epsilon +1) & \epsilon  (\epsilon +1) \\
 D_{18} & 2 & -2 & -2 & 2 & 2 & -2 & 2 & -2 & -2 & 2 & 0 & 0 & 0 & 0 & 0 & 0
   & 0 & 0 & \epsilon  (\epsilon +1) & -\epsilon  (\epsilon +1) \\
 D_{19} & 6 & 6 & -2 & -2 & -2 & -2 & -2 & -2 & 2 & 2 & 0 & 0 & 0 & 0 & 0 &
   0 & 0 & 0 & 0 & 0 \\
 D_{20} & 6 & -6 & 2 & -2 & -2 & 2 & -2 & 2 & -2 & 2 & 0 & 0 & 0 & 0 & 0 & 0
   & 0 & 0 & 0 & 0\\
   \hline
\end{array}
\end{equation}
}
\par
The most attractive feature of this particular subgroup of the Universal Classifying group is given by the 12
three-dimensional representations $D_5,\dots,D_{16}$. Any time we find one of these irreps in the decomposition with
respect to $\mathrm{GF_{192}}$ of any  of the 37 irreps of $\mathrm{G_{1536}}$, we can utilize the corresponding
Arnold-Beltrami one-forms as fluxes in a $2$-brane solution of supergravity. Indeed in order to construct a solution of supergravity
we are supposed to find an ansatz for the triplet of vector fields, $\mathbf{A}^\Lambda$ appearing in the bosonic spectrum of the theory. As we show in section \ref{pulcinella}, splitting the $7$ coordinates in a group of three spanning the brane world volume and a group of four $\{U,X,Y,Z\}$ transverse to the brane and further identifying the last three $\{X,Y,Z\} \equiv {\mathbf{X}}$ with those of a three torus $\mathrm{T^3}$, a convenient general ansatz is the following one:
\begin{equation}\label{golattus}
    \mathbf{A}^\Lambda \, = \, e^{2 \,\mu \, U} \, \mathcal{E}^\Lambda_{\phantom{\Lambda}I} \, \mathbf{Y}^I( {\mathbf{X}})
\end{equation}
where $\mathbf{Y}^A( {\mathbf{X}})$ is a basis set of solutions of Beltrami equation on the three-torus
(\ref{formaduale}) with eigenvalue $\mu$ and $\mathcal{E}^\Lambda_{\phantom{\Lambda}I}$ is an embedding matrix ($3
\times \mathrm{d}_\mu$) where $\mathrm{d}_\mu$ is the degeneracy of the eigenvalue $\mu$. The key point is that minimal
$D=7$ supergravity has a global symmetry $\mathrm{SO(3)}$ with respect to which the triplet of vector fields transform
in the defining three-dimensional representation. In the gauged version of the theory the same global symmetry
$\mathrm{SO(3)}$ is promoted to a local one and the three vector fields become an $\mathrm{SO(3)}$-gauge connection.
Consider now the action of a global discrete symmetry group $\Gamma$, like for instance $\mathrm{GF_{192}}$, on the
Arnold-Beltrami one-forms $\mathbf{Y}^I( {\mathbf{X}})$. We have:
\begin{equation}\label{carillus}
    \forall \, \gamma \, \in \, \Gamma \quad : \quad  \mathbf{Y}^I(\, \gamma \cdot  {\mathbf{X}}) \, = \, \mathbf{Y}^J(  {\mathbf{X}})  \, \mathrm{O}_J^{\phantom{J}I}(\gamma) \,
\end{equation}
where, by an appropriate choice of the basis elements $\mathbf{Y}^I(  {\mathbf{X}})$ the matrix $\mathrm{O}^I_{\phantom{I}J}(\gamma)$ can always be made orthogonal:
\begin{equation}\label{oronallo}
    \mathrm{O}(\gamma) \, \in \, \mathrm{O}(\mathrm{d}_\mu) \quad \Rightarrow \quad \Gamma \, \subset \, \mathrm{O}(\mathrm{d}_\mu)
\end{equation}
As stressed in the above equation it follows that the discrete group $\Gamma$ is always a subgroup of the orthogonal group in a dimension equal to the multiplicity of the Beltrami eigenvalue. On the other hand $\Gamma \subset \mathrm{SO(3)}$, since all the
considered $\Gamma$ have $3$-dimensional orthogonal representations. In this way the embedding matrix $\mathcal{E}^\Lambda_{\phantom{\Lambda}I}$ turns out to be an intertwining operator between irreducible representations of $\Gamma$. By Schur's Lemma, there are only  a few relevant type of cases:
\par
\begin{equation}\label{roncisvalle}
    \begin{array}{lcl}
       D_x(\Gamma,3) & \stackrel{\mathcal{E}}{\Leftrightarrow} & D_x(\Gamma,3) \\
      \left(
        \begin{array}{c|c}
         D_x(\Gamma,1) & 0 \\
         \hline
          0 & D_y(\Gamma,2) \\
        \end{array}
      \right)
        & \stackrel{\mathcal{E}}{\Leftrightarrow}& \left(
        \begin{array}{c|c}
         D_x(\Gamma,1) & 0 \\
         \hline
          0 & D_y(\Gamma,2) \\
        \end{array}
      \right) \\
       \left(
          \begin{array}{ccc}
            D_x(\Gamma,1) & 0 & 0 \\
            0& D_y(\Gamma,1)& 0 \\
            0 & 0 & D_z(\Gamma,1) \\
          \end{array}
        \right)
        & \stackrel{\mathcal{E}}{\Leftrightarrow} &  \left(
          \begin{array}{ccc}
            D_x(\Gamma,1) & 0 & 0 \\
            0& D_y(\Gamma,1)& 0 \\
            0 & 0 & D_z(\Gamma,1) \\
          \end{array}
        \right)
     \end{array}
\end{equation}
where $\mathcal{E}$ intertwines between identical one-dimensional, two-dimensional and three-dimensional representations. Actually if we restrict our attention to cases that admit an uplifting to the gauged theory, the triplet representation of $\Gamma$ must be compatible with the adjoint of $\mathrm{SO(3)}$. Hence the only admitted cases are those where the final $3 \times 3$ matrix has determinant one. The most natural choice  is clearly that where $\mathcal{E}$ intertwines between two identical irreducible, faithful representations of $\Gamma$ and for this reason $\Gamma \, = \, \mathrm{GF_{192}}$ is an inspiring choice. It is a rather large subgroup of the Universal Classifying Group whose numerous irreducible three-dimensional representations are almost ubiquitous in all point orbit constructions of Arnold-Beltrami one-forms.
For instance a quick survey of the branching rules presented in \cite{arnolderie} and of the assignments to
$\mathrm{G_{1536}}$-irreps of the Arnold-Beltrami one--forms constructed with point group orbits of lengths $6$, $8$ and
$12$ (this information is also  presented in \cite{arnolderie}) reveals that using just only these type of momenta we can
already construct triplets of Beltrami vector fields in the following eight irreducible three-dimensional
representations of $\mathrm{GF_{192}}$: $\quad \mathrm{D_5,D_6,D_7,D_8,D_9,D_{11},D_{13},D_{12},D_{12},D_{16}}$. In this paper
we will not consider all such constructions neither we will attempt a classification of all $2$-brane solutions. We
just confine ourselves to four examples that will illustrate  the main features of this new playing ground for
brane-physics.
\subsection{A triplet of Arnold--Beltrami fields in the representation $\mathrm{D_{12}}\left[\mathrm{GF_{192}},3\right]$
 from the octahedral orbit of $ {\mathbf{k}}=\{1,0,0\}$}
\label{k100gf192}
 The first example that we consider corresponds to the celebrated ABC-flow that, in the hydrodynamical
 literature, has been the focus of many investigations over the last 40 years \cite{arnoldorussopapero}, \cite{Childress}.
 \par
 According to the results of \cite{arnolderie}, if we construct the Arnold-Beltrami one-forms starting from the orbit
 of length six generated by the momentum $ {\mathbf{k}}=\{1,0,0\}$, we obtain a set of 6 one--forms that transform in
 the representation $\mathrm{D_{23}}\left[\mathrm{G_{1536}},6\right]$ of the Universal Classifying Group. The branching
 rule of such a representation with respect to the considered subgroup $\mathrm{GF_{192}}$ is the following one:
 \begin{equation}\label{d23split}
    \mathrm{D_{23}}\left[\mathrm{G_{1536}},6\right]\, = \,
  \mathrm{ D_{12}}\left[\mathrm{GF_{192}},3\right]+\mathrm{D_{15}}\left[\mathrm{GF_{192}},3\right]
 \end{equation}
A generic one-form in the triplet representation $\mathrm{ D_{12}}\left[\mathrm{GF_{192}},3\right]$ exactly corresponds
to one of the ABC-flows, the parameters $\mathrm{A,B,C}$ being the components of a generic vector in this
representation.
\par
Projecting onto the representation $\mathrm{ D_{12}}\left[\mathrm{GF_{192}},3\right]$ we find the following basis of
 one--forms:
\begin{equation}
\mathrm{ D_{12}}\left[\mathrm{GF_{192}},3\right]\, \ni\,\mathbf{Y}^I \, = \,\left\{\begin{array}{rcl}
  \mathbf{Y}^1\left(X,Y,Z\right) &=& 2\, \cos (2 \pi  Z) \mathrm{d}X-2 \,\mathrm{d}Y \sin (2 \pi  Z) \\
 \mathbf{Y}^2\left(X,Y,Z\right) &=& 2\, \cos (2 \pi  Y) \mathrm{d}X+2\,\mathrm{d}Z \sin (2 \pi  Y))\\
 \mathbf{Y}^3\left(X,Y,Z\right)&=& 2\, \cos (2 \pi  X) \mathrm{d}Y-2\, \mathrm{d}Z \sin (2 \pi  X)\\
 \end{array}\right.
\end{equation}
that satisfy Beltrami equation with eigenvalue $\mu\, = \, 2\pi$:
\begin{equation}\label{autovalloUno}
    \star \, \mathrm{d} \mathbf{Y}^I \, = \, \mu \, \mathbf{Y}^I \quad ; \quad \mu \, = \, 2\pi \,
    \sqrt{\mathbf{k} \cdot \mathbf{k} } \, = \,\, 2\pi
\end{equation}
 The components of these one--forms are easily extracted utilizing the definition:
\begin{equation}\label{compentati}
    \mathbf{Y}^I \, = \, \mathbf{Y}^I_{i} \, \mathrm{d}X^i \quad ; \quad \mathrm{d}X^i \, = \, \left\{\mathrm{d}X,\mathrm{d}Y,\mathrm{d}Z \right\}
\end{equation}
and the explicit action of the $\mathrm{GF_{192}}$ group on these one-forms is specified once we give the explicit action
of the three generators $S,Z,T$. This latter is the following one:
\begin{equation}\label{gomitolone}
    \begin{array}{rccclcccc}
       S & : & \{X,Y,Z\} & \to & \left\{Y,X,\frac{1}{4}-Z\right\} &\Rightarrow & \mathrm{D_{12}}[S] & = & \left(
\begin{array}{lll}
 -1 & 0 & 0 \\
 0 & 0 & 1 \\
 0 & 1 & 0
\end{array}
\right) \\
\null&\null&\null&\null&\null&\null&\null&\null&\null\\
       Z & : & \{X,Y,Z\} & \to & \left\{X,Y,Z+\frac{1}{2}\right\} & \Rightarrow & \mathrm{D_{12}[Z]} & = & \left(
\begin{array}{lll}
 -1 & 0 & 0 \\
 0 & 1 & 0 \\
 0 & 0 & 1
\end{array}
\right) \\
\null&\null&\null&\null&\null&\null&\null&\null&\null\\
       T & : & \{X,Y,Z\} & \to & \left\{Y+\frac{1}{4},Z+\frac{3}{4},X+\frac{1}{2}\right\} & \Rightarrow & \mathrm{D_{12}}[T]
       & = & \left(
\begin{array}{lll}
 0 & -1 & 0 \\
 0 & 0 & -1 \\
 -1 & 0 & 0
\end{array}
\right)
     \end{array}
\end{equation}
\begin{eqnarray}
\mathbf{Y}\left(Y,X,\frac{1}{4}-Z\right)  &=&  \mathbf{Y}\left(X,Y,Z\right)\cdot\mathrm{D_{12}}[S]  \nonumber\\
 \mathbf{Y}\left(X,Y,Z+\frac{1}{2}\right)&=& \mathbf{Y}\left(X,Y,Z\right)\cdot\mathrm{D_{12}}[Z] \nonumber\\
  \mathbf{Y}\left(Y+\frac{1}{4},Z+\frac{3}{4},X+\frac{1}{2}\right)&=&  \mathbf{Y}\left(X,Y,Z\right)\cdot\mathrm{D_{12}}[T]
  \label{transformoD12}
\end{eqnarray}
Next we calculate the matrix of scalar products of the basis one-forms:
\begin{equation}\label{consapel}
    gY^{IJ} \, \equiv \, \langle \mathbf{Y}^I \, , \, \mathbf{Y}^J \rangle \, \equiv \,  \sum_{i=1}^3\, \mathbf{Y}^I_i \mathbf{Y}^J_i
\end{equation}
and we find:
\begin{equation}\label{gYrobus}
 gY^{IJ} \, = \,   \left(
\begin{array}{lll}
 4 & 4 \cos (2 \pi  Y) \cos (2 \pi  Z) & -4 \cos (2 \pi  X) \sin (2 \pi  Z) \\
 4 \cos (2 \pi  Y) \cos (2 \pi  Z) & 4 & -4 \sin (2 \pi  X) \sin (2 \pi  Y) \\
 -4 \cos (2 \pi  X) \sin (2 \pi  Z) & -4 \sin (2 \pi  X) \sin (2 \pi  Y) & 4
\end{array}
\right)
\end{equation}
The trace of this matrix, which plays an important role in the $2$-brane construction is in this case a constant:
\begin{eqnarray}\label{tracciatus}
     \mbox{Tr}\left(gY_{IJ}\right) &\equiv & \Lambda_{\mathbf{1,0,0}}^{\mathrm{D_{12}}(\mathrm{GF_{192}},3)} \, + \,
     \mathfrak{J}_{\mathbf{1,0,0}}^{\mathrm{D_{12}}(\mathrm{GF_{192}},3)}\left(X,Y,Z\right)\nonumber\\
\Lambda_{\mathbf{1,0,0}}^{\mathrm{D_{12}}(\mathrm{GF_{192}},3)} & = & 12\nonumber\\
\mathfrak{J}_{\mathbf{1,0,0}}^{\mathrm{D_{12}}(\mathrm{GF_{192}},3)}\left(X,Y,Z\right)& = & 0
\end{eqnarray}
\subsection{A singlet of Arnold--Beltrami fields in the representation $\mathrm{D_{1}}\left[\mathrm{GS_{24}},1\right]$
 from the octahedral orbit of $ {\mathbf{k}}=\{1,0,0\}$}
 \label{k100gs24}
 The second example that we consider corresponds to the case of the  AAA-flow, namely to the subcase of the ABC-flows that is
 invariant under the subgroup $\mathrm{GS_{24}}\subset \mathrm{GF_{192}}$. As explained in \cite{arnolderie},
 under the subgroup $\mathrm{GS_{24}}$, which is isomorphic to the octahedral group and is thoroughly described in
 appendix \ref{coniugatoGS24}, the irreducible representation $\mathrm{D_{12}}(\mathrm{GF_{192}},3)$ branches as follows:
\begin{equation}\label{sparatus}
    \mathrm{D_{12}}(\mathrm{GF_{192}},3)\, = \,\mathrm{D_{1}}(\mathrm{GS_{24}},1)\oplus\mathrm{D_{3}}(\mathrm{GS_{24}},2)
\end{equation}
This means that from the construction considered in section \ref{k100gf192} of  Arnold-Beltrami fields associated with
the orbit  of the momentum vector $ {\mathbf{k}}=\{1,0,0\}$,   we can extract a vector field that is in the singlet
representation of $\mathrm{GS_{24}}$.
\par
Projecting onto this singlet, according with eq.(\ref{roncisvalle}) we obtain the following basis of one--forms:
\begin{equation}
\mathrm{ D_{1}}\left[\mathrm{GS_{24}},1\right]\, \ni\,\mathbf{Y}^I \, = \,\left\{\begin{array}{rcl}
  \mathbf{Y}^1\left(X,Y,Z\right) &=& 2  {dY} \cos [2 \pi  X]+2  {dX} \cos [2 \pi  Y]+2  {dX}
   \cos [2 \pi  Z]\\
   &&-2  {dZ} \sin [2 \pi  X]+2  {dZ} \sin [2 \pi
    Y]-2  {dY} \sin [2 \pi  Z]\\
 \mathbf{Y}^2\left(X,Y,Z\right) &=& 0\\
 \mathbf{Y}^3\left(X,Y,Z\right)&=& 0\\
 \end{array}\right.
\end{equation}
satisfying Beltrami equation with eigenvalue $\mu\, = \, 2\pi$:
\begin{equation}\label{autovalloUnoBis}
    \star \, \mathrm{d} \mathbf{Y}^I \, = \, \mu \, \mathbf{Y}^I \quad ; \quad \mu \, = \, 2\pi \,
    \sqrt{\mathbf{k} \cdot \mathbf{k} } \, = \,\, 2\pi
\end{equation}
 The components of these one--forms are easily extracted utilizing the definition:
\begin{equation}\label{compentati2}
    \mathbf{Y}^I \, = \, \mathbf{Y}^I_{i} \, \mathrm{d}X^i \quad ; \quad \mathrm{d}X^i \, = \, \left\{\mathrm{d}X,\mathrm{d}Y,\mathrm{d}Z \right\}
\end{equation}
The explicit action of the $\mathrm{GS_{24}}$ group on these one-forms is specified once we give the explicit action of
the two generators $S,T$, satisfying the defining relations:
\begin{equation}\label{relazionis2}
    S^2 \, = \, T^3 \, = \, (S\,T)^4 \, = \, \mathbf{e}
\end{equation}
Two such generators can be chosen as follows:
\begin{eqnarray}
\label{picciotto}
  S &\equiv& \left\{4_6,0,0,\frac{3}{2}\right\} \\
  T &\equiv& \left\{2_8,\frac{3}{2},\frac{1}{2},0\right\}
\end{eqnarray}
The action of these latter is the following one:
\begin{equation}\label{gomitolone2}
    \begin{array}{rccclcccc}
       S & : & \{X,Y,Z\} & \to & \left\{Y,X,\frac{3}{4}-Z\right\} &\Rightarrow & \mathrm{D_{1}}[S] & = & 1 \\
\null&\null&\null&\null&\null&\null&\null&\null&\null\\
       T & : & \{X,Y,Z\} & \to & \left\{Y+\frac{3}{4},Z+\frac{1}{4},X\right\} & \Rightarrow & \mathrm{D_{1}}[T]
       & = &  1\\
       \end{array}
\end{equation}
\begin{eqnarray}
\mathbf{Y}\left(Y,X,\frac{3}{4}-Z\right)  &=&  \mathbf{Y}\left(X,Y,Z\right)  \nonumber\\
  \mathbf{Y}\left(Y+\frac{3}{4},Z+\frac{1}{4},X\right)&=&  \mathbf{Y}\left(X,Y,Z\right)
  \label{transformoD1GS}
\end{eqnarray}
Next we calculate the matrix of scalar products of the basis one-forms:
\begin{equation}\label{consapel2bis}
    gY^{IJ} \, \equiv \, \langle \mathbf{Y}^I \, , \, \mathbf{Y}^J \rangle \, \equiv \,  \sum_{i=1}^3\, \mathbf{Y}^I_i \mathbf{Y}^J_i
\end{equation}
and we find:
\begin{equation}\label{gYrobus2}
  gY^{IJ} \, = \,\left(\begin{array}{ccc}
                   \Lambda_{\mathbf{1,0,0}}^{\mathrm{D_1}(\mathrm{GS_{24}},1)} \, + \,
     \mathfrak{J}_{\mathbf{1,0,0}}^{\mathrm{D_1}(\mathrm{GS_{24}},1)}\left(X,Y,Z\right) & 0 & 0 \\
                   0 & 0 & 0 \\
                   0 & 0 & 0
                 \end{array}\right)
\end{equation}
The trace of this matrix, which plays an important role in the $2$-brane construction is obviously equal to the first
and unique non vanishing element of the matrix (\ref{gYrobus2}):
\begin{eqnarray}\label{tracciatus2}
     \mbox{Tr}\left(gY_{IJ}\right) &\equiv & \Lambda_{\mathbf{1,0,0}}^{\mathrm{D_1}(\mathrm{GS_{24}},1)} \, + \,
     \mathfrak{J}_{\mathbf{1,0,0}}^{\mathrm{D_1}(\mathrm{GS_{24}},1)}\left(X,Y,Z\right)\nonumber\\
\Lambda_{\mathbf{1,0,0}}^{\mathrm{D_1}(\mathrm{GS_{24}},1)} & = & 12\label{lamgs24}\\
\mathfrak{J}_{\mathbf{1,0,0}}^{\mathrm{D_1}(\mathrm{GS_{24}},1)}\left(X,Y,Z\right)&
= & -4 \cos [2 \pi  (X-Y)]+4 \cos [2 \pi  (X+Y)]+4 \cos [2 \pi (Y-Z)]\nonumber\\
&&+4
   \cos [2 \pi  (Y+Z)]+4 \sin [2 \pi  (X-Z)]-4 \sin [2 \pi  (X+Z)]\label{jjgs24}
\end{eqnarray}
Defining the three-dimensional Laplacian:
\begin{equation}\label{laplacianusT3}
    \Delta_{\mathrm{T^3}} \, \equiv \, \frac{\partial^2}{\partial X^2} \, + \,\frac{\partial^2}{\partial Y^2}
    \, + \,\frac{\partial^2}{\partial Z^2}
\end{equation}
we can verify that the function $\mathfrak{J}_{\mathbf{1,0,0}}^{\mathrm{D_1}(\mathrm{GS_{24}},1)}\left(X,Y,Z\right)$
defined in eq.(\ref{jjgs24}) is invariant\footnote{It suffices to check invariance under the two transformations $S$,
and $T$.} under the full group $\mathrm{GS_{24}}$ and that it is an eigenfunction of $\Delta_{\mathrm{T^3}}$ with
eigenvalue $\lambda \, = \, - \, 2\,\mu^2$:
\begin{equation}\label{eigenfunziaD1GS}
    \Delta_{\mathrm{T^3}}\, \mathfrak{J}_{\mathbf{1,0,0}}^{\mathrm{D_1}(\mathrm{GS_{24}},1)}\left(X,Y,Z\right)
    \, = \, \lambda  \,
    \mathfrak{J}_{\mathbf{1,0,0}}^{\mathrm{D_1}(\mathrm{GS_{24}},1)}\left(X,Y,Z\right)\quad ; \quad \lambda
    \, = \, - \, 8\pi^2 \, = \, -\, 2\,\mu^2
\end{equation}
\subsection{A triplet of Arnold--Beltrami fields in the representation $\mathrm{D_{7}}\left[\mathrm{GF_{192}},3\right]$
 from the octahedral orbit of $ {\mathbf{k}}=\{2,0,0\}$}
 \label{k200d7}
 The third example that we consider is extracted from the orbit of length six
 generated by the momentum $ {\mathbf{k}}=\{2,0,0\}$. From this construction  we obtain a set of six one--forms that
 transform in the representation $\mathrm{D_{19}}\left[\mathrm{G_{1536}},6\right]$ of the Universal Classifying Group.
 As given in \cite{arnolderie}, the branching rule of such a representation with respect to
 the considered subgroup $\mathrm{GF_{192}}$ is the following one:
 \begin{equation}\label{d19split}
    \mathrm{D_{19}}\left[\mathrm{G_{1536}},6\right]\, = \,
  \mathrm{ D_{7}}\left[\mathrm{GF_{192}},3\right]+\mathrm{D_{8}}\left[\mathrm{GF_{192}},3\right]
 \end{equation}
Projecting onto the representation $\mathrm{ D_{7}}\left[\mathrm{GF_{192}},3\right]$ we find the following basis of
one--forms:
\begin{equation}
\mathrm{ D_{7}}\left[\mathrm{GF_{192}},3\right]\, \ni\,\mathbf{Y}^I \, = \,\left\{\begin{array}{rcl}
  \mathbf{Y}^1\left(X,Y,Z\right) &=& 2 (\cos (4 \pi  Y)+\cos (4 \pi  Z)) \mathrm{d}X+2 \mathrm{d}Z \sin (4 \pi  Y)
  -2 \mathrm{d}Y \sin (4 \pi  Z) \\
 \mathbf{Y}^2\left(X,Y,Z\right) &=& 2 (\cos (4 \pi  X)-\cos (4 \pi  Z)) \mathrm{d}Y
 -2 (\mathrm{d}Z \sin (4 \pi  X)+\mathrm{d}X \sin (4 \pi  Z))\\
 \mathbf{Y}^3\left(X,Y,Z\right)&=& 2 (\cos (4 \pi  X)+\cos (4 \pi  Y)) \mathrm{d}Z
 +2 \mathrm{d}Y \sin (4 \pi  X)-2 \mathrm{d}X \sin (4 \pi  Y)\\
 \end{array}\right.\nonumber\\
 \label{fieldD7}
\end{equation}
that satisfy Beltrami equation with eigenvalue $\mu\, = \, 4\pi$:
\begin{equation}\label{autovalloDue}
    \star \, \mathrm{d} \mathbf{Y}^I \, = \, \mu \, \mathbf{Y}^I\quad ; \quad \mu \, = \, 2\pi \,
    \sqrt{\mathbf{k} \cdot \mathbf{k} } \, = \,\, 4\pi
\end{equation}
 The components of these one--forms are easily
extracted  from:
\begin{equation}\label{compentati4}
    \mathbf{Y}^I \, = \, \mathbf{Y}^I_{i} \, \mathrm{d}X^i \quad ; \quad \mathrm{d}X^i \, = \, \left\{\mathrm{d}X,\mathrm{d}Y,\mathrm{d}Z \right\}
\end{equation}
In this representation the explicit action of  the three generators $S,Z,T$ is the following one:
\begin{equation}\label{gomitolone4}
    \begin{array}{rccclcccc}
       S & : & \{X,Y,Z\} & \to & \left\{Y,X,\frac{1}{4}-Z\right\} &\Rightarrow & \mathrm{D_{7}}[S] & = & \left(
\begin{array}{lll}
 0 & 1 & 0 \\
 1 & 0 & 0 \\
 0 & 0 & -1
\end{array}
\right) \\
\null&\null&\null&\null&\null&\null&\null&\null&\null\\
       Z & : & \{X,Y,Z\} & \to & \left\{X,Y,Z+\frac{1}{2}\right\} & \Rightarrow & \mathrm{D_{7}[Z]} & = & \left(
\begin{array}{lll}
 1 & 0 & 0 \\
 0 & 1 & 0 \\
 0 & 0 & 1
\end{array}
\right) \\
\null&\null&\null&\null&\null&\null&\null&\null&\null\\
       T & : & \{X,Y,Z\} & \to & \left\{Y+\frac{1}{4},Z+\frac{3}{4},X+\frac{1}{2}\right\} & \Rightarrow & \mathrm{D_{7}}[T]
       & = & \left(
\begin{array}{lll}
 0 & 0 & -1 \\
 1 & 0 & 0 \\
 0 & -1 & 0
\end{array}
\right)
     \end{array}
\end{equation}
and we have:
\begin{eqnarray}
\mathbf{Y}\left(Y,X,\frac{1}{4}-Z\right)  &=&  \mathbf{Y}\left(X,Y,Z\right)\cdot \mathrm{D_{7}}[S]  \nonumber\\
 \mathbf{Y}\left(X,Y,Z+\frac{1}{2}\right)&=&  \mathbf{Y}\left(X,Y,Z\right)\cdot \mathrm{D_{7}}[Z] \nonumber\\
  \mathbf{Y}\left(Y+\frac{1}{4},Z+\frac{3}{4},X+\frac{1}{2}\right)&=&  \mathbf{Y}\left(X,Y,Z\right)\cdot
  \mathrm{D_{7}}[T]\label{transformoD7}
\end{eqnarray}
Next we calculate the matrix of scalar products of the basis one-forms:
\begin{equation}\label{consapel3}
    gY^{IJ} \, \equiv \, \langle \mathbf{Y}^I \, , \, \mathbf{Y}^J \rangle \, \equiv \,  \sum_{i=1}^3\, \mathbf{Y}^I_i \mathbf{Y}^J_i
\end{equation}
We do not display this matrix since it is too large, but we write its trace which is the most important item entering
the $2$-brane construction.
\begin{eqnarray}
     \mbox{Tr}\left(gY_{IJ}\right) &\equiv & \Lambda_{\mathbf{2,0,0}}^{\mathrm{D_7}(\mathrm{GF_{192}},3)} \, + \,
     \mathfrak{J}_{\mathbf{2,0,0}}^{\mathrm{D_7}(\mathrm{GF_{192}},3)}\left(X,Y,Z\right)\nonumber\\
\Lambda_{\mathbf{2,0,0}}^{\mathrm{D_7}(\mathrm{GF_{192}},3)} & = & 24\nonumber\\
\mathfrak{J}_{\mathbf{2,0,0}}^{\mathrm{D_7}(\mathrm{GF_{192}},3)}\left(X,Y,Z\right)& = &  8 \left(\cos [4 \pi  X] \left(\cos [4 \pi Y]-\cos [4 \pi  Z]\right)+\cos [4 \pi  Y]
   \cos [4 \pi  Z]\right) \label{tracciatus3}
\end{eqnarray}
Then we can verify that the function
$\mathfrak{J}_{\mathbf{2,0,0}}^{\mathrm{D_7}(\mathrm{GF_{192}},3)}\left(X,Y,Z\right)$ is invariant\footnote{It suffices
to check invariance under the three transformations $S$, $Z$ and $T$.} under the full group $\mathrm{GF_{192}}$ and
that it is an eigenfunction of $\Delta_{\mathrm{T^3}}$ with eigenvalue $\lambda \, = \, - \, 2\,\mu^2$:
\begin{equation}\label{eigenfunziaD2}
    \Delta_{\mathrm{T^3}}\, \mathfrak{J}_{\mathbf{2,0,0}}^{\mathrm{D_7}(\mathrm{GF_{192}},3)}\left(X,Y,Z\right)
    \, = \, \lambda \,
    \mathfrak{J}_{\mathbf{2,0,0}}^{\mathrm{D_7}(\mathrm{GF_{192}},3)}\left(X,Y,Z\right) \quad ; \quad \lambda \, =
    \, - \, 32\pi^2 \, = \, -\, 2\,\mu^2
\end{equation}
\subsection{A triplet of Arnold--Beltrami fields in the representation $\mathrm{D_{9}}\left[\mathrm{GF_{192}},3\right]$
 from the octahedral orbit of $ {\mathbf{k}}=\{1,1,0\}$}
 \label{k110d9}
 The third example that we consider is extracted from the orbit of length twelve
 generated by the momentum $ {\mathbf{k}}=\{1,1,0\}$. From this construction  we obtain a set of 12 one--forms that
 transform in the representation $\mathrm{D_{32}}\left[\mathrm{G_{1536}},6\right]$ of the Universal Classifying Group.
 According to \cite{arnolderie}, the branching rule of such a representation with respect to the considered subgroup $\mathrm{GF_{192}}$ is the following one:
 \begin{equation}\label{d32split}
    \mathrm{D_{32}}\left[\mathrm{G_{1536}},6\right]\, = \,
  \mathrm{ D_{9}}\left[\mathrm{GF_{192}},3\right]+\mathrm{D_{13}}\left[\mathrm{GF_{192}},3\right]
  +\mathrm{D_{19}}\left[\mathrm{GF_{192}},6\right]
 \end{equation}
Projecting onto the representation $\mathrm{ D_{9}}\left[\mathrm{GF_{192}},3\right]$ we find the following basis of
one--forms:
\begin{eqnarray}
\mathbf{Y}^I\left(X,Y,Z\right) &\in & \mathrm{ D_{9}}\left[\mathrm{GF_{192}},3\right]\nonumber\\
  \mathbf{Y}^1\left(X,Y,Z\right) &=& \sqrt{2} \Big(\left(\mathrm{d}Z-\mathrm{d}Y\right) \sin [2 \pi  (Y+Z)]-\left(\mathrm{d}Y+\mathrm{d}Z\right)
  \sin [2 \pi
   (Y-Z)]\Big)-4 \,\mathrm{d}X \sin [2 \pi  Y] \sin [2 \pi  Z] \nonumber\\
 \mathbf{Y}^2\left(X,Y,Z\right) &=& \sqrt{2} \Big(\sqrt{2} \left(\left(-\mathrm{d}Y-\mathrm{d}Z\right)
   \sin [2 \pi  (Y-Z)]+\left(\mathrm{d}Z-\mathrm{d}Y\right) \sin
   [2 \pi  (Y+Z)]\right)\nonumber\\
   && -4 \, \mathrm{d}X \sin [2 \pi Y] \sin [2 \pi  Z]\Big)
 \nonumber\\
 \mathbf{Y}^3\left(X,Y,Z\right)&=& \sqrt{2} \Big( \,2 \,\mathrm{d}Z \, (\cos [2 \pi
   (X-Y)]+\cos [2 \pi  (X+Y)])\,
    \nonumber\\
   && +\sqrt{2} ((\mathrm{d}X+\mathrm{d}Y) \sin [2
   \pi  (X-Y)]+(\mathrm{d}Y-\mathrm{d}X) \sin [2 \pi
   (X+Y)])\Big)
 \nonumber\\
 \label{fieldD9}
\end{eqnarray}
that satisfy Beltrami equation with eigenvalue $\mu\, = \, 2 \sqrt{2} \pi$:
\begin{equation}\label{autovalloTre}
    \star \, \mathrm{d} \mathbf{Y}^I \, = \, \mu \, \mathbf{Y}^I\quad ; \quad \mu \, = \, 2  \pi \,
    \sqrt{\mathbf{k} \cdot \mathbf{k} } \, = \,\, 2 \sqrt{2} \pi
\end{equation}
 The components of these one--forms are easily
extracted from:
\begin{equation}\label{compentati4Bis}
    \mathbf{Y}^I \, = \, \mathbf{Y}^I_{i} \, \mathrm{d}X^i \quad ; \quad \mathrm{d}X^i \, = \, \left\{\mathrm{d}X,\mathrm{d}Y,\mathrm{d}Z \right\}
\end{equation}
In this representation the explicit action of  the three generators $S,Z,T$ is the following one:
\begin{equation}\label{gomitolone3}
    \begin{array}{rccclcccc}
       S & : & \{X,Y,Z\} & \to & \left\{Y,X,\frac{1}{4}-Z\right\} &\Rightarrow & \mathrm{D_{9}}[S] & = & \left(
\begin{array}{lll}
 0 & -\frac{1}{2} & 0 \\
 -2 & 0 & 0 \\
 0 & 0 & -1
\end{array}
\right) \\
\null&\null&\null&\null&\null&\null&\null&\null&\null\\
       Z & : & \{X,Y,Z\} & \to & \left\{X,Y,Z+\frac{1}{2}\right\} & \Rightarrow & \mathrm{D_{9}[Z]} & = & \left(
\begin{array}{lll}
 -1 & 0 & 0 \\
 0 & -1 & 0 \\
 0 & 0 & 1
\end{array}
\right) \\
\null&\null&\null&\null&\null&\null&\null&\null&\null\\
       T & : & \{X,Y,Z\} & \to & \left\{Y+\frac{1}{4},Z+\frac{3}{4},X+\frac{1}{2}\right\} & \Rightarrow & \mathrm{D_{9}}[T]
       & = & \left(
\begin{array}{lll}
 0 & -\frac{1}{2} & 0 \\
 0 & 0 & 1 \\
 -2 & 0 & 0
\end{array}
\right)
     \end{array}
\end{equation}
and we have:
\begin{eqnarray}
\mathbf{Y}\left(Y,X,\frac{1}{4}-Z\right)  &=&  \mathbf{Y}\left(X,Y,Z\right)\cdot \mathrm{D_{9}}[S]  \nonumber\\
 \mathbf{Y}\left(X,Y,Z+\frac{1}{2}\right)&=&  \mathbf{Y}\left(X,Y,Z\right)\cdot \mathrm{D_{9}}[Z] \nonumber\\
  \mathbf{Y}\left(Y+\frac{1}{4},Z+\frac{3}{4},X+\frac{1}{2}\right)&=&  \mathbf{Y}\left(X,Y,Z\right)\cdot
  \mathrm{D_{9}}[T]\label{transformoD9}
\end{eqnarray}
Next we calculate the matrix of scalar products of the basis one-forms:
\begin{equation}\label{consapel2}
    gY^{IJ} \, \equiv \, \langle \mathbf{Y}^I \, , \, \mathbf{Y}^J \rangle \, \equiv \,  \sum_{i=1}^3\, \mathbf{Y}^I_i \mathbf{Y}^J_i
\end{equation}
We do not display this matrix since it is too large, but we write its trace which is the most important item entering
the $2$-brane construction:
\begin{eqnarray}\label{tracciatus4}
     \mbox{Tr}\left(gY_{IJ}\right) &\equiv & \Lambda_{\mathbf{1,1,0}}^{\mathrm{D_9}(\mathrm{GF_{192}},3)} \, + \,
     \mathfrak{J}_{\mathbf{1,1,0}}^{\mathrm{D_9}(\mathrm{GF_{192}},3)}\left(X,Y,Z\right)\nonumber\\
\Lambda_{\mathbf{1,1,0}}^{\mathrm{D_9}(\mathrm{GF_{192}},3)} & = & 48\nonumber\\
\mathfrak{J}_{\mathbf{1,1,0}}^{\mathrm{D_9}(\mathrm{GF_{192}},3)}\left(X,Y,Z\right)& = &  0
\end{eqnarray}
\section{D=7 two-branes with Arnold Beltrami Fluxes in the transverse directions}
\label{twobranastoria}
After the long preparation of the previous sections we can finally come to the central issue addressed in  the present
paper, namely the construction of two-brane solutions of $D=7$ minimal supergravity with Arnold-Beltrami fluxes in the
transverse space. Initially, without making direct reference to supergravity, we consider the general form of a
$p$-brane action as it is described in many places in the literature\footnote{For a concise but comprehensive summary
we refer the reader to chapter 7, Volume Two of \cite{pietrobook} and to all the papers there cited.} and we focus on
the the case $p\, = \, 2$ in $D=7$. Our first concern  is the elementary $2$-brane solution in $D=7$. We show that this
latter exists for all values of the exponential coupling parameter $a$ whose definition is recalled below. Each value
of $a$ corresponds to a different value of the dimensional reduction invariant parameter $\Delta$ whose definition is
also recalled below. Obviously $D=7$ supergravity corresponds to a unique value of $\Delta$ which can be determined by
comparing the general brane-action with the bosonic  action of minimal $D=7$ supergravity, as it was constructed in the
original papers of thirty years ago, namely in \cite{PvNT} and \cite{bershoffo1}. It turns out that for supergravity
the value of $\Delta$ is the magic one $\Delta \, = \, 4$ for which the solution becomes particularly simple and
elegant and typically preserves one half of the supersymmetries.
\par
Subsequently, on the background of the $2$-brane solution we switch on fluxes of an additional triplet of vector
fields, in this way completing  the bosonic field content of minimal $D=7$ supergravity. In presence of a topological
interaction term between the triplet of gauge fields and the $3$-form which defines the $2$-brane, we show that the
additional fluxes can be fitted into the framework of an exact solution if  they are Arnold Beltrami vector fields
satisfying Beltrami equation (\ref{formaduale}). The only conditions for the existence of such a solution is $\Delta \,
= \, 4$ plus a precise relation between the coefficients of the kinetic terms in the lagrangian and the coefficient of
the topological interaction term. Clearly we expect that such a relation should be satisfied by the coefficients of
minimal $D=7$ supergravity and by comparison with \cite{PvNT},\cite{bershoffo1}, it is shown that this is indeed the
case. Thus we arrive at the very much inspiring conclusion that $2$-branes with Arnold-Beltrami fluxes in the
transverse space are just a special feature of minimal $D=7$ supergravity. Deriving the number of Killing spinors
possessed by each solution, analyzing the relation of these latter with the discrete symmetry groups introduced by the
AB-fluxes and establishing the corresponding Maxwell-Chern-Simons gauge theories on the world volume are obviously the
next steps in a vast programme of investigations that opens up as a consequence of the result we present here.
\par
In order to accomplish this programme we need a firm control on  the Lagrangian and the supersymmetry transformation
rules of this remarkable supergravity, both in its gauged and in its ungauged versions. For this reason, within the
framework of a larger collaboration \cite{d7collabo}, we have started  a thorough reconstruction of  minimal $D=7$
supergravity in the rheonomic approach \cite{castdauriafre}\footnote{For a modernized and shorter review see also the
book \cite{pietrobook}.} and the result of this reconstruction will form the framework for our further investigation of
$2$-branes with Arnold-Beltrami fluxes.
\subsection{The general form of a $2$-brane action in $D=7$}
In the mostly minus metric that we utilize, the correct form of the
action in $D=7$ admitting an electric $2$-brane solution is the
following one:
\begin{eqnarray}
  \mathcal{A}_{2brane} &=& \int \, d^7x \, \mathcal{L}_{2brane} \nonumber\\
  \mathcal{L}_{2brane} &=&  \mbox{det} V \, \left(- R[g] \, - \, \ft 14 \,\partial^\mu \varphi \, \partial_\mu\varphi
  \, + \, \ft {1}{96} \, e^{-a\,\varphi} \, \mathbf{F}_{\lambda\mu\nu\rho}\,\mathbf{F}^{\lambda\mu\nu\rho}\right)\label{braneaction}
\end{eqnarray}
where $a$ is a free parameter, $\varphi$ denotes the dilaton field
with a canonically normalized kinetic term\footnote{Note that in the
notations adopted in this paper and in all the literature on
rheonomic supergravity the normalization of the curvature scalar and
of the Ricci tensor is one half of the normalization used in most
textbooks of General Relativity. Hence the relative normalization of
the Einstein term $R[g]$ and of the dilaton term $\partial^\mu
\varphi \, \partial_\mu\varphi$ is $\ft 14$ and not $\ft 12$.} and
\footnote{Note also that in the notations of all the literature on
rheonomic supergravity the components of the form $Q^{[p]}\,=
\,\mathrm{d} \Omega^{[p-1]}$ are defined with strength one, namely
$Q_{\lambda_1 \dots \lambda_{p}} \, = \, \ft{1}{p!} \left(
\partial_{\lambda_1} \Omega_{\lambda_2 \dots \lambda_{p}}\, + \, (p!
-1)\mbox{-terms} \right)$.}:
\begin{equation}\label{turacciolo}
    \mathbf{F}_{\lambda\mu\nu\rho} \, \equiv \, \partial_{[\lambda } \, \mathbf{A}_{\mu\nu\rho]}
\end{equation}
is the field-strength of the three-form $\mathbf{A}^{[3]}$ which
couples to the world volume of the two-brane.
\par
The field equations following from (\ref{braneaction}) can be put
into the following convenient form:
\begin{eqnarray}
  \Box_{cov} \, \varphi  &=& \frac{a}{48} e^{-a \varphi}\, \mathbf{F}_{\lambda\mu\nu\rho}\,
  \mathbf{F}^{\lambda\mu\nu\rho} \label{stdilato}\\
  \mathrm{d} \star\left[ e^{-a\varphi} \, \star\mathbf{F}^{[4]}\right] &=& 0 \label{st4form}\\
  \mbox{Ric}_{\mu\nu}&=& \frac{1}{4}\partial_\mu\varphi\, \partial_\nu\varphi \, + \, S_{\mu\nu}
  \label{stEinstein}\\
  S_{\mu\nu}  &=& - \, \frac{1}{24} e^{-a\varphi} \left(\mathbf{F}_{\mu...}
  \,\mathbf{F}_{\nu}^{\phantom{\nu}...}\, - \, \ft {3}{20} \, g_{\mu\nu} \, \mathbf{F}_{....}\,\mathbf{F}^{....}\right)
\end{eqnarray}
and they admit the following exact electric $2$-brane solution:
\begin{eqnarray}
  ds^2 &=& H(y)^{-\frac{8}{5\Delta}}\, d\xi^\mu\otimes d\xi^\nu \, - \, H(y)^{\frac{12}{5\Delta}}
  \, dy^I\otimes dy^J \, \delta_{IJ} \nonumber\\
  \phi  &=& - \frac{2a}{\Delta} \, \log \, H(y) \nonumber\\
  \mathbf{F}_{[4]} &=& \mathrm{d} \left[ H(y)^{-1} \, \frac{1}{3!} \,\mathrm{d}\xi^\mu\wedge
  \mathrm{d}\xi^\nu \wedge \mathrm{d}\xi^\rho \, \epsilon_{\mu\nu\rho}  \right] \label{branamet}
\end{eqnarray}
where, according to the main idea put forward in the introduction (see fig.\ref{metafora}) the seven coordinates have
been separated into two sets, the first set of three $\xi^\mu$ ($\mu=0,1,2$) spanning the $2$-brane world volume, the
second set of four $y^{I}$ ($I=3,4,5,6$) spanning the transverse space to the brane. In the above solution $H(y)$ is
any harmonic function living on the $4$-dimensional transverse space to the brane whose metric is assumed to be flat:
\begin{equation}\label{harmonica}
 \Box_{\mathbb{R}^4} \, H(y) \, \equiv \,   \sum_{I=1}^4\frac{\partial^2}{\partial (y^I)^2} H(y) \, = \, 0
\end{equation}
and the parameters $a$ and $\Delta$ are related by the celebrated formula\footnote{Once again compare with Chapter 7 of
book \cite{pietrobook} and consider all the references therein. In particular consider \cite{mbrastelle} where the very
definition of the index $\Delta$ was introduced.}:
\begin{equation}\label{simonbocca}
\Delta \, = \, a^2 \, + \, 2 \frac{d \, \tilde{d}}{D-2} \, = \, a^2
\, + \, \frac{12}{5}
\end{equation}
which follows from $d=3, \quad\tilde{d}\,=\,2$ and $D=7$. Physically
$d$ is the dimension of the electric $2$-brane world volume, while
$\tilde{d}$ is the dimension of the world-sheet spanned by the
magnetic string which is dual to the $2$-brane.
\par
In section \ref{compaTPvN} we will discuss the relation of the brane action (\ref{braneaction}) with the bosonic action
of minimal ungauged $D=7$ supergravity and show that the specific coefficients of the kinetic terms appearing in this
latter determine the value of $\Delta$. Indeed the supersymmetry of the action imposes $\Delta \, = \, 4$. In a future
publication \cite{d7collabo}, after the rheonomic reconstruction of the theory is completed  we will  discuss the
Killing spinors admitted by the solution (\ref{branamet}) and by its extension with Arnold-Beltrami fluxes.
\subsection{The two-brane with Arnold Beltrami Fluxes}
\label{pulcinella} As a next step we generalize the two-brane action (\ref{braneaction}) introducing also a triplet of
one-form fields $\mathbf{A}^{\Lambda}$, ($\Lambda \, = \, 1,2,3$) whose field strengths we denote $\mathbf{F}^{\Lambda}
\, \equiv \, \mathrm{d}\mathbf{A}^{\Lambda}$. In this way we complete the field-content of minimal $D=7$ supergravity.
Explicitly we write the new bosonic action:
\begin{eqnarray}
  \mathcal{A}^{flux}_{2brane} &=& \int \, d^7x \, \mathcal{L}^{flux}_{2brane} \nonumber\\
  \mathcal{L}^{flux}_{2brane} &=&  \mbox{det} V \, \left(- R[g] \, - \, \ft 14 \,\partial^\mu \varphi \, \partial_\mu\varphi
  \, + \, \ft {1}{96} \, e^{-a\,\varphi} \, \mathbf{F}_{\lambda\mu\nu\rho}\,\mathbf{F}^{\lambda\mu\nu\rho}\right.\nonumber\\
  &&\left. + \, \ft{\omega}{8}\, e^{\ft{a}{2} \varphi} \, \mathbf{F}^\Lambda_{\lambda\mu\ }\,\mathbf{F}^{\Lambda|\lambda\mu}\right)
  \, - \, \kappa \,\, \mathbf{F}_{\lambda_1\dots\lambda_4}
  \, \mathbf{F}^\Lambda_{\lambda_5\lambda_6} \, \mathbf{A}^\Lambda_{\lambda_7} \, \epsilon^{\lambda_1\dots \lambda_7}\label{fluxbraneaction}
\end{eqnarray}
where we have introduced two new real parameters $\omega$ and
$\kappa$. Crucial for the consistent insertion of fluxes is the
topological interaction term with coefficient $\kappa$.
\par
The modified field equations associated with the new action
(\ref{fluxbraneaction}) can be written in the following way:
\begin{eqnarray}
  \Box_{cov} \, \varphi  &=& \frac{a}{48} e^{-a \varphi}\, \mathbf{F}_{\lambda\mu\nu\rho}\,\mathbf{F}^{\lambda\mu\nu\rho}
  \, - \, \omega \frac{a}{8} e^{a \varphi}\, \mathbf{F}^\Lambda_{\lambda\mu}\,\mathbf{F}^{\Lambda|\lambda\mu}\label{stdilatoflux}\\
  \mathrm{d} \star\left[ e^{-a\varphi} \, \star\mathbf{F}^{[4]}\right] &=& 1152 \, \kappa \, \,\mathbf{F}^\Lambda \, \wedge \,
  \mathbf{F}^\Lambda \label{st4formflux}\\
  \mathrm{d} \star\left[ e^{\ft a 2 \varphi} \, \star\mathbf{F}^{\Lambda}\right] &=& 8 \, \frac{\kappa}{\omega} \, \,\mathbf{F}^{[4]} \,
  \wedge \, \mathbf{F}^\Lambda \label{st2formflux}\\
  \mbox{Ric}_{\mu\nu}&=& \frac{1}{4}\partial_\mu\varphi\, \partial_\nu\varphi \, + \, S^{[4]}_{\mu\nu}
  \, + \, S^{[2]}_{\mu\nu}\label{stEinsteinflux}\\
  S^{[4]}_{\mu\nu}  &=& - \, \frac{1}{24} e^{-a\varphi} \left(\mathbf{F}_{\mu...} \,\mathbf{F}_{\nu}^{\phantom{\nu}...}
  \, - \, \ft {3}{20} \, g_{\mu\nu} \, \mathbf{F}_{....}\,\mathbf{F}^{....}\right) \label{st4stressflux}\\
  S^{[2]}_{\mu\nu}  &=& - \, \omega \, \frac{1}{4} e^{\ft a 2 \varphi} \left(\mathbf{F}^\Lambda_{\mu.}
  \,\mathbf{F}_{\nu}^{\Lambda|\phantom{\nu}.}\, - \, \ft {1}{10} \, g_{\mu\nu} \, \mathbf{F}^\Lambda_{..}
  \,\mathbf{F}^{\Lambda|..}\right)\label{st2stressflux}
\end{eqnarray}
We plan to solve them with the same ansatz as we had in the previous
case for the metric, the dilaton and the $4$-form, introducing  also
a non trivial  $\mathbf{F}^\Lambda$ in the transverse space spanned
by the coordinates $y$, namely we set:
\begin{eqnarray}
  ds^2 &=& H(y)^{-\frac{8}{5\Delta}}\, d\xi^\mu\otimes d\xi^\nu \, - \, H(y)^{\frac{12}{5\Delta}}
  \, dy^I\otimes dy^J \, \delta_{IJ} \nonumber\\
  \phi  &=& - \frac{2a}{\Delta} \, \log \, H(y) \nonumber\\
  \mathbf{F}_{[4]} &=& \mathrm{d} \left[ H(y)^{-1} \, \frac{1}{3!} \,\mathrm{d}\xi^\mu\wedge \mathrm{d}\xi^\nu \wedge
   \mathrm{d}\xi^\rho \, \epsilon_{\mu\nu\rho}  \right] \nonumber\\
  \mathbf{F}^{\Lambda} &=&\mathrm{d} \left[ \mathbf{W}^\Lambda_I(y) \, dy^I \right]
  \label{branametflux}
\end{eqnarray}
The question remains, what  should we choose for the one-form fields
$\mathbf{W}^\Lambda_I(y)$ and what will be the modified differential
equation satisfied by the function $H(y)$?
\subsubsection{Arnold Beltrami vector fields on the torus $\mathrm{T}^3$ as fluxes}
The first step in order to answer the two questions posed at the end
of the previous subsection consists of a change of topology. So far
the transverse space to the two-brane was chosen flat and non
compact, namely $\mathbb{R}^4$. We maintain it flat but we
compactify three of its dimensions by identifying them with those of
a three-torus $\mathrm{T}^3$. In other words we perform the
replacement:
\begin{equation}\label{corinto1}
    \mathbb{R}^4 \, \rightarrow \, \mathbb{R}\otimes \mathrm{T}^3
\end{equation}
Secondly, on the abstract $\mathrm{T}^3$-torus we utilize the flat metric consistent with octahedral symmetry, namely
according to the setup of \cite{arnolderie} and eq.(\ref{tritorustop}) we identify:
\begin{equation}\label{torellocubico}
    \mathrm{T}^3 \, \simeq \, \frac{\mathbb{R}^3}{\Lambda_{cubic}}
\end{equation}
where $\Lambda_{cubic}$ denotes the cubic lattice, \textit{i.d.} the
abelian group of discrete translations of the euclidian
three-coordinates $\left\{ X,Y,Z\right\}$, defined below:
\begin{equation}\label{traslodiscreto}
\Lambda_{cubic} \, \ni \, \gamma_{n_1,n_2,n_3} \quad : \quad
\left\{ X,Y,Z\right\}\, \rightarrow \, \left\{
X+n_1,Y+n_2,Z+n_3\right\} \quad ; \quad n_{1,2,3} \, \in \,
\mathbb{Z}
\end{equation}
Functions on $\mathrm{T}^3$ are periodic functions of $X,Y,Z$, with
respect to the translations (\ref{traslodiscreto}).
\par
According to (\ref{corinto1}) we split the four coordinates $y^I$ as
follows:
\begin{equation}\label{kastriula}
    y^I \, = \, \left\{\underbrace{U}_{\in \,\mathbb{R}}\, ,\, \underbrace{X,Y,Z}_{\equiv \,\mathbf{X} \,\in \,\mathrm{T}^3}\right\}
\end{equation}
After these preparations we are ready to implement the ansatz anticipated in eq.(\ref{golattus}) which leads to
specialize the (\ref{branametflux}) ans\"atze in the following way:
\begin{equation}\label{corrido}
    \mathbf{F}^{\Lambda} \, = \, \mathcal{E }^\Lambda_{\phantom{\Lambda}I} \, \mathrm{d} \left[ e^{2\, \mu U}
    \, \mathbf{Y}^{I}\left(\mathbf{X}\right)  \right]
\end{equation}
where $\mathbf{Y}^{A}\left(\mathbf{X}\right)$ denotes a basis of solutions of Beltrami equation (\ref{formaduale})
pertaining to eigenvalue $\mu$ and the \textit{embedding matrix} $\mathcal{E }^\Lambda_{\phantom{\Lambda}I}$ is the
already discussed intertwining matrix (\ref{roncisvalle}).
\par
\par
Collecting all the results of the above discussion we arrive at a definite and explicit  ansatz for a $2$-brane
solution of the field equations presented in eq.s (\ref{stdilatoflux}-\ref{st2stressflux}).
\par
Explicitly we have:
\begin{eqnarray}
  ds^2 &=& H(y)^{-\frac{8}{5\Delta}}\, d\xi^\mu\otimes d\xi^\nu \, - \, H(y)^{\frac{12}{5\Delta}}
  \, dy^I\otimes dy^J \, \delta_{IJ} \nonumber\\
  \phi  &=& - \frac{2a}{\Delta} \, \log \, H(y) \nonumber\\
  \mathbf{F}_{[4]} &=& \mathrm{d} \left[ H(y)^{-1} \, \frac{1}{3!} \,\mathrm{d}\xi^\mu\wedge \mathrm{d}\xi^\nu \wedge
   \mathrm{d}\xi^\rho \, \epsilon_{\mu\nu\rho}  \right] \nonumber\\
  \mathbf{F}^{\Lambda} &=&\lambda \, \mathrm{d} \left[ e^{2\pi\,\mu \,U} \,
  \mathbf{Y}^{\Lambda}({\mathbf{X}}) \right]
\label{branametfluxBis}
\end{eqnarray}
where, relying on Schur's lemma, the embedding matrix has been reduced to $\mathcal{E }^\Lambda_{\phantom{\Lambda}I}\,
= \, \lambda \, \delta^\Lambda_I$ and and the $\mathbf{Y}^{\Lambda}({\mathbf{X}})$  are a triplet of Arnold-Beltrami one
forms satisfying Beltrami equation with eigenvalue $\mu$:
\begin{equation}\label{beltramino}
    \star \, \mathrm{d}\mathbf{Y}^{\Lambda} \, = \, \mu \, \mathbf{Y}^{\Lambda}
\end{equation}
and transforming in a three-dimensional irreducible\footnote{If the representation of $\Gamma$ is reducible, according to Schur's lemma we will have $\mathcal{E }^\Lambda_{\phantom{\Lambda}I}\, =\, \left(\begin{array}{c|c|c}
                                                  \lambda_1 &0 & 0 \\
                                                  \hline
                                                  0 & \lambda_2 & 0 \\
                                                  \hline
                                                  0& 0&\lambda_3
                                                \end{array}\right) $ or $\mathcal{E }^\Lambda_{\phantom{\Lambda}I}\, =\, \left(\begin{array}{c|c}
                                                  \lambda_1 &0  \\
                                                  \hline
                                                  0 & \lambda_2  \times \mathbf{1}_{\mathrm{2\times2}}\\
                                                \end{array}\right) $.}  representation of some subgroup $\Gamma \subset
\mathrm{G_{1536}}$ of the Universal Classifying Group. For instance,  $\mathbf{Y}^{\Lambda}$ can be one of the triplets
discussed in sections \ref{k100gf192}, \ref{k100gs24},\ref{k110d9},\ref{k200d7}.
\par
Inserting the ansatz (\ref{branametfluxBis}) into the field eq.s (\ref{stdilatoflux}-\ref{st2stressflux}) we reach the
following conclusion. \textit{If and only if} the following two conditions on the lagrangian parameters are fulfilled:
\begin{eqnarray}
  \Delta &=& 4 \nonumber \\
 \sqrt{\Delta } \, \omega \, - \, 768\,
   \kappa &=& 0  \label{consistenza}
\end{eqnarray}
then all field equations are identically satisfied provided the function  $H(y)$ obeys the following  differential
equation:
\begin{eqnarray}\label{kalinus1}
  \Box_{\mathbb{R} \times \mathrm{T}^3} H(U,\mathbf{X}) & = & \, - \, \frac{\lambda^2}{24} \, \mu^2 \, e^{2\,\mu \, U} \, \left(\Lambda^{\mathrm{D}(\Gamma)}_{\mathbf{k}}\, + \,\mathfrak{J} ^{\mathrm{D}(\Gamma)}_{\mathbf{k}}(\mathbf{X}) \right) \end{eqnarray}
  where we have introduced the notation
  \begin{equation}
   \Box_{\mathbb{R} \times \mathrm{T}^3} \, \equiv \,
  \left(\frac{\partial^2}{\partial U^2} \, + \, \Box_{\mathrm{T}^3}\right) \label{mixedlaplat}
   \end{equation}
and where:
\begin{eqnarray}\label{corleonis}
    \Lambda^{\mathrm{D}(\Gamma)}_{\mathbf{k}}\, + \,\mathfrak{J} ^{\mathrm{D}(\Gamma)}_{\mathbf{k}}(\mathbf{X}) & \equiv & \mbox{Tr} \, {gY}(\mathbf{X}) \nonumber\\
    {gY}^{\Lambda\Sigma }(\mathbf{X}) & =& \langle \mathbf{Y}_{\mathbf{k}}^\Lambda \, , \, \mathbf{Y}_{\mathbf{k}}^\Sigma \rangle \, = \, \sum_{i=1}^3 \mathbf{Y}_i ^\Lambda(\mathbf{X}) \, \mathbf{Y}_i^\Sigma(\mathbf{X})
\end{eqnarray}
In eq.(\ref{corleonis}) we have separated, in the trace of the scalar product matrix ${gY}^{\Lambda\Sigma }$ a constant part named $\Lambda^{\mathrm{D}(\Gamma)}_{\mathbf{k}}$ from a point-dependent part $\mathfrak{J} ^{\mathrm{D}(\Gamma)}_{\mathbf{k}}(\mathbf{X})$ which in the examples presented in this paper happens to satisfy the following equation:
\begin{equation}\label{kalinus2}
    \Box_{\mathrm{T}^3}\, \mathfrak{J} ^{\mathrm{D}(\Gamma)}_{\mathbf{k}}(\mathbf{X})
     \, = \, - \, 2 \, \mu^2 \, \mathfrak{J} ^{\mathrm{D}(\Gamma)}_{\mathbf{k}}(\mathbf{X})
\end{equation}
namely $\mathfrak{J} ^{\mathrm{D}(\Gamma)}_{\mathbf{k}}(\mathbf{X})$ is an eigenfunction of the Laplace operator on the torus with the above specified eigenvalue. Furthermore, 
in the above equations the symbol $\mathrm{D}(\Gamma)$ refers to the representation of the subgroup $\Gamma \subset \mathrm{G_{1536}}$ that  is utilized to define the triplet of Arnold-Beltrami fields and ${\mathbf{k}}$ refers to the momentum orbit in the dual lattice from which these latter are constructed according to the discussion of section \ref{algoritmo}.
When eq. (\ref{kalinus2}) holds true, which is not always the case, a significant simplification occurs in the solution of the inhomogeneous equation (\ref{kalinus1}). Indeed under this condition we are enabled to write a nice compact formula for the a particular solution of (\ref{kalinus1}) which yields a function $H(y)$ endowed with the appropriate  boundary condition for asymptotic flatness of the metric and  invariant, by construction, under the discrete group $\Gamma$. Indeed if we  set:
\begin{equation}\label{genesolut}
    H(U,\mathbf{X}) \, = \, 1 \, - \, \frac{\lambda^2}{96} \, e^{2\, \mu \, U} \, \left ( \Lambda^{\mathrm{D}(\Gamma)}_{\mathbf{k}} \, + \, 2 \, \mathfrak{J}^{\mathrm{D}(\Gamma)}_{\mathbf{k}}(\mathbf{X})\right)
\end{equation}
eq.(\ref{kalinus1}) is satisfied and for $U\to -\infty$ the metric in (\ref{branametfluxBis}) tends to the flat metric.
\par
 As we said,  eq.(\ref{genesolut}) contains only a particular solution of the inhomogeneous eq.(\ref{kalinus1}). To this particular solution, in principle, we might add the general solution of the harmonic homogeneous equation $\Box_{\mathbb{R} \times \mathrm{T}^3} \, H_0(U,\mathbf{X}) \, = \,0$. The general form of such harmonic function is 
 \begin{equation}
 H_0(U,\mathbf{X})\, =
 \, \sum_{\mu, d_\mu} \, a_{\mu,d_\mu} \, \exp[\sqrt{2}\mu\,U] \,J^{d_\mu}_\mu(\mathbf{X})
 \end{equation} 
  where $J^{d_\mu}_\mu(\mathbf{X})$ is an eigenfunction of the Laplace operator on the $T^3$ torus pertaining to the eigenvalue $-2\mu^2$ just as in eq.(\ref{kalinus2}) (the index $d_\mu$ spans the eigenspace that has always some degeneracy). Yet, given the full spectrum of the operator $\Box_{\mathrm{T}^3}$ we should restrict the coefficients $a_{\mu,d_\mu}$ in such a way as to obtain a harmonic function that is invariant under the considered discrete group $\Gamma$ and this typically involves an extensive analysis which can be done only case by case. At the moment we skip this analysis since for the purposes of the present paper we can restrict ourselves to the simple particular solution (\ref{genesolut}).
\par
Let us further observe that the function $\Lambda^{\mathrm{D}(\Gamma)}_{\mathbf{k}} \, + \, 2 \, \mathfrak{J}^{\mathrm{D}(\Gamma)}_{\mathbf{k}}(\mathbf{X})$ which is $\Gamma$ invariant is always a limited function taking values in a finite interval:
\begin{eqnarray}\label{candeladisego}
    N_- \, \le &\Lambda^{\mathrm{D}(\Gamma)}_{\mathbf{k}} \, + \, 2 \, \mathfrak{J}^{\mathrm{D}(\Gamma)}_{\mathbf{k}}(\mathbf{X})& \le \, N_+\label{arianna}\\
\end{eqnarray}
Hence if we choose the parameter $\lambda$ as follows:
\begin{equation}\label{normalusco}
    \lambda \, = \, \sqrt{\frac{96}{N_+}}
\end{equation}
we obtain:
\begin{equation}\label{genesolutB}
    H(U,\mathbf{X}) \, = \, 1 \, - \, \frac{1}{N_+} \, e^{2\, \mu \, U} \, \left ( \Lambda^{\mathrm{D}(\Gamma)}_{\mathbf{k}} \, + \, 2 \, \mathfrak{J}^{\mathrm{D}(\Gamma)}_{\mathbf{k}}(\mathbf{X})\right)
\end{equation}
which is positive definite in the interval $U\in \left[ -\infty \, , \, 0\right]$ and for $U=0$ has a zero in the $T^3$
locus where $\mathfrak{J}^{\mathrm{D}(\Gamma)}_{\mathbf{k}}(\mathbf{X})$ attains its maximal value $m_+ \, = \, N_+ \,
+ \, \Lambda^{\mathrm{D}(\Gamma)}_{\mathbf{k}}$. This observation concludes our general discussion of $2$-branes with
Arnold Beltrami fluxes. Before examining the details of the four examples provided by the  Arnold-Beltrami triplets
introduced in sections \ref{k100gf192},\ref{k100gs24},\ref{k200d7},\ref{k110d9} let us stress that  in section
\ref{compaTPvN} by means of comparison with the action of minimal $D=7$ supergravity constructed by Townsend and van
Nieuwenhuizen in \cite{PvNT} we show that the relation (\ref{consistenza}) is indeed satisfied by the coefficients of
the supergravity lagrangian. Hence we arrive at the quite relevant conclusion that differently from pure $2$-branes
that can be solutions of any Lagrangian of the same type, not necessarily supersymmetric, the flux two-branes with
Arnold Beltrami fluxes are solutions only of supergravity.
\subsubsection{The $2$-brane with Arnold Beltrami fluxes in the representation $\mathrm{D_{12}(GF_{192},3)}$}
\label{soluzia1} This example is obtained from the use of the Arnold-Beltrami one-forms discussed in section
\ref{k100gf192}. From equation (\ref{tracciatus}) we know that $\Lambda^{\mathrm{D_{12}(GF_{192},3)
}}_{\mathbf{1,0,0}}\, = \, 12$ and $\mathfrak{J}^{\mathrm{D_{12}(GF_{192},3) }}_{\mathbf{1,0,0}}\, = \, 0$: hence
conclude that:
\begin{eqnarray}
\label{fighettus1}
  \lambda &=& 2\sqrt{2} \\
  H(U,\mathbf{X}) &=& 1\, - \, e^{4\, \pi \, U}
\end{eqnarray}
Correspondingly we obtain the very simple solution for all the bosonic fields:
\begin{eqnarray}
    ds^2 & = & \frac{d\xi^\mu\otimes d\xi^\nu \, \eta_{\mu\nu} \, - \, \left(1-e^{
   4 \pi  U}\right)
   \left(dU^2+dX^2+dY^2+dZ^2\right)}{\left(1-e^{4 \pi
   U}\right)^{2/5}}\label{metrica1} \\
   \varphi & = & -\sqrt{\frac{2}{5}} \log \left(1-e^{4
   \pi  U}\right)\label{dilatone1}\\
   \mathbf{F}_{[4]} & = & -\, 16 \, \pi  \, \frac{ e^{4 \pi  U}  }{\left(1-e^{4 \pi  U}\right)^2}\, d\xi^\mu\wedge
   d\xi^\nu\wedge d\xi^\rho \, \epsilon_{\mu\nu\rho} \,\wedge
   dU\label{fforma4}\\
   \mathbf{F}_{[2]}^1&=&8 \sqrt{2} e^{2 \pi  U} \pi  \left(\sin [2
   \pi  Z] ({dX}\wedge
   {dZ}-{dU}\wedge
   {dY})+\cos [2 \pi  Z]
   ({dU}\wedge
   {dX}+{dY}\wedge
   {dZ})\right)\label{fforma21}\\
    \mathbf{F}_{[2]}^2&=&8 \sqrt{2} e^{2 \pi  U} \pi  \left (\sin [2
   \pi  Y] ( {dU}\wedge
    {dZ}+ {dX}\wedge
    {dY})+\cos [2 \pi  Y]
   ( {dU}\wedge
    {dX}+ {dY}\wedge
    {dZ})\right)\label{fforma22}\\
    \mathbf{F}_{[2]}^3&=&8 \sqrt{2} e^{2 \pi  U} \pi  \left(\cos [2
   \pi  X] ( {dU}\wedge
    {dY}- {dX}\wedge
    {dZ})-\sin [2 \pi  X]
   ( {dU}\wedge
    {dZ}+ {dX}\wedge
    {dY})\right)\label{fforma23}\\
\end{eqnarray}

\subsubsection{The $2$-brane with Arnold Beltrami fluxes in the representation $\mathrm{D_{1}(GS_{24},1)}$}
\label{soluzia2} 
\begin{figure}[!hbt]
\begin{center}
\iffigs
\includegraphics[height=70mm]{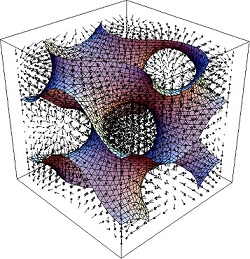}
\else
\end{center}
 \fi
\caption{\it  A  visualization of the function $\mathfrak{J}(X,Y,Z)$ for the $2$-brane solution which is invariant with respect
to discrete group $\mathrm{GS_{24}}$  is provided by plotting equal level surfaces of the function, namely the two dimensional surfaces defined by $\mathfrak{J}(X,Y,Z) \, = \, \ell $. In the same box we plot also the gradient field $\nabla\mathfrak{J}(X,Y,Z)$, namely the vector field orthogonal to the level surfaces. The two information combined provide a sort of visualizations of the function.} \label{JJGS24}
 \iffigs
 \hskip 1cm \unitlength=1.1mm
 \end{center}
  \fi
\end{figure}

This example is obtained from the use of the Arnold-Beltrami one-forms discussed in section
\ref{k100gs24}. From equation (\ref{lamgs24}) we know that $\Lambda^{\mathrm{D_{1}(GS_{24},1) }}_{\mathbf{1,0,0}}\, =
\, 12$ and $\mathfrak{J}^{\mathrm{D_{1}(GS_{24},1) }}_{\mathbf{1,0,0}}\, \ne \, 0$. The minimum and the maximum of this
function are $\pm 12$:
\begin{equation}\label{maxmin}
    - \, 12 \, \le \, \mathfrak{J}^{\mathrm{D_{1}(GS_{24},1) }}_{\mathbf{1,0,0}}(\mathbf{X}) \, \le \, 12
\end{equation}
Hence conclude that:
\begin{eqnarray}
\label{fighettus3}
  \lambda &=& \sqrt{\frac{8}{3}} \\
  H(U,\mathbf{X}) &=& 1\, - \, \frac{1}{36} \, e^{4\, \pi \, U}\left(12 \, + \, 2 \, \mathfrak{J}^{\mathrm{D_{1}(GS_{24},1) }}_{\mathbf{1,0,0}}(\mathbf{X}) \right)
\end{eqnarray}
In this case the explicit form of the solution  has no longer the same simplicity as in the previous case. Therefore it is better to write it
more implicitly as it follows:
\begin{eqnarray}
  ds^2 &=&  H(U,\mathbf{X})^{-\frac{2}{5}}\, d\xi^\mu\otimes d\xi^\nu \, \eta_{\mu\nu}\,  - \, H(U,\mathbf{X})^{\frac{3}{5}}
  \, \left(dU^2\,+ \, dX^2 \, + \, dY^2 \, + \, dZ^2 \right) \nonumber\\
  \varphi  &=& - \sqrt{\frac{2}{5}} \, \log \, H(U,\mathbf{X})\nonumber\\
  \mathbf{F}_{[4]} &=& \mathrm{d} \left[ H(U,\mathbf{X})^{-1} \, \frac{1}{3!} \,\mathrm{d}\xi^\mu\wedge
  \mathrm{d}\xi^\nu \wedge \mathrm{d}\xi^\rho \, \epsilon_{\mu\nu\rho}  \right]\nonumber\\
  \mathbf{F}_{[2]}^1 & = & \mathrm{d}\left[ 4 \sqrt{\frac{2}{3}} e^{2 \pi  U}\Bigl ( \cos [2 \pi  Y] dX+\cos (2 \pi  Z)
   dX  \right. \nonumber\\
   && \left.+\cos [2 \pi  X] \, dY\, -\, dZ \, \sin[2 \pi  X]+dZ \sin [2 \pi  Y]-dY \,\sin [2 \pi  Z] \Bigr) \right]\nonumber\\
   \mathbf{F}_{[2]}^2 & = & 0 \nonumber\\
    \mathbf{F}_{[2]}^3 & = & 0
   \label{branosaGS24}
\end{eqnarray}
The entire analytic structure of this brane solution is encoded in the function $\mathfrak{J}^{\mathrm{D_{1}(GS_{24},1) }}$. Being a function of three-variables it is difficult to visualize its behavior. One possibility is provided by the contour plots which are visualized
in fig. \ref{JJGS24} .
\subsubsection{The $2$-brane with Arnold Beltrami fluxes in the representation $\mathrm{D_{7}(GF_{192},3)}$}
\label{soluzia3} 
\begin{figure}[!hbt]
\begin{center}
\iffigs
\includegraphics[height=60mm]{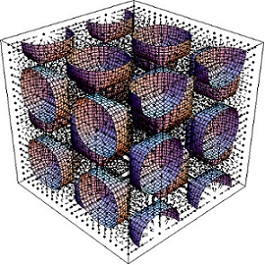}
\else
\end{center}
 \fi
\caption{\it  A  visualization of the function $\mathfrak{J}(X,Y,Z)$ for the $2$-brane  invariant with respect to
discrete group $\mathrm{GF_{192}}$ and assigned to the representation $\mathrm{D_7}$  is provided by plotting equal
level surfaces of the function, namely the two dimensional surfaces defined by $\mathfrak{J}(X,Y,Z) \, = \, \ell $. In
the same box we plot also the gradient field $\nabla\mathfrak{J}(X,Y,Z)$, namely the vector field orthogonal to the
level surfaces. The two information combined provide a sort of visualizations of the function.} \label{JJD7GF192}
 \iffigs
 \hskip 1cm \unitlength=1.1mm
 \end{center}
  \fi
\end{figure}

This example is obtained from the use of the Arnold-Beltrami one-forms discussed in section
\ref{k200d7}. From equation (\ref{tracciatus3}) we know that $\Lambda^\mathrm{D_{7}(GF_{192},3)}_{\mathbf{2,0,0}}\, =
\, 24$ and $\mathfrak{J}^\mathrm{D_{7}(GF_{192},3)}_{\mathbf{2,0,0}}\, \ne \, 0$. The minimum and the maximum of this
function are displayed below :
\begin{equation}\label{maxmink2}
    - \, 24 \, \le \, \mathfrak{J}^\mathrm{D_{7}(GF_{192},3)}_{\mathbf{2,0,0}}(\mathbf{X}) \, \le \, 8
\end{equation}
Hence conclude that:
\begin{eqnarray}
\label{fighettus2}
  \lambda &=& \sqrt{\frac{12}{5}} \\
  H(U,\mathbf{X}) &=& 1\, - \, \frac{1}{40} \, e^{4\, \pi \, U}\left(24 \, + \, 2 \, \mathfrak{J}^\mathrm{D_{7}(GF_{192},3)}_{\mathbf{2,0,0}}(\mathbf{X}) \right)
\end{eqnarray}
Once again the explicit form of the solution  is not simply looking. Therefore it is better to write it implicitly as it follows:
\begin{eqnarray}
  ds^2 &=&  H(U,\mathbf{X})^{-\frac{2}{5}}\, d\xi^\mu\otimes d\xi^\nu \, \eta_{\mu\nu} \, - \, H(U,\mathbf{X})^{\frac{3}{5}}
  \, \left(dU^2\,+ \, dX^2 \, + \, dY^2 \, + \, dZ^2 \right) \nonumber\\
  \varphi  &=& - \sqrt{\frac{2}{5}} \, \log \, H(U,\mathbf{X})\nonumber\\
  \mathbf{F}_{[4]} &=& \mathrm{d} \left[ H(U,\mathbf{X})^{-1} \, \frac{1}{3!} \,\mathrm{d}\xi^\mu\wedge
  \mathrm{d}\xi^\nu \wedge \mathrm{d}\xi^\rho \, \epsilon_{\mu\nu\rho}  \right]\nonumber\\
  \mathbf{F}^1_{[2]} &=& \sqrt{\frac{12}{5}}\,\mathrm{d} \left[e^{2\, \pi \, U}\,\left( 2 (\cos [4 \pi  Y]+\cos [4 \pi  Z]) \mathrm{d}X+2 \mathrm{d}Z \sin [4 \pi  Y]
  -2 \mathrm{d}Y \sin [4 \pi  Z]\right)\right] \nonumber\\
 \mathbf{F}^2_{[2]} &=& \sqrt{\frac{12}{5}}\,\mathrm{d} \left[e^{2\, \pi \, U}\,\left( 2 (\cos [4 \pi  X]-\cos [4 \pi  Z]) \mathrm{d}Y
 -2 (\mathrm{d}Z \sin [4 \pi  X]+\mathrm{d}X \sin [4 \pi  Z])\right)\right]\nonumber\\
 \mathbf{F}^3_{[3]}&=& \sqrt{\frac{12}{5}}\,\mathrm{d} \left[e^{2\, \pi \, U}\,\left( 2 (\cos [4 \pi  X]+\cos [4 \pi  Y]) \mathrm{d}Z
 +2 \mathrm{d}Y \sin [4 \pi  X]-2 \mathrm{d}X \sin [4 \pi  Y]\right)\right] \label{branosaGF192d7}
\end{eqnarray}
As in the previous case the entire analytic structure of this brane solution is encoded in the function
$\mathfrak{J}^\mathrm{D_{7}(GF_{192},3)}_{\mathbf{2,0,0}}$. A visualization of this function is provided  in fig.
\ref{JJD7GF192} .
\par
We do not dwell on the fourth example of section \ref{k110d9} since the feature
$\mathfrak{J}^{\mathrm{D_9}(\mathrm{GF_{192})}}_{\mathbf{1,1,0}}(X,Y,Z)\,= \,0$ yields a $2$-brane solution with
exactly the same structure (apart from numerical factors) as the $2$-brane solution of section \ref{soluzia1}.
\section{Comparison with the bosonic action of Minimal $D=7$ supergravity according to
the \textit{TPvN} construction.} \label{compaTPvN} As promised above,  in this appendix we make a comparison between
the action (\ref{fluxbraneaction}) and the bosonic action of minimal $D=7$ Supergravity as it was derived in
\cite{PvNT}, which, for brevity we name \textit{TPvN}. The goal is that of verifying whether the constraint
(\ref{consistenza}) is verified by the coefficients in the \textit{TPvN} action.
\par
Since the authors of \cite{PvNT} use the Dutch conventions for tensor calculus with imaginary time, the comparison of
the lagrangians at the level of signs is difficult, yet at the level of absolute values of the coefficients it is
possible, by means of several rescalings. First we observe that the normalization of the Einstein term in eq.(2) of
\textit{TPvN} is the same, if we take into account the already stressed $\ft 12$ difference in the definition of the
Ricci tensor and scalar curvature. Secondly we note the normalization of the dilaton kinetic term in eq.(2) of
\textit{TPvN}, namely $\ft 12$ becomes that of the action (\ref{fluxbraneaction}), namely $\ft 14$ if we define:
\begin{equation}\label{subillus1}
    \phi_{TPvN} \, = \, \ft{1}{\sqrt{2}} \, \varphi
\end{equation}
A check that this is the correct identification arises from
inspection of the dilaton factor in front of the three-form kinetic
term. Using eq.(3) of \textit{TPvN}, we see that according to this
construction such a factor is:
\begin{equation}\label{dilfactus}
    \exp \left[- \, \ft{4}{\sqrt{5}} \, \phi_{TPvN} \right] \, = \, \exp \left[- 2 \,
    \ft{2}{\sqrt{5}} \, \varphi \right]
\end{equation}
This confirms the value $a \, = -\, 2 \, \ft{2}{\sqrt{5}}$ leading
to the miraculous value $\Delta \, = \, 4$ of the dimensional
reduction invariant. Secondly we consider the necessary rescalings
for the $\mathbf{A}^{[3]}$ and $\mathbf{A}^{\Lambda}$ gauge fields.
Taking into account the different strengths of the exterior
derivatives (see unnumbered eq.s of \cite{PvNT} in between eq.(1)
and (2)) we see that in order to match the normalizations of
(\ref{fluxbraneaction}) we have to define:
\begin{eqnarray}
  {A}^{TPvN}_{\lambda\mu\nu} &=& \ft {1}{4\sqrt{2}} \, \mathbf{A}^{[3]}_{\lambda\mu\nu}
  \quad \Rightarrow \quad F^{TPvN}_{\lambda\mu\nu\rho} \, = \, \ft {1}{\sqrt{2}}
  \mathbf{F}_{\lambda\mu\nu\rho} \nonumber\\
  {A}^{\Lambda|TPvN}_\mu &=& \sqrt{\ft {\omega}{8}} \, \mathbf{A}^{\Lambda}_\mu
  \quad \Rightarrow \quad F^{\Lambda|TPvN}_{\lambda\mu} \, = \, \sqrt{\ft {\omega}{2} } \,\mathbf{F}^\Lambda_{\lambda\mu}
  \label{minotaurotto}
\end{eqnarray}
with these redefinitions  we can calculate the value of $\kappa$
according to \textit{TPvN}. We find:
\begin{equation}\label{gelsomino}
    \ft{1}{48 \, \sqrt{2}} \, F^{TPvN}_{\mu\nu\rho\sigma}F^{\Lambda|TPvN}_{\kappa\lambda} A^{\Lambda|TPvN}_\tau \,
    \epsilon^{\mu\nu\rho\sigma\lambda\kappa\tau} \, = \, \ft{\omega}{384} \,
    \mathbf{F}_{\mu\nu\rho\sigma}\mathbf{F}^{\Lambda}_{\kappa\lambda} \mathbf{A}^{\Lambda}_\tau \,
    \epsilon^{\mu\nu\rho\sigma\lambda\kappa\tau}
\end{equation}
which implies:
\begin{equation}\label{conservadipomodoro}
    \kappa \, = \, \ft{\omega}{384}
\end{equation}
The above values satisfies the consistency condition (\ref{consistenza}), when $\Delta=4$, which has already been
verified above. This shows that Arnold Beltrami flux branes are solutions of minimal $D=7$ supergravity and of no other
theory of the same type which is not supersymmetric.
\par
As we stressed in previous pages the next important step is the derivation of Killing spinors and the analysis of
preserved supersymmetries. We postpone this task until we have reconstructed the entire theory within the  rheonomy
framework \cite{d7collabo}.
\section{Conclusions}
\label{concludone} The motivations of this \textit{Sentimental Journey from Hydrodynamics to Supergravity} have been
extensively discussed in the introduction and will not be repeated here. The hidden link between Beltrami equation and
supersymmetry was not suspected: now it has become a matter of fact. The main exciting consequence of this unveiling,
already outlined in our introduction, is the injection of a vast variety of discrete symmetries into the brane-world.
Such a richness of symmetries and the link with supersymmetry have to be exploited. We plane to exploit them as soon as
the parallel work on $D=7$ supergravity will be finished. For the moment we can just compile a list of tasks to be
accomplished which constitute our agenda for the nearest future:
\begin{enumerate}
  \item Study of the Killing spinor equation and derivation of the supersymmetries preserved by each Arnold--Beltrami
  flux--brane.
  \item Study the field content of the world-volume gauge theory $\mathcal{GT}_3$ associated with each brane.
  Describe the transmission of the discrete symmetries to $\mathcal{GT}_3$.
  \item Study the topological twist of $\mathcal{GT}_3$ and possibly calculate its partition function\cite{conantonio}.
  \item Construct the $\kappa$-supersymmetric world-volume action of these branes.
  \item Explore aspects of the Gauge/Gravity correspondence in this new setup.
\end{enumerate}
The \textit{Sentimental Journey} has just started and we hope it can continue and contribute to new understanding.
\section*{Aknowledgements}
During the completion of this work we had a few very important and clarifying discussions with our colleagues and
friends, L. Andrianopoli, L. Castellani, R. D'Auria, S. Ferrara, P.A. Grassi, A. Sagnotti and M. Trigiante. We thank them warmheartedly. The
work of A.S. was partially supported by the RFBR Grants No. 16-52-12012 -NNIO-a, No.  15-52-05022-Arm-a and by the DFG Grant LE 838/12-2.
\newpage
\appendix
\section{The Group $\mathrm{GF_{192}}$ and its subgroup $\mathrm{GS_{24}}$ } \label{coniugatoGF192}
\label{DNAofGF192}
 In this section we list all the elements of the space group $\mathrm{GF_{192}}$, organized into
their $20$ conjugacy classes and of its subgroup $\mathrm{GS_{24}}$ which is isomorphic to the octahedral group
$\mathrm{O_{24}}$.
\subsection{$\mathrm{GF_{192}}$}

\paragraph{Conjugacy class $\mathcal{C}_{1}\left(\mathrm{GF_{192}}\right)$: $\#$ of elements = $1$}
\begin{equation}\label{GF192classe1}
\begin{array}{l}
 \left\{1_1,0,0,0\right\}
\end{array}
\end{equation}
\paragraph{Conjugacy class $\mathcal{C}_{2}\left(\mathrm{GF_{192}}\right)$: $\#$ of elements = $1$}
\begin{equation}\label{GF192classe2}
\begin{array}{l}
 \left\{1_1,1,1,1\right\}
\end{array}
\end{equation}
\paragraph{Conjugacy class $\mathcal{C}_{3}\left(\mathrm{GF_{192}}\right)$: $\#$ of elements = $3$}
\begin{equation}\label{GF192classe3}
\begin{array}{lll}
 \left\{1_1,0,0,1\right\} & \left\{1_1,0,1,0\right\} &
   \left\{1_1,1,0,0\right\}
\end{array}
\end{equation}
\paragraph{Conjugacy class $\mathcal{C}_{4}\left(\mathrm{GF_{192}}\right)$: $\#$ of elements = $3$}
\begin{equation}\label{GF192classe4}
\begin{array}{lll}
 \left\{1_1,0,1,1\right\} & \left\{1_1,1,0,1\right\} &
   \left\{1_1,1,1,0\right\}
\end{array}
\end{equation}
\paragraph{Conjugacy class $\mathcal{C}_{5}\left(\mathrm{GF_{192}}\right)$: $\#$ of elements = $3$}
\begin{equation}\label{GF192classe5}
\begin{array}{lll}
 \left\{3_1,0,0,1\right\} & \left\{3_2,1,1,0\right\} &
   \left\{3_3,1,1,1\right\}
\end{array}
\end{equation}
\paragraph{Conjugacy class $\mathcal{C}_{6}\left(\mathrm{GF_{192}}\right)$: $\#$ of elements = $3$}
\begin{equation}\label{GF192classe6}
\begin{array}{lll}
 \left\{3_1,0,0,0\right\} & \left\{3_2,1,0,0\right\} &
   \left\{3_3,0,1,1\right\}
\end{array}
\end{equation}
\paragraph{Conjugacy class $\mathcal{C}_{7}\left(\mathrm{GF_{192}}\right)$: $\#$ of elements = $3$}
\begin{equation}\label{GF192classe7}
\begin{array}{lll}
 \left\{3_1,1,1,1\right\} & \left\{3_2,0,1,1\right\} &
   \left\{3_3,1,0,0\right\}
\end{array}
\end{equation}
\paragraph{Conjugacy class $\mathcal{C}_{8}\left(\mathrm{GF_{192}}\right)$: $\#$ of elements = $3$}
\begin{equation}\label{GF192classe8}
\begin{array}{lll}
 \left\{3_1,1,1,0\right\} & \left\{3_2,0,0,1\right\} &
   \left\{3_3,0,0,0\right\}
\end{array}
\end{equation}
\paragraph{Conjugacy class $\mathcal{C}_{9}\left(\mathrm{GF_{192}}\right)$: $\#$ of elements = $6$}
\begin{equation}\label{GF192classe9}
\begin{array}{lll}
 \left\{3_1,0,1,1\right\} & \left\{3_1,1,0,1\right\} &
   \left\{3_2,0,1,0\right\} \\
 \left\{3_2,1,1,1\right\} & \left\{3_3,1,0,1\right\} &
   \left\{3_3,1,1,0\right\}
\end{array}
\end{equation}
\paragraph{Conjugacy class $\mathcal{C}_{10}\left(\mathrm{GF_{192}}\right)$: $\#$ of elements = $6$}
\begin{equation}\label{GF192classe10}
\begin{array}{lll}
 \left\{3_1,0,1,0\right\} & \left\{3_1,1,0,0\right\} &
   \left\{3_2,0,0,0\right\} \\
 \left\{3_2,1,0,1\right\} & \left\{3_3,0,0,1\right\} &
   \left\{3_3,0,1,0\right\}
\end{array}
\end{equation}
\paragraph{Conjugacy class $\mathcal{C}_{11}\left(\mathrm{GF_{192}}\right)$: $\#$ of elements = $12$}
\begin{equation}\label{GF192classe11}
\begin{array}{llll}
 \left\{4_1,\frac{1}{2},0,0\right\} & \left\{4_1,\frac{1}{2},1,1\right\} &
   \left\{4_2,\frac{3}{2},0,0\right\} & \left\{4_2,\frac{3}{2},1,1\right\} \\
 \left\{4_3,0,0,\frac{1}{2}\right\} & \left\{4_3,1,1,\frac{1}{2}\right\} &
   \left\{4_4,\frac{1}{2},\frac{3}{2},\frac{1}{2}\right\} &
   \left\{4_4,\frac{3}{2},\frac{3}{2},\frac{3}{2}\right\} \\
 \left\{4_5,\frac{1}{2},\frac{1}{2},\frac{3}{2}\right\} &
   \left\{4_5,\frac{3}{2},\frac{1}{2},\frac{1}{2}\right\} &
   \left\{4_6,0,0,\frac{3}{2}\right\} & \left\{4_6,1,1,\frac{3}{2}\right\}
\end{array}
\end{equation}
\paragraph{Conjugacy class $\mathcal{C}_{12}\left(\mathrm{GF_{192}}\right)$: $\#$ of elements = $12$}
\begin{equation}\label{GF192classe12}
\begin{array}{llll}
 \left\{4_1,\frac{1}{2},0,1\right\} & \left\{4_1,\frac{1}{2},1,0\right\} &
   \left\{4_2,\frac{3}{2},0,1\right\} & \left\{4_2,\frac{3}{2},1,0\right\} \\
 \left\{4_3,0,1,\frac{1}{2}\right\} & \left\{4_3,1,0,\frac{1}{2}\right\} &
   \left\{4_4,\frac{1}{2},\frac{3}{2},\frac{3}{2}\right\} &
   \left\{4_4,\frac{3}{2},\frac{3}{2},\frac{1}{2}\right\} \\
 \left\{4_5,\frac{1}{2},\frac{1}{2},\frac{1}{2}\right\} &
   \left\{4_5,\frac{3}{2},\frac{1}{2},\frac{3}{2}\right\} &
   \left\{4_6,0,1,\frac{3}{2}\right\} & \left\{4_6,1,0,\frac{3}{2}\right\}
\end{array}
\end{equation}
\paragraph{Conjugacy class $\mathcal{C}_{13}\left(\mathrm{GF_{192}}\right)$: $\#$ of elements = $12$}
\begin{equation}\label{GF192classe13}
\begin{array}{llll}
 \left\{4_1,\frac{3}{2},0,0\right\} & \left\{4_1,\frac{3}{2},1,1\right\} &
   \left\{4_2,\frac{1}{2},0,0\right\} & \left\{4_2,\frac{1}{2},1,1\right\} \\
 \left\{4_3,0,0,\frac{3}{2}\right\} & \left\{4_3,1,1,\frac{3}{2}\right\} &
   \left\{4_4,\frac{1}{2},\frac{1}{2},\frac{1}{2}\right\} &
   \left\{4_4,\frac{3}{2},\frac{1}{2},\frac{3}{2}\right\} \\
 \left\{4_5,\frac{1}{2},\frac{3}{2},\frac{3}{2}\right\} &
   \left\{4_5,\frac{3}{2},\frac{3}{2},\frac{1}{2}\right\} &
   \left\{4_6,0,0,\frac{1}{2}\right\} & \left\{4_6,1,1,\frac{1}{2}\right\}
\end{array}
\end{equation}
\paragraph{Conjugacy class $\mathcal{C}_{14}\left(\mathrm{GF_{192}}\right)$: $\#$ of elements = $12$}
\begin{equation}\label{GF192classe14}
\begin{array}{llll}
 \left\{4_1,\frac{3}{2},0,1\right\} & \left\{4_1,\frac{3}{2},1,0\right\} &
   \left\{4_2,\frac{1}{2},0,1\right\} & \left\{4_2,\frac{1}{2},1,0\right\} \\
 \left\{4_3,0,1,\frac{3}{2}\right\} & \left\{4_3,1,0,\frac{3}{2}\right\} &
   \left\{4_4,\frac{1}{2},\frac{1}{2},\frac{3}{2}\right\} &
   \left\{4_4,\frac{3}{2},\frac{1}{2},\frac{1}{2}\right\} \\
 \left\{4_5,\frac{1}{2},\frac{3}{2},\frac{1}{2}\right\} &
   \left\{4_5,\frac{3}{2},\frac{3}{2},\frac{3}{2}\right\} &
   \left\{4_6,0,1,\frac{1}{2}\right\} & \left\{4_6,1,0,\frac{1}{2}\right\}
\end{array}
\end{equation}
\paragraph{Conjugacy class $\mathcal{C}_{15}\left(\mathrm{GF_{192}}\right)$: $\#$ of elements = $12$}
\begin{equation}\label{GF192classe15}
\begin{array}{llll}
 \left\{5_1,0,0,\frac{1}{2}\right\} & \left\{5_1,1,1,\frac{1}{2}\right\} &
   \left\{5_2,\frac{1}{2},\frac{3}{2},\frac{3}{2}\right\} &
   \left\{5_2,\frac{3}{2},\frac{3}{2},\frac{1}{2}\right\} \\
 \left\{5_3,\frac{1}{2},\frac{1}{2},\frac{1}{2}\right\} &
   \left\{5_3,\frac{3}{2},\frac{1}{2},\frac{3}{2}\right\} &
   \left\{5_4,0,0,\frac{3}{2}\right\} & \left\{5_4,1,1,\frac{3}{2}\right\} \\
 \left\{5_5,\frac{1}{2},0,1\right\} & \left\{5_5,\frac{1}{2},1,0\right\} &
   \left\{5_6,\frac{3}{2},0,1\right\} & \left\{5_6,\frac{3}{2},1,0\right\}
\end{array}
\end{equation}
\paragraph{Conjugacy class $\mathcal{C}_{16}\left(\mathrm{GF_{192}}\right)$: $\#$ of elements = $12$}
\begin{equation}\label{GF192classe16}
\begin{array}{llll}
 \left\{5_1,0,0,\frac{3}{2}\right\} & \left\{5_1,1,1,\frac{3}{2}\right\} &
   \left\{5_2,\frac{1}{2},\frac{1}{2},\frac{3}{2}\right\} &
   \left\{5_2,\frac{3}{2},\frac{1}{2},\frac{1}{2}\right\} \\
 \left\{5_3,\frac{1}{2},\frac{3}{2},\frac{1}{2}\right\} &
   \left\{5_3,\frac{3}{2},\frac{3}{2},\frac{3}{2}\right\} &
   \left\{5_4,0,0,\frac{1}{2}\right\} & \left\{5_4,1,1,\frac{1}{2}\right\} \\
 \left\{5_5,\frac{3}{2},0,1\right\} & \left\{5_5,\frac{3}{2},1,0\right\} &
   \left\{5_6,\frac{1}{2},0,1\right\} & \left\{5_6,\frac{1}{2},1,0\right\}
\end{array}
\end{equation}
\paragraph{Conjugacy class $\mathcal{C}_{17}\left(\mathrm{GF_{192}}\right)$: $\#$ of elements = $12$}
\begin{equation}\label{GF192classe17}
\begin{array}{llll}
 \left\{5_1,0,1,\frac{1}{2}\right\} & \left\{5_1,1,0,\frac{1}{2}\right\} &
   \left\{5_2,\frac{1}{2},\frac{3}{2},\frac{1}{2}\right\} &
   \left\{5_2,\frac{3}{2},\frac{3}{2},\frac{3}{2}\right\} \\
 \left\{5_3,\frac{1}{2},\frac{1}{2},\frac{3}{2}\right\} &
   \left\{5_3,\frac{3}{2},\frac{1}{2},\frac{1}{2}\right\} &
   \left\{5_4,0,1,\frac{3}{2}\right\} & \left\{5_4,1,0,\frac{3}{2}\right\} \\
 \left\{5_5,\frac{1}{2},0,0\right\} & \left\{5_5,\frac{1}{2},1,1\right\} &
   \left\{5_6,\frac{3}{2},0,0\right\} & \left\{5_6,\frac{3}{2},1,1\right\}
\end{array}
\end{equation}
\paragraph{Conjugacy class $\mathcal{C}_{18}\left(\mathrm{GF_{192}}\right)$: $\#$ of elements = $12$}
\begin{equation}\label{GF192classe18}
\begin{array}{llll}
 \left\{5_1,0,1,\frac{3}{2}\right\} & \left\{5_1,1,0,\frac{3}{2}\right\} &
   \left\{5_2,\frac{1}{2},\frac{1}{2},\frac{1}{2}\right\} &
   \left\{5_2,\frac{3}{2},\frac{1}{2},\frac{3}{2}\right\} \\
 \left\{5_3,\frac{1}{2},\frac{3}{2},\frac{3}{2}\right\} &
   \left\{5_3,\frac{3}{2},\frac{3}{2},\frac{1}{2}\right\} &
   \left\{5_4,0,1,\frac{1}{2}\right\} & \left\{5_4,1,0,\frac{1}{2}\right\} \\
 \left\{5_5,\frac{3}{2},0,0\right\} & \left\{5_5,\frac{3}{2},1,1\right\} &
   \left\{5_6,\frac{1}{2},0,0\right\} & \left\{5_6,\frac{1}{2},1,1\right\}
\end{array}
\end{equation}
\paragraph{Conjugacy class $\mathcal{C}_{19}\left(\mathrm{GF_{192}}\right)$: $\#$ of elements = $32$}
{\scriptsize
\begin{equation}\label{GF192classe19}
\begin{array}{llllllll}
 \left\{2_1,\frac{1}{2},\frac{1}{2},0\right\} &
   \left\{2_1,\frac{1}{2},\frac{3}{2},1\right\} &
   \left\{2_1,\frac{3}{2},\frac{1}{2},1\right\} &
   \left\{2_1,\frac{3}{2},\frac{3}{2},0\right\} &
   \left\{2_2,\frac{1}{2},\frac{1}{2},0\right\} &
   \left\{2_2,\frac{1}{2},\frac{3}{2},1\right\} &
   \left\{2_2,\frac{3}{2},\frac{1}{2},1\right\} &
   \left\{2_2,\frac{3}{2},\frac{3}{2},0\right\} \\
 \left\{2_3,0,\frac{1}{2},\frac{3}{2}\right\} &
   \left\{2_3,0,\frac{3}{2},\frac{1}{2}\right\} &
   \left\{2_3,1,\frac{1}{2},\frac{1}{2}\right\} &
   \left\{2_3,1,\frac{3}{2},\frac{3}{2}\right\} &
   \left\{2_4,0,\frac{1}{2},\frac{1}{2}\right\} &
   \left\{2_4,0,\frac{3}{2},\frac{3}{2}\right\} &
   \left\{2_4,1,\frac{1}{2},\frac{3}{2}\right\} &
   \left\{2_4,1,\frac{3}{2},\frac{1}{2}\right\} \\
 \left\{2_5,0,\frac{1}{2},\frac{1}{2}\right\} &
   \left\{2_5,0,\frac{3}{2},\frac{3}{2}\right\} &
   \left\{2_5,1,\frac{1}{2},\frac{3}{2}\right\} &
   \left\{2_5,1,\frac{3}{2},\frac{1}{2}\right\} &
   \left\{2_6,0,\frac{1}{2},\frac{3}{2}\right\} &
   \left\{2_6,0,\frac{3}{2},\frac{1}{2}\right\} &
   \left\{2_6,1,\frac{1}{2},\frac{1}{2}\right\} &
   \left\{2_6,1,\frac{3}{2},\frac{3}{2}\right\} \\
 \left\{2_7,\frac{1}{2},\frac{1}{2},1\right\} &
   \left\{2_7,\frac{1}{2},\frac{3}{2},0\right\} &
   \left\{2_7,\frac{3}{2},\frac{1}{2},0\right\} &
   \left\{2_7,\frac{3}{2},\frac{3}{2},1\right\} &
   \left\{2_8,\frac{1}{2},\frac{1}{2},1\right\} &
   \left\{2_8,\frac{1}{2},\frac{3}{2},0\right\} &
   \left\{2_8,\frac{3}{2},\frac{1}{2},0\right\} &
   \left\{2_8,\frac{3}{2},\frac{3}{2},1\right\}
\end{array}
\end{equation}
}
\paragraph{Conjugacy class $\mathcal{C}_{20}\left(\mathrm{GF_{192}}\right)$: $\#$ of elements = $32$}
{\scriptsize
\begin{equation}\label{GF192classe20}
\begin{array}{llllllll}
 \left\{2_1,\frac{1}{2},\frac{1}{2},1\right\} &
   \left\{2_1,\frac{1}{2},\frac{3}{2},0\right\} &
   \left\{2_1,\frac{3}{2},\frac{1}{2},0\right\} &
   \left\{2_1,\frac{3}{2},\frac{3}{2},1\right\} &
   \left\{2_2,\frac{1}{2},\frac{1}{2},1\right\} &
   \left\{2_2,\frac{1}{2},\frac{3}{2},0\right\} &
   \left\{2_2,\frac{3}{2},\frac{1}{2},0\right\} &
   \left\{2_2,\frac{3}{2},\frac{3}{2},1\right\} \\
 \left\{2_3,0,\frac{1}{2},\frac{1}{2}\right\} &
   \left\{2_3,0,\frac{3}{2},\frac{3}{2}\right\} &
   \left\{2_3,1,\frac{1}{2},\frac{3}{2}\right\} &
   \left\{2_3,1,\frac{3}{2},\frac{1}{2}\right\} &
   \left\{2_4,0,\frac{1}{2},\frac{3}{2}\right\} &
   \left\{2_4,0,\frac{3}{2},\frac{1}{2}\right\} &
   \left\{2_4,1,\frac{1}{2},\frac{1}{2}\right\} &
   \left\{2_4,1,\frac{3}{2},\frac{3}{2}\right\} \\
 \left\{2_5,0,\frac{1}{2},\frac{3}{2}\right\} &
   \left\{2_5,0,\frac{3}{2},\frac{1}{2}\right\} &
   \left\{2_5,1,\frac{1}{2},\frac{1}{2}\right\} &
   \left\{2_5,1,\frac{3}{2},\frac{3}{2}\right\} &
   \left\{2_6,0,\frac{1}{2},\frac{1}{2}\right\} &
   \left\{2_6,0,\frac{3}{2},\frac{3}{2}\right\} &
   \left\{2_6,1,\frac{1}{2},\frac{3}{2}\right\} &
   \left\{2_6,1,\frac{3}{2},\frac{1}{2}\right\} \\
 \left\{2_7,\frac{1}{2},\frac{1}{2},0\right\} &
   \left\{2_7,\frac{1}{2},\frac{3}{2},1\right\} &
   \left\{2_7,\frac{3}{2},\frac{1}{2},1\right\} &
   \left\{2_7,\frac{3}{2},\frac{3}{2},0\right\} &
   \left\{2_8,\frac{1}{2},\frac{1}{2},0\right\} &
   \left\{2_8,\frac{1}{2},\frac{3}{2},1\right\} &
   \left\{2_8,\frac{3}{2},\frac{1}{2},1\right\} &
   \left\{2_8,\frac{3}{2},\frac{3}{2},0\right\}
\end{array}
\end{equation}
}
\subsection{The Group $\mathrm{GS_{24}}$}
\label{coniugatoGS24}
 In this section we list all the elements of the space group $\mathrm{GS_{24}}$, organized into
their $5$ conjugacy classes that, in this case are arranged according to the order which is customary in
crystallography for the proper octahedral  group.
\paragraph{Conjugacy class $\mathcal{C}_{1}\left(\mathrm{GS_{24}}\right)$: $\#$ of elements = $1$}
\begin{equation}\label{GS24classe1}
\begin{array}{l}
 \left\{1_1,0,0,0\right\}
\end{array}
\end{equation}
\paragraph{Conjugacy class $\mathcal{C}_{2}\left(\mathrm{GS_{24}}\right)$: $\#$ of elements = $8$}
\begin{equation}\label{GS24classe2}
\begin{array}{llll}
 \left\{2_1,\frac{3}{2},\frac{1}{2},1\right\} &
   \left\{2_2,\frac{1}{2},\frac{3}{2},1\right\} &
   \left\{2_3,1,\frac{1}{2},\frac{1}{2}\right\} &
   \left\{2_4,0,\frac{3}{2},\frac{3}{2}\right\} \\
 \left\{2_5,1,\frac{3}{2},\frac{1}{2}\right\} &
   \left\{2_6,0,\frac{1}{2},\frac{3}{2}\right\} &
   \left\{2_7,\frac{1}{2},\frac{3}{2},0\right\} &
   \left\{2_8,\frac{3}{2},\frac{1}{2},0\right\}
\end{array}
\end{equation}
\paragraph{Conjugacy class $\mathcal{C}_{3}\left(\mathrm{GS_{24}}\right)$: $\#$ of elements = $3$}
\begin{equation}\label{GS24classe3}
\begin{array}{lll}
 \left\{3_1,1,1,1\right\} & \left\{3_2,0,1,1\right\} &
   \left\{3_3,1,0,0\right\}
\end{array}
\end{equation}
\paragraph{Conjugacy class $\mathcal{C}_{4}\left(\mathrm{GS_{24}}\right)$: $\#$ of elements = $6$}
\begin{equation}\label{GS24classe4}
\begin{array}{lll}
 \left\{4_1,\frac{1}{2},1,1\right\} &
   \left\{4_2,\frac{3}{2},1,1\right\} &
   \left\{4_3,1,1,\frac{1}{2}\right\} \\
 \left\{4_4,\frac{1}{2},\frac{3}{2},\frac{1}{2}\right\} &
   \left\{4_5,\frac{3}{2},\frac{1}{2},\frac{1}{2}\right\}
   & \left\{4_6,0,0,\frac{3}{2}\right\}
\end{array}
\end{equation}
\paragraph{Conjugacy class $\mathcal{C}_{5}\left(\mathrm{GS_{24}}\right)$: $\#$ of elements = $6$}
\begin{equation}\label{GS24classe5}
\begin{array}{lll}
 \left\{5_1,0,1,\frac{3}{2}\right\} &
   \left\{5_2,\frac{3}{2},\frac{1}{2},\frac{3}{2}\right\}
   &
   \left\{5_3,\frac{1}{2},\frac{3}{2},\frac{3}{2}\right\}
   \\
 \left\{5_4,1,0,\frac{1}{2}\right\} &
   \left\{5_5,\frac{3}{2},0,0\right\} &
   \left\{5_6,\frac{1}{2},0,0\right\}
\end{array}
\end{equation}

\end{document}